\PassOptionsToPackage{table}{xcolor}
\documentclass[lettersize,journal]{IEEEtran}
\usepackage{amsmath,amsfonts}
\usepackage{algorithmic}
\usepackage{algorithm}
\usepackage{array}
\usepackage{caption} 
\usepackage[caption=false,font=normalsize,labelfont=sf,textfont=sf]{subfig}
\usepackage{textcomp}
\usepackage{stfloats}
\usepackage{url}
\usepackage{verbatim}
\usepackage{graphicx}
\usepackage{lipsum}
\usepackage{enumitem} 
\usepackage{svg}
\usepackage{float}
\usepackage{soul} 
\usepackage{multirow}
\usepackage{booktabs}
\newcommand{\cmark}{\ding{51}} 
\newcommand{\xmark}{\ding{55}} 
\definecolor{lightgray}{gray}{0.9}
\usepackage{pifont} 
\usepackage{threeparttable}
\usepackage{cite}
\usepackage{cleveref}
\usepackage{epstopdf}

\begin{document}

\title{VLMaterial: Vision-Language Model-Based Camera-Radar Fusion for Physics-Grounded Material Identification}

\author{Jiangyou Zhu and He Chen
        
\thanks{The authors are with the Department of Inforation Engineering, The Chinese University of Hong Kong, Hong Kong SAR, China (e-mail: zj124@ie.cuhk.edu.hk; he.chen@ie.cuhk.edu.hk).}}



\maketitle

\begin{abstract}
Accurate material recognition is a fundamental capability for intelligent perception systems to interact safely and effectively with the physical world. For instance, in domestic service scenarios, distinguishing between visually similar objects, such as a glass cup versus a plastic one, is critical for safety but remains a significant challenge for vision-based methods due to specular reflections, transparency, and visual deception. While millimeter-wave (mmWave) radar offers robust material sensing capabilities impervious to lighting conditions, existing camera-radar fusion approaches are often limited to closed-set categories and lack semantic interpretability. In this paper, we introduce VLMaterial, a training-free framework that fuses vision-language models (VLMs) with domain-specific radar knowledge for physics-grounded material identification. First, we propose a dual-pipeline architecture: an optical pipeline uses the segment anything model and VLM for material candidate proposals, while an electromagnetic characterization pipeline extracts the intrinsic dielectric constant from radar signals via an effective peak reflection cell area (PRCA) method and weighted vector synthesis. Second, we employ a context-augmented generation (CAG) strategy to equip the VLM with radar-specific physical knowledge, enabling it to interpret electromagnetic parameters as stable references. Third, an adaptive fusion mechanism is introduced to intelligently integrate outputs from both sensors by resolving cross-modal conflicts based on uncertainty estimation. We evaluated VLMaterial in over 120 real-world experiments involving 41 diverse everyday objects and 4 typical visually deceptive counterfeits across varying environments. Experimental results demonstrate that VLMaterial achieves a recognition accuracy of 96.08\%, delivering performance on par with state-of-the-art closed-set benchmarks while eliminating the need for extensive task-specific data collection and training.
\end{abstract}


\section{Introduction}
\label{introduction}
\IEEEPARstart{O}{bject} recognition is a fundamental capability in human-centered assistive domains and core technology for autonomous systems in both household environments and industrial facilities \cite{zou2023object}. 
Driven by artificial intelligence (AI) advances, vision-based object recognition has improved significantly, with vision-language models (VLMs) advancing both object identification and material composition \cite{zhang2024vision, sapkota2025object}. 
Despite their impressive recognition performance, VLMs are inherently constrained by their reliance on RGB images, whose imaging principle encodes only geometric and chromatic cues, making them susceptible to misidentifying visually identical objects.
To illustrate this, consider a common household scenario where a system integrated with a VLM attempts to heat a cup of milk. The system fail to discriminate between a glass cup and a plastic one, particularly when they share an identical appearance.  Such misidentification could present a serious safety risk, as only the glass cup is microwave-safe. 
Motivated by this challenge, we conducted a preliminary study to evaluate the recognition capabilities of state-of-the-art (SOTA) VLMs (see \Cref{Feasibility} for details). Our observations confirm that these models excel at category recognition such as cups, bottles, boxes, etc., but struggle with material identification when relying solely on visual data. This finding naturally leads us to ask: \textit{Can we introduce a supplementary sensing modality to support vision in uncovering latent physical properties?}

Nowadays, a wide range of sensors for material recognition have been explored, including infrared emissivity \cite{li2021intelligent, wani2022lowlight}, acoustic \cite{sun2023akte}, tactile \cite{liu2024material}, specialized optical such as time-of-flight (ToF) cameras \cite{ yang2024object} and event cameras \cite{kim2021n}, achieving strong benchmark performance \cite{perez2024beyond, duan2022soda} with recent computer vision advances further improving material recognition \cite{gao2024physically}, and a diverse set of radio frequency (RF)-based sensing technologies also enables material identification.
For example, longer wavelengths are utilized by technologies like radio frequency identification (RFID) \cite{xie2019tagtag}, Wi-Fi \cite{yan2024wi}, Bluetooth, and long range radio (LoRa). Conversely, broad bandwidths, characteristic of technologies such as ultra-wideband (UWB) \cite{zheng2021siwa,yang2025nlos}, millimeter wave (mmWave) \cite{chen2025material,zhu2025can}, and Terahertz (THz) \cite{yazdnian2025robotera}. 
However, each modality has distinct drawbacks: infrared is weather-sensitive, acoustic is noise-prone, tactile risks surface damage, and optical is constrained by lighting conditions, while traditional RF trades penetration for resolution. In contrast, mmWave radar \cite{yeo2016radarcat, wu2020msense} delivers both high resolution and long range with all-weather capability, yet still lacks object geometric information.

Therefore, multimodal fusion arises naturally \cite{cheng2023intelligent}, with camera-radar fusion standing out for its complementary strengths in appearance and physical characteristics \cite{shuai2021millieye, deng2025fuselid}. 
To date, systems such as MilliEye \cite{shuai2021millieye}, RODNet \cite{wang2021rodnet}, Craft \cite{kim2023craft} and CRFUSION \cite{xiao2025crfusion} focus on object detection, all using neural networks for black-box \cite{aghababaeyan2023black} processing of radar reflection data (e.g., range-doppler map \cite{chen2025material}, channel sample vector \cite{skaria2022machine}). These approaches are confined to closed-set categories and simple geometries, requiring training.  
From a training-free view, LLMaterial \cite{zhu2025can} shows large language model (LLM)-based radar material identification is feasible but constrained by controlled environments, lacks 60 GHz dielectric references in its retrieval-augmented generation (RAG) \cite{gao2023retrieval} knowledge base. 
After extensive investigation, we summarize material identification sensors and representative works in \Cref{Comparison_all_papers}.



\begin{table*}[!t]
  \centering
  \caption{Comparison of Object Recognition Modalities.}
  \label{Comparison_all_papers}
  \resizebox{\textwidth}{!}{%
  \begin{tabular}{l|l|c|c|c|c|c}
    \toprule
    \textbf{Work} & \textbf{Modality} & \textbf{Contactless} & \textbf{Category} & \textbf{Material} & \textbf{C. \& M.} & \textbf{Training-free} \\
    \midrule
    Li et al. \cite{li2021intelligent} & \multirow{2}{*}{Infrared} & \cmark & \xmark & \cmark & \xmark & \xmark\\ 
    Wani et al. \cite{wani2022lowlight} &    & \cmark & \cmark & \xmark & \xmark & \xmark\\
    \midrule
    Akte-Liquid \cite{sun2023akte} & Acoustic & \cmark & \xmark & \cmark & \xmark & \xmark\\
    \midrule
    Liu et al. \cite{liu2024material} & Tactile & \xmark & \xmark & \cmark & \xmark & \xmark\\
    \midrule
    Yang et al. \cite{yang2024object} & \multirow{3}{*}{Camera} & \cmark & \xmark & \cmark & \xmark & \xmark\\
    Beyond Appearances \cite{perez2024beyond} &    & \cmark & \xmark & \cmark & \xmark & \xmark\\
    PG-InstructBLIP \cite{gao2024physically} &    & \cmark & \cmark & \cmark & \cmark & \xmark\\
    \midrule
    Xie et al. \cite{xie2019tagtag} (RFID) & \multirow{6}{*}{RF-based} & \cmark & \xmark & \cmark & \xmark & \xmark\\ 
    Wi-painter \cite{yan2024wi} (WiFi) &  & \cmark & \xmark & \cmark & \xmark & \xmark\\
    SiWa \cite{zheng2021siwa} (UWB) &    & \cmark & \xmark & \cmark & \xmark & \xmark\\ 
    mSense \cite{wu2020msense} (mmWave) &    & \cmark & \xmark & \cmark & \xmark & \xmark\\ 
    LLMaterial \cite{zhu2025can} (mmWave) &    & \cmark & \xmark & \cmark & \xmark & \cmark\\ 
    Robotera \cite{yazdnian2025robotera} (THz) &    & \cmark & \xmark & \cmark & \xmark & \xmark\\ 
    \midrule
    millieye \cite{shuai2021millieye} & \multirow{5}{*}{Camera \& Radar} & \cmark & \cmark & \xmark & \xmark & \xmark\\ 
    RODNet \cite{wang2021rodnet} &    & \cmark & \cmark & \xmark & \xmark & \xmark\\
    Craft \cite{kim2023craft} &    & \cmark & \cmark & \xmark & \xmark & \xmark\\
    CRFUSION \cite{xiao2025crfusion} &    & \cmark & \cmark & \cmark & \cmark & \xmark\\
    \rowcolor{gray!20}
    \textbf{VLMaterial (ours)}  &    & \cmark & \cmark & \cmark & \cmark & \cmark\\
    \bottomrule
  \end{tabular}%
  }
  \captionsetup{justification=raggedright,singlelinecheck=false}
  \caption*{\footnotesize Note: C. \& M. means category and material.}
\end{table*}

Motivated by the advanced capabilities of VLMs, this leads to our central research question: \textit{Can a VLM substantially enhance material recognition by synergizing visual and radar modalities to infer electromagnetic properties, thereby enabling accurate, physics-grounded identification?} Answering this question is non-trivial as it requires addressing both VLM limitations and radar-derived characteristics. First, VLMs lack radar-specific knowledge in their pre-training, requiring raw radar data to be transformed into a VLM-understandable representation. Second, raw radar signals are high-dimensional and redundant, requiring identification of electromagnetic characteristics most indicative of intrinsic material properties.



In this paper, we present VLMaterial, a new framework fuses advanced VLMs with domain-specific knowledge for physics-grounded camera-radar material identification. VLMaterial integrates an optical characterization pipeline for RGB processing, with a complementary radar sensing stream to capture intrinsic physical properties. Our contributions are:
\begin{itemize}[leftmargin=0pt, itemsep=0pt, topsep=0pt]
    \item We propose an electromagnetic characterization pipeline that transforms raw radar signals into compact representations of intrinsic material properties by explicitly leveraging the dielectric constant, an inherent property unaffected by geometry. Specifically, we introduce the peak reflection cell area (PRCA) to overcome the geometric dependence of radar cross section (RCS) measurements. Building on this, we first isolate the target reflection via distance gating. The signal is then enhanced using beamforming and finally extracted through a weighted vector synthesis approach. 
    \item We design an adaptive fusion mechanism to integrate visual and radar modalities for material identification. This mechanism incorporates an intersection strategy for consistent predictions and an uncertainty-aware adaptive conflict resolution mechanism to address conflict cases. This decision-level strategy dynamically balances both modalities by calculating confidence scores for visual and radar inputs based on factors that affect their signal quality.
    \item We perform extensive validation on 41 everyday objects and 4 visually deceptive counterfeit items across 6 categories and 7 common materials. Our results demonstrate that VLMaterial achieves a recognition accuracy of 96.08\%. This performance not only surpasses the 96\% accuracy of the SOTA closed-set model, CRFUSION, but also dramatically outperforms the training-free baseline LLMaterial, which achieves an accuracy of only 19.69\%. Furthermore, VLMaterial exceeds the capabilities of unimodal baselines, specifically VLM-only with 78.74\% accuracy and radar-only with 41.73\% accuracy, demonstrating its superior performance.
\end{itemize}

\begin{figure*}[!t]
  \centering
  \includegraphics[width=\textwidth]{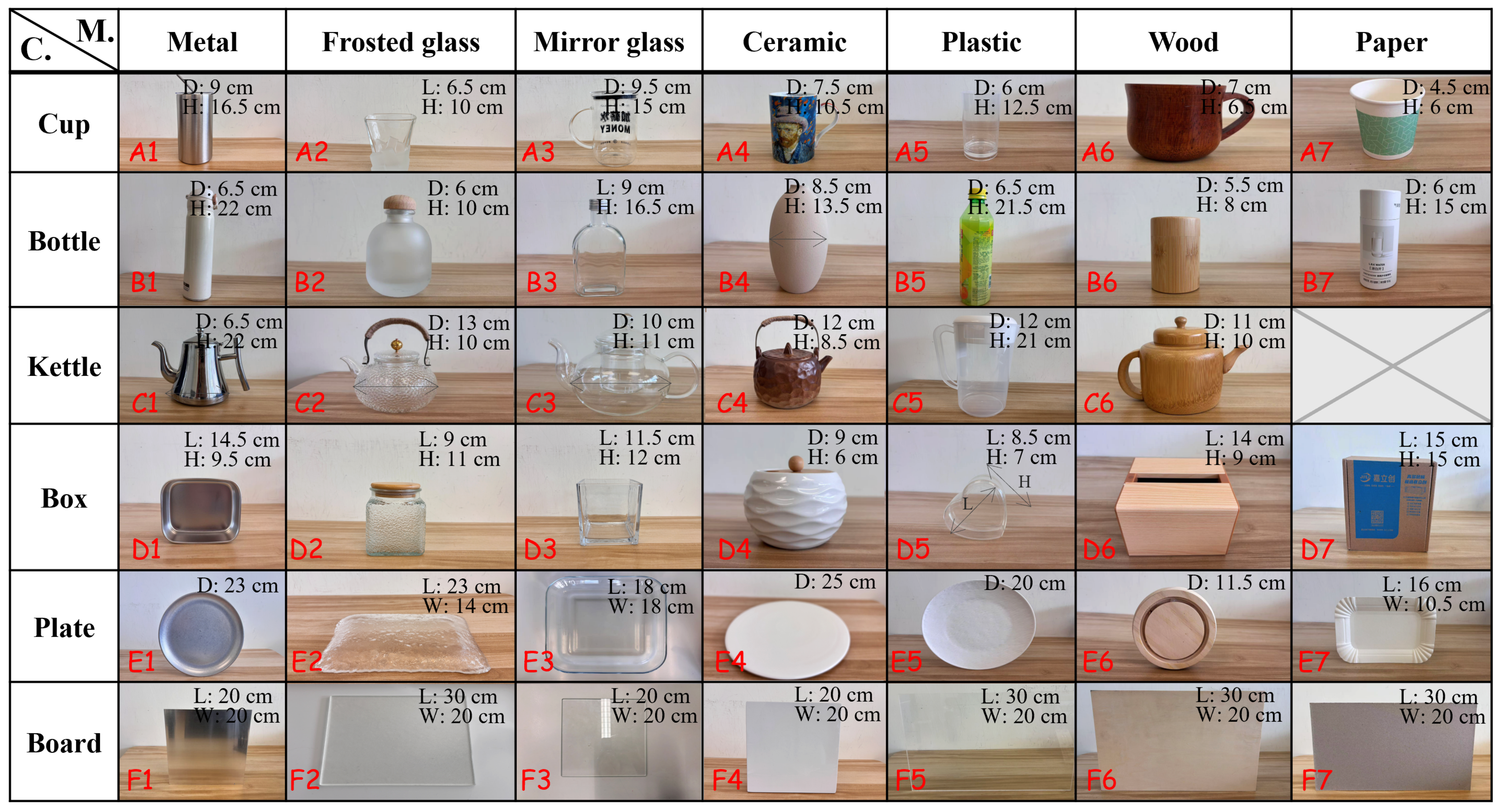}
  \caption{All typical daily objects: 41 items across 6 materials and 7 categories (the paper kettle was excluded due to its rarity). For clarity, images are annotated with dimensions (where L, W, H, and D represent Length, Width, Height, and Diameter, respectively) to provide scale context, with index number marked in the bottom-left (size not provided to the VLM).}
  \label{all_targets}
\end{figure*}

\section{Preliminary Study and Challenges}
\label{motivation}

To investigate the feasibility of VLMs in material identification, we randomly collected 41 everyday objects covering six materials and seven categories, as shown in \Cref{all_targets}.

\subsection{Evaluation of VLM's Capability in Material Identification}
\label{Feasibility}


\begin{figure*}[!t]
  \centering
  \includegraphics[width=0.98\textwidth]{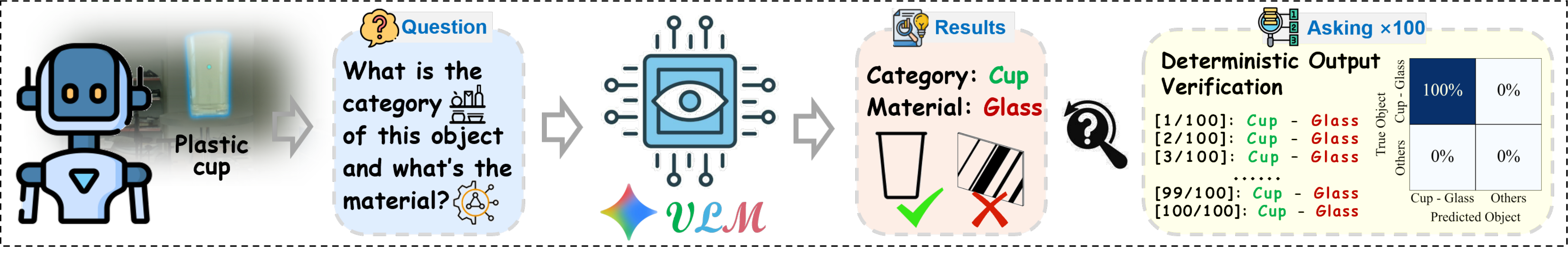}
  \caption{Validating inference stability via deterministic output verification.}
  \label{veri_100times}
\end{figure*}

We first employ segment anything model (SAM) \cite{kirillov2023segment} to isolate the target object within each RGB image. Subsequently, the image is processed by VLM prompted with \textit{“What is this object and what is the material?”}, inferring the object category and listing potential material candidates. 
For robustness against single-trial inference randomness, the VLM was queried 100 times for each object. As illustrated in \Cref{veri_100times}, the VLM demonstrated deterministic yet erroneous behavior. For instance, it consistently misclassified a transparent plastic cup as glass across all 100 trials. This deterministic error is visualized in the confusion matrix at the end of \Cref{veri_100times}, confirming that the VLM's predictions are stable and reproducible.


\begin{figure*}[!t]
  \centering
  \includegraphics[width=\linewidth]{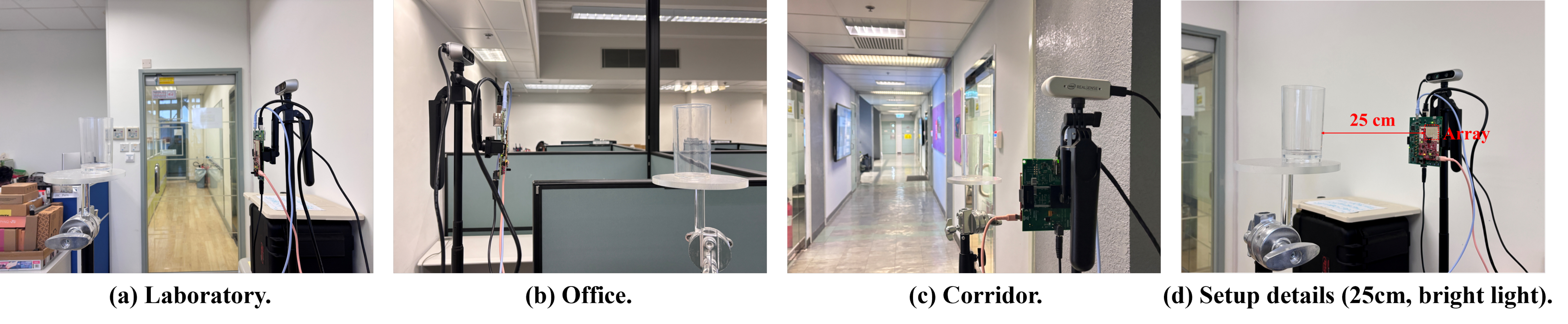}
  \caption{Experimental setups for data collection in different scenarios and detailed equipment configuration.}
    \label{three_scenarios}
\end{figure*}

Then, we evaluated all 41 objects using Gemini-3-Pro, GPT-5, and Qwen3-VL. In a laboratory scenario, all models achieved 100\% object categorization accuracy but struggled with material identification: Gemini-3-Pro reached 75.61\%, followed by GPT-5 (73.17\%) and Qwen3-VL (68.29\%).
In short, current VLMs recognize object categories well but lack accurate material perception. 
In the following, we employ the top-performing Gemini-3-Pro for the following experiments.

Unlike existing studies that simplify setups or isolate objects in controlled spaces, we maintained a realistic environment by collecting data across three distinct scenarios: a laboratory (\Cref{three_scenarios}a), an office (\Cref{three_scenarios}b), and a corridor (\Cref{three_scenarios}c).

\begin{figure*}[!t]
  \centering
  \subfloat[Cups.]{
    \includegraphics[width=0.14\textwidth]{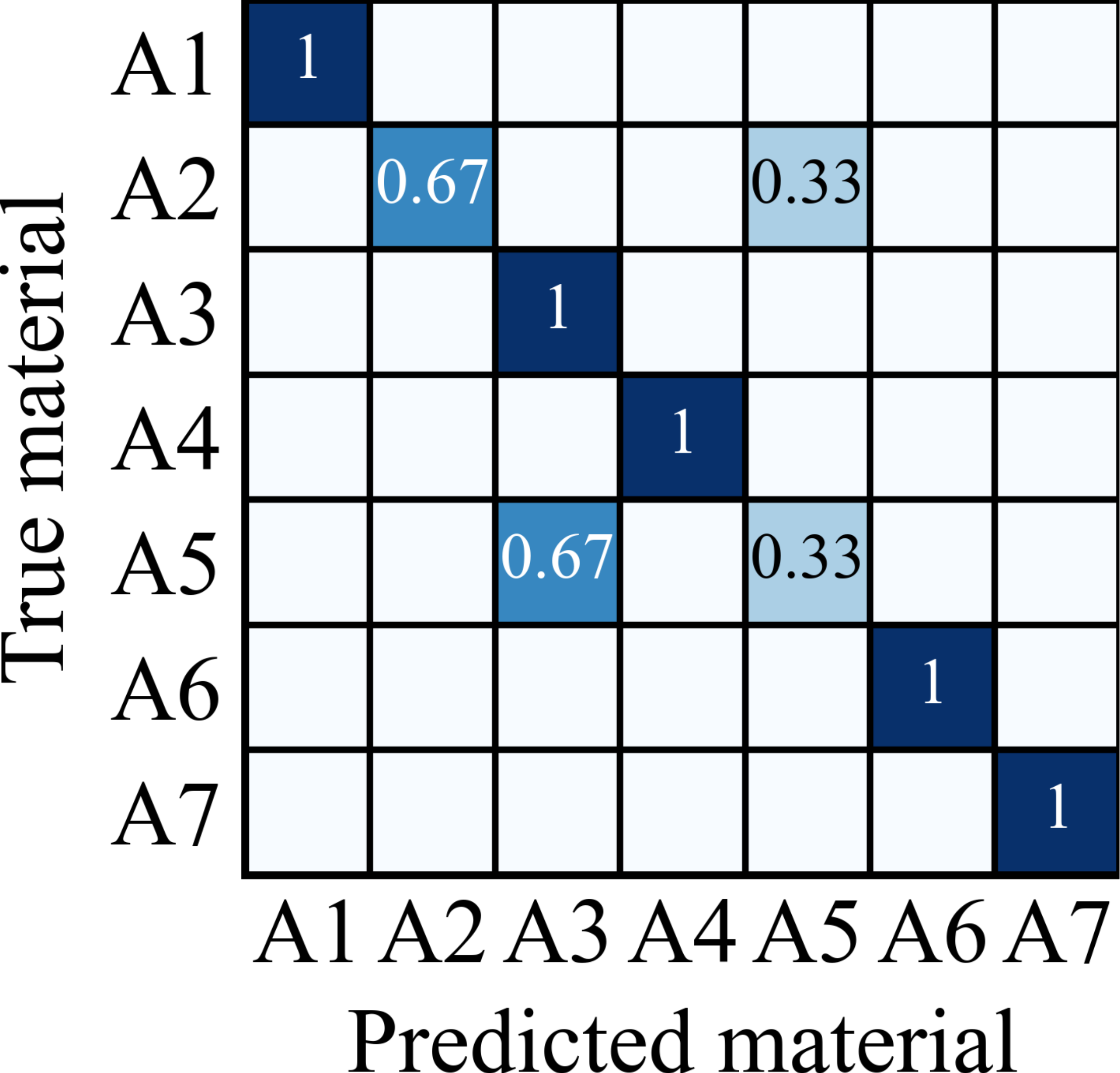}
    \label{VLM_only_41_v1-1}
  }\hfill
  \subfloat[Bottles.]{
    \includegraphics[width=0.14\textwidth]{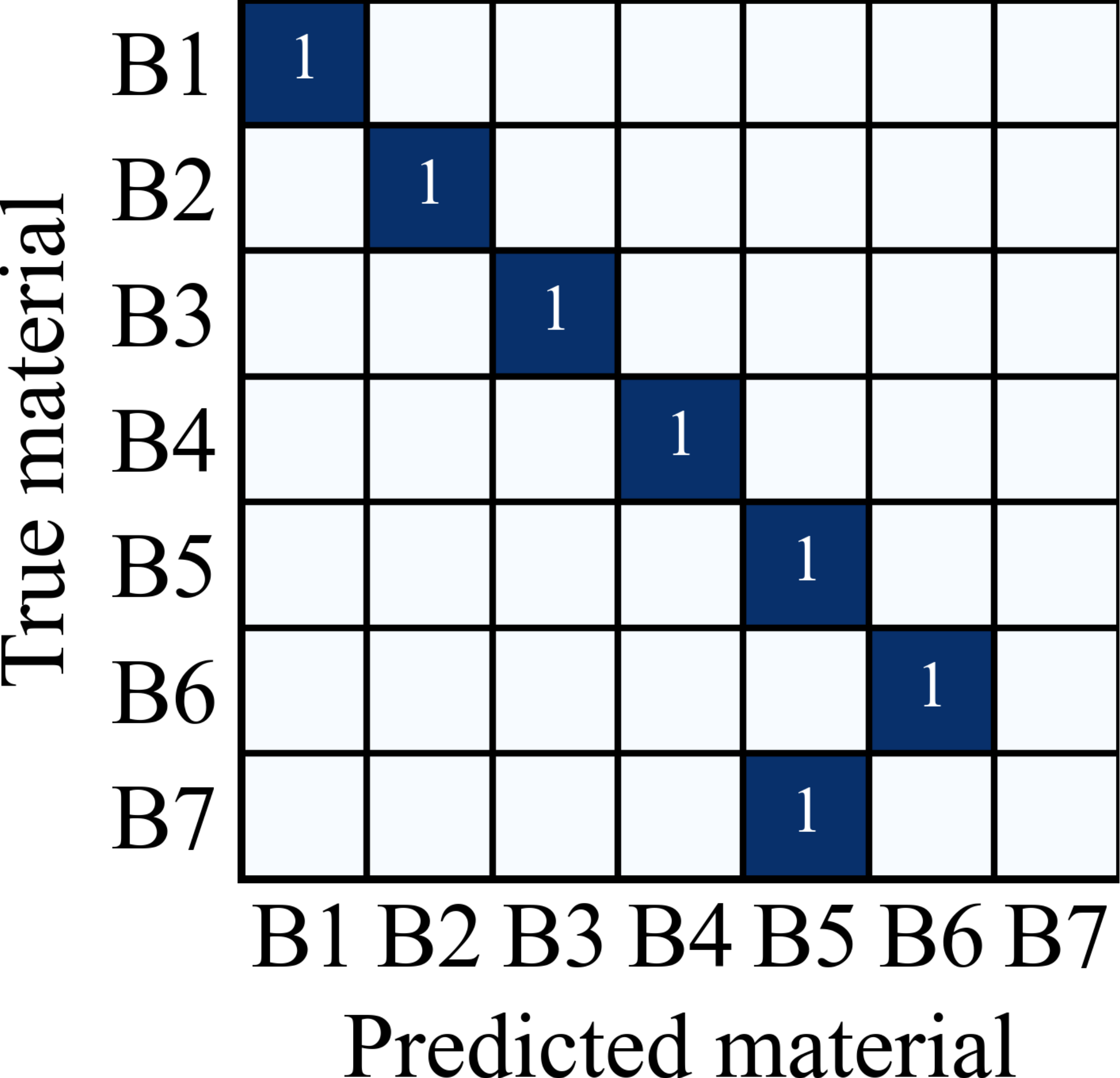}
    \label{VLM_only_41_v1-2}
  }\hfill
  \subfloat[Kettles.]{
    \includegraphics[width=0.14\textwidth]{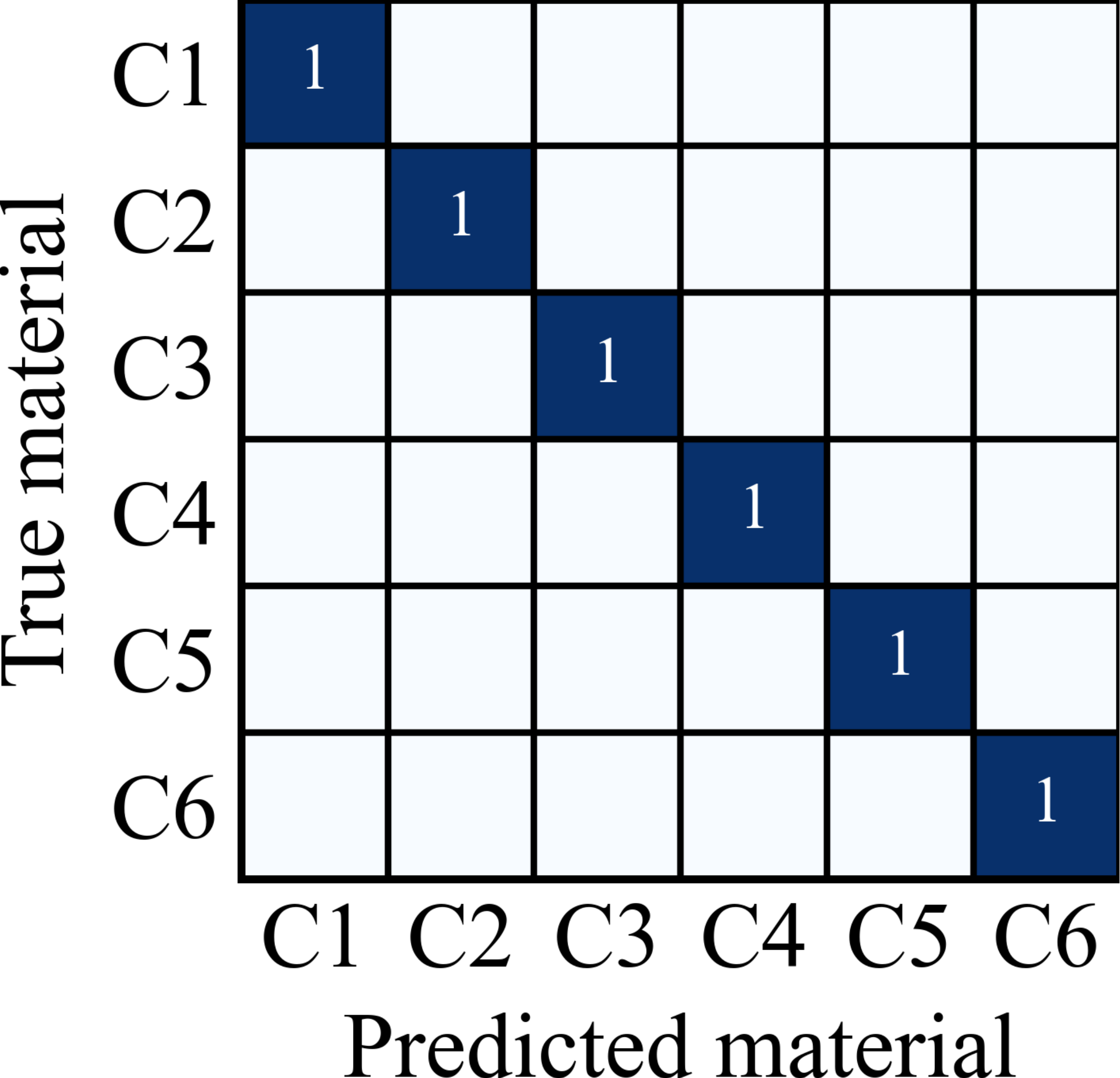}
    \label{VLM_only_41_v1-3}
  }\hfill
  \subfloat[Boxes.]{
    \includegraphics[width=0.14\textwidth]{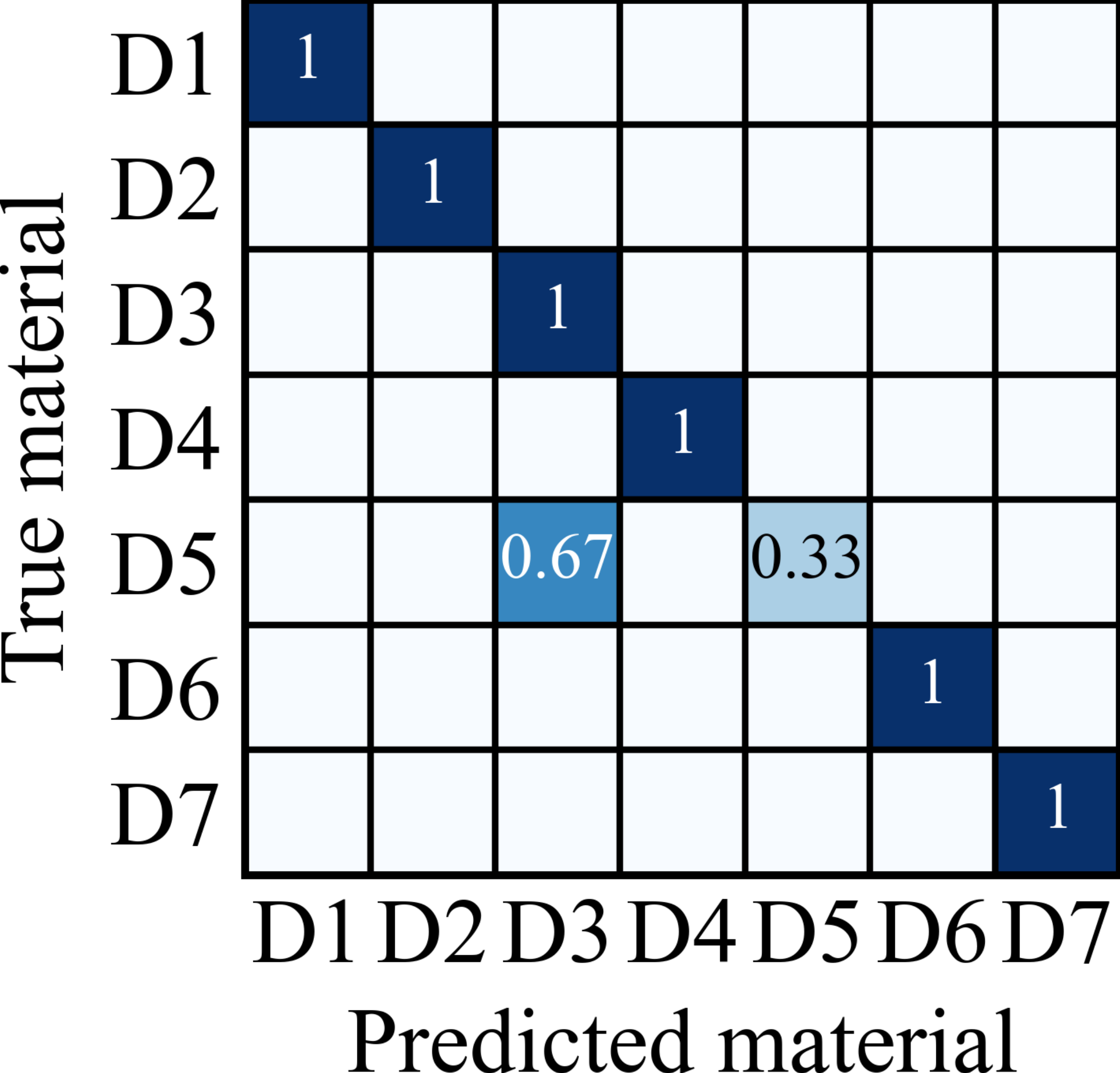}
    \label{VLM_only_41_v1-4}
  }\hfill
  \subfloat[Plates.]{
    \includegraphics[width=0.14\textwidth]{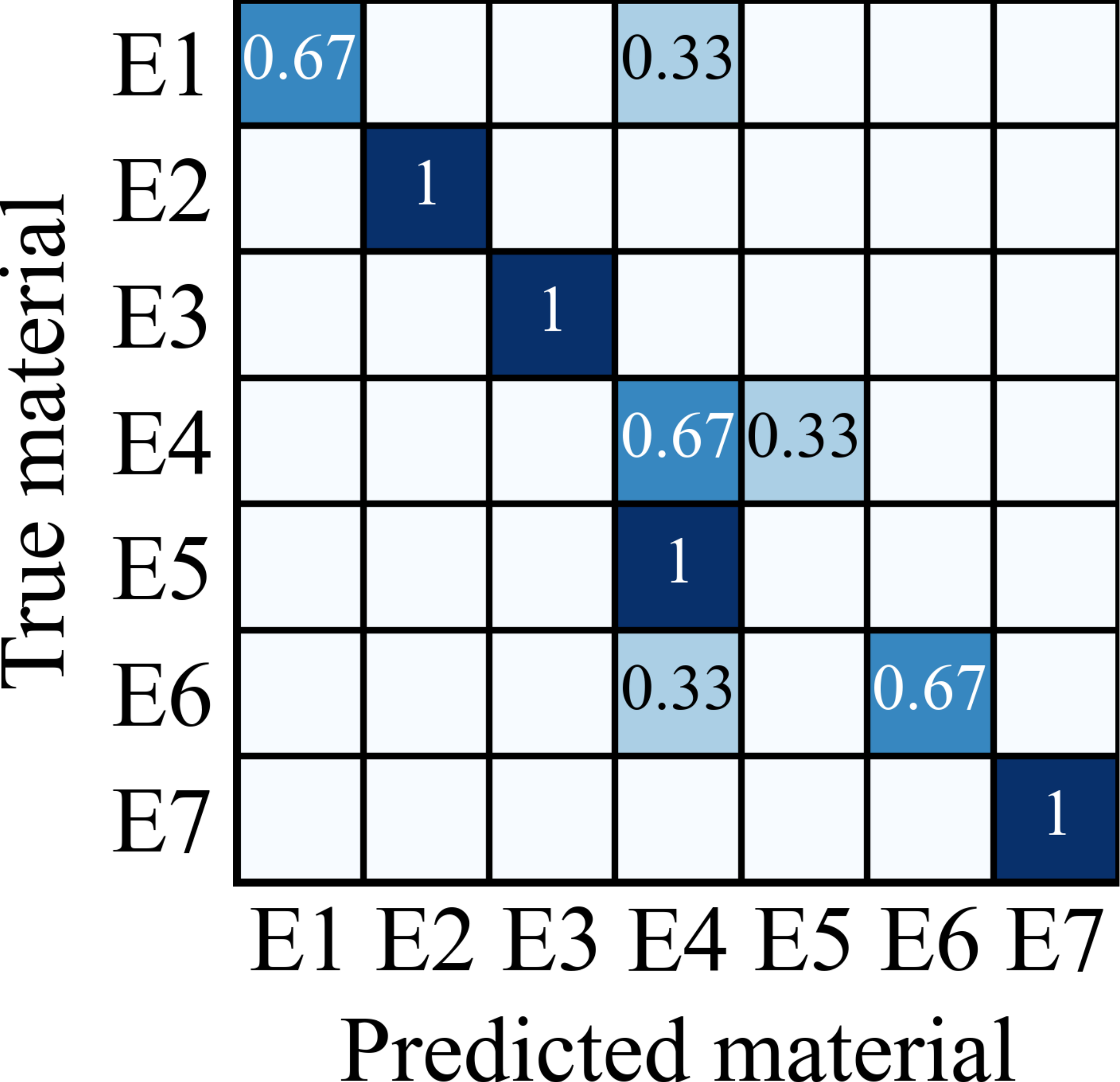}
    \label{VLM_only_41_v1-5}
  }\hfill
  \subfloat[Boards.]{
    \includegraphics[width=0.14\textwidth]{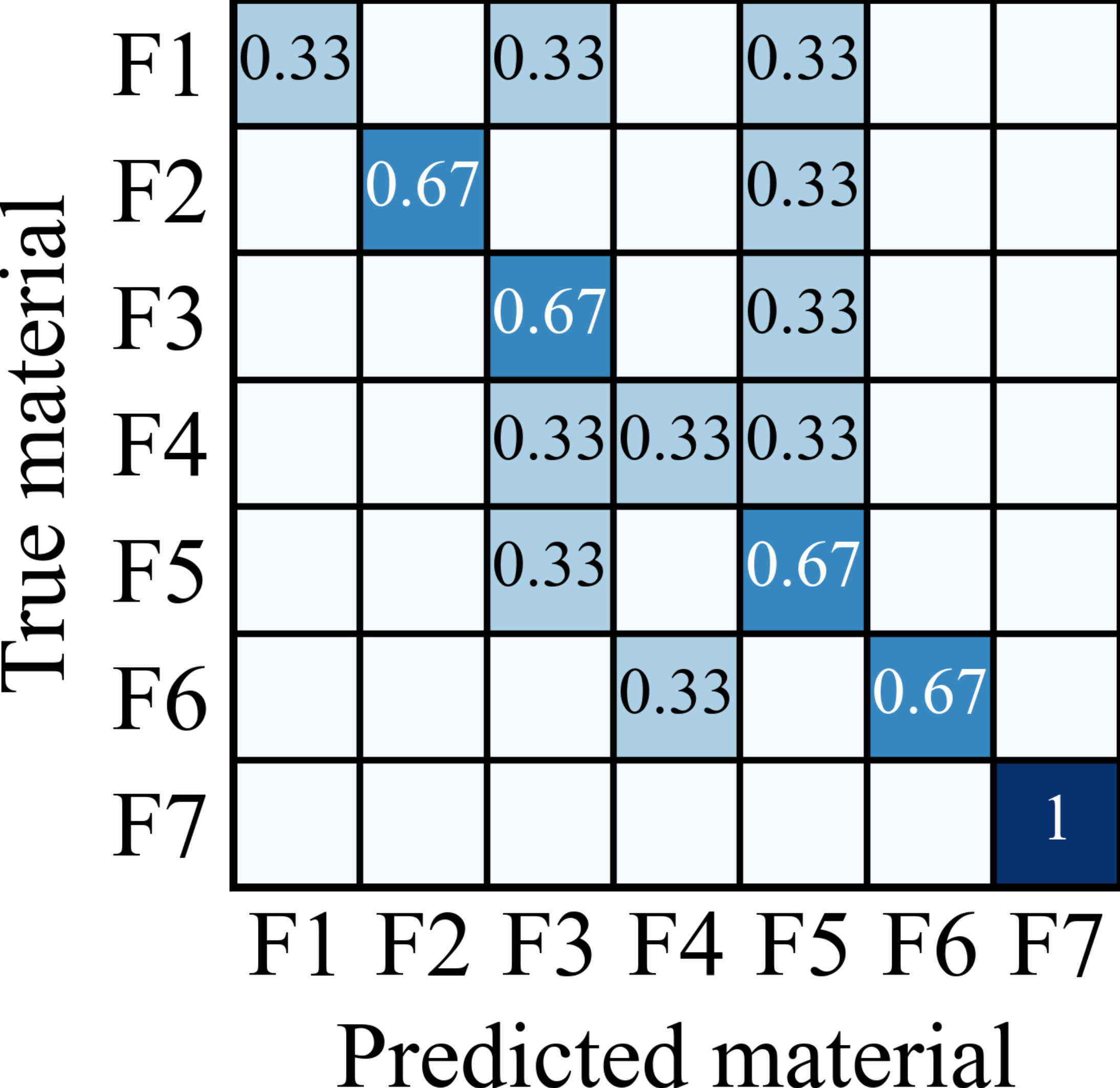}
    \label{VLM_only_41_v1-6}
  }\\
  \caption{Overall object material recognition (VLM-only).}
  \label{VLM_only_41}
\end{figure*}

As shown in \Cref{VLM_only_41}, visual-only material recognition faces significant challenges. First, due to specular reflections, metallic objects were frequently misidentified as ceramic (e.g., E1), plastic or mirror glass (e.g., F1). Similarly, strong reflections caused ceramic to be mistaken for mirror glass (e.g., F4) and plastics to be misclassified as mirror glass (e.g., A5, D5, F5).
Furthermore, high visual similarity posed a major obstacle to accurate distinction. For instance, frosted glass (e.g., A2, F2), mirror glass (e.g., F3), and ceramic (e.g., F4) were erroneously identified as plastic. Conversely, plastic (e.g., E5) and wood (e.g., E6, F6) were wrongly classified as ceramic. Additionally, specially processed paper was misidentified as plastic (e.g., B7) due to its deceptive appearance.

Analyzing these failure cases, we categorize VLM-only errors into three types: (1) \textbf{Specular Reflections}, where high reflectivity masks surface textures and confuses reflected environments with material features (e.g., metal vs. mirror glass); (2) \textbf{Visual Similarity}, where indistinguishable colors or textures (e.g., frosted glass vs. plastic) lack discriminative features; (3) \textbf{Visually Deceptive Objects}, where materials resemble others (e.g., plastic-like paper) and mislead the VLM (see \Cref{microbenchmarks} for details).

\begin{figure*}[!t]
  \centering
  \includegraphics[width=\textwidth]{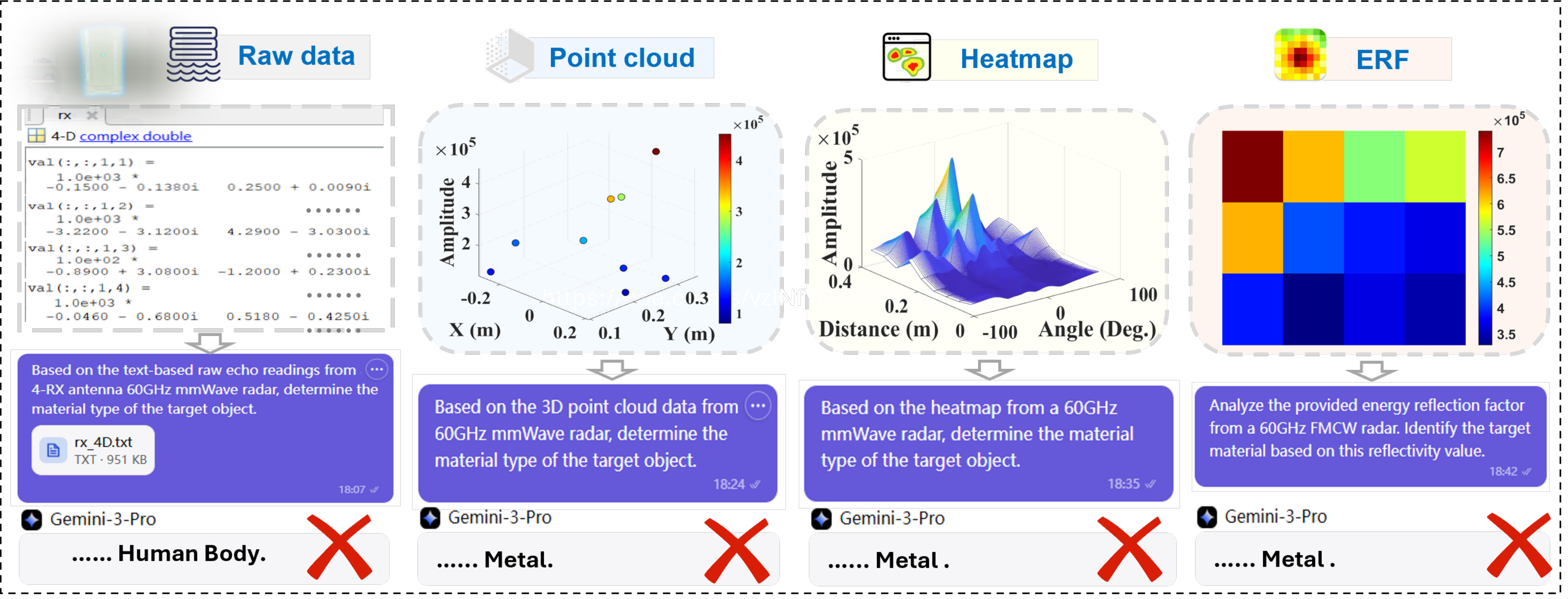}
  \caption{Evaluation of VLMs on four radar data representations (Raw data, Point cloud, Heatmap, and ERF).}
  \label{four_raw_inputs}
\end{figure*}

Therefore, vision-only VLM recognition is unreliable. We find that visually similar objects often differ in latent physical properties, so another modality is needed to reveal intrinsic material properties.
Despite mmWave radar's potential, existing radar-based processing methods are ill-suited for our goal of physics-grounded recognition, as shown in \Cref{four_raw_inputs}. Directly inputting raw radar data with brief explanation is infeasible due to the modality gap with VLM encoders. Using radar point clouds proves insufficient due to extreme data sparsity, while pre-processed radar images (e.g., heatmaps) lack the semantic alignment required for VLMs to understand them. Furthermore, prior works utilizing reflection intensity (e.g., ERF) often rely on multi-channel arrays to construct feature maps for closed-set classification. This creates a domain gap that VLMs cannot interpret, limiting these methods to predefined categories and preventing real-world generalization.

\begin{figure*}
  \includegraphics[width=\textwidth]{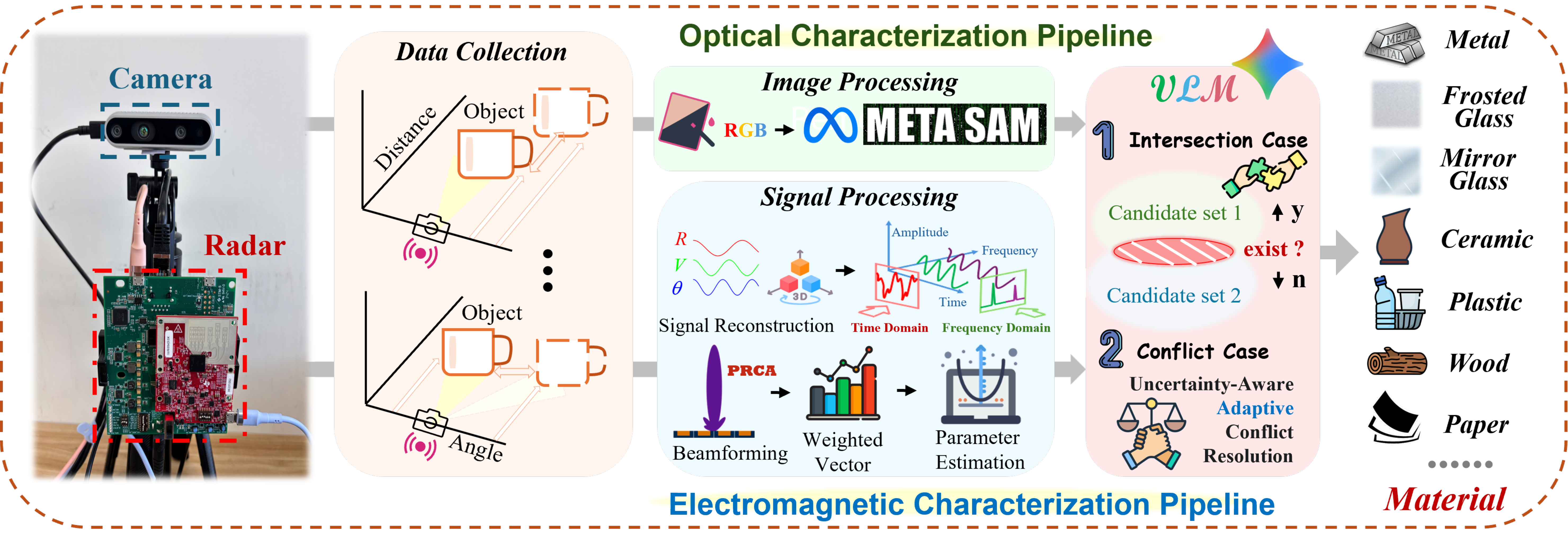}
  \caption{Overview of VLMaterial.}
  \label{VLMaterial_workflow}
\end{figure*}

\subsection{Challenges in Radar-based Material Recognition}
Based on prior work and our preliminary experiments, we summarize several critical challenges for robust physics-grounded material recognition.

(1) \textit{Extracting Intrinsic Material Properties from Reflection Signals}. Accurately isolating reflection signals for intrinsic material properties requires precise 3D target localization, decoupling RCS from object geometry, and rigorous calibration of hardware imperfections and phase deviations.

(2) \textit{Identifying Universal and Interpretable Electromagnetic Parameters}. To enable meaningful VLM interaction, we need basic electromagnetic parameters that are universally applicable and natively interpretable by VLMs, requiring raw data transformation into semantically grounded representations.

(3) \textit{Bridging the Semantic Gap between Raw Radar Data and VLMs}. Raw radar signals are inherently abstract and redundant, making direct VLM interpretation difficult. Addressing this requires both mitigating redundancy and constructing a domain knowledge system to compensate for the model's lack of radar-specific expertise.

(4) \textit{Implementing Adaptive Multimodal Mechanism}. Effective fusion requires integrating both modalities and resolving conflicts by intelligently weighing each output based on confidence or context to ensure reliable recognition.

The paper is organized as follows. \Cref{radar_sensing} addresses radar characterization, dielectric constant extraction, and context-augmented generation (CAG)-based domain knowledge. \Cref{adaptive_fusion} presents an adaptive fusion mechanism. \Cref{results_analysis} validates VLMaterial experimentally, followed by limitations in \Cref{limitations_discussion} and conclusions in \Cref{conclusion}.

\section{System design}
\label{radar_sensing}

In this section, we introduce the architecture of VLMaterial, a system designed for robust material identification by fusing camera and radar sensing with CAG-enhanced VLM reasoning, as illustrated in \mbox{\Cref{VLMaterial_workflow}}. The framework comprises two parallel streams: an optical characterization pipeline that employs SAM to segment the RGB image, followed by a VLM to propose potential material candidates, and an electromagnetic characterization pipeline that analyzes radar reflections to recognize material based on electromagnetic properties. 

\subsection{Electromagnetic Characterization Pipeline}
\label{Electromagnetic_Pipeline}
\begin{figure*}[!t]
    \centering
    \includegraphics[width=\linewidth]{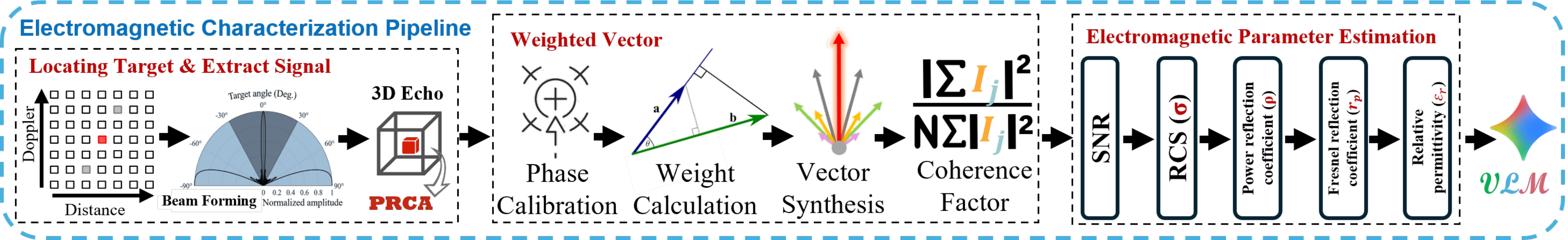}
    \caption{Electromagnetic characterization pipeline.}
    \label{electromagnetic_pipeline}
\end{figure*}

This part details our signal preprocessing pipeline, as depicted in \Cref{electromagnetic_pipeline}. Based on the reconstructed 3D radar signal spanning range, Doppler, and angle dimensions, we locate and extract the target signal by generating range-Doppler (RD) through a 2D-fast Fourier transform (2D-FFT) \cite{brigham1988fast} and digital beamforming. To compensate for unavoidable positional deviations in practice, we propose a new weighted vector synthesis method that refines the extracted signal through phase calibration, adaptive weight calculation, and coherent vector synthesis, leveraging a coherence factor to significantly enhance the SNR. This high-quality signal then enables precise derivation of key electromagnetic properties, including the relative dielectric constant, for robust material recognition.

\subsubsection{Signal preprocessing}

We first generate a high-resolution range-Doppler (RD) map via multichannel accumulation and derive the range-angle (RA) map by processing fast-time and spatial dimensions. These maps provide distance, velocity, and angle for accurate localization, as shown in \Cref{RD_RA}.

To enhance angle estimation precision, then we employ beamforming to improve direction of arrival (DoA) estimation, overcoming the wide native half-power beamwidth (HPBW) caused by the array's physical aperture. As shown in \Cref{array_BF}, this enhances directional signals while suppressing noise. Once the target's signal is extracted based on its distance and angle, we focus on electromagnetic parameter estimation, where the received echo power follows the radar equation \cite{barton2013radar}:

\begin{figure*}[!t]
\centering
\begin{minipage}{0.48\linewidth}
    \centering
    \includegraphics[width=0.48\linewidth]{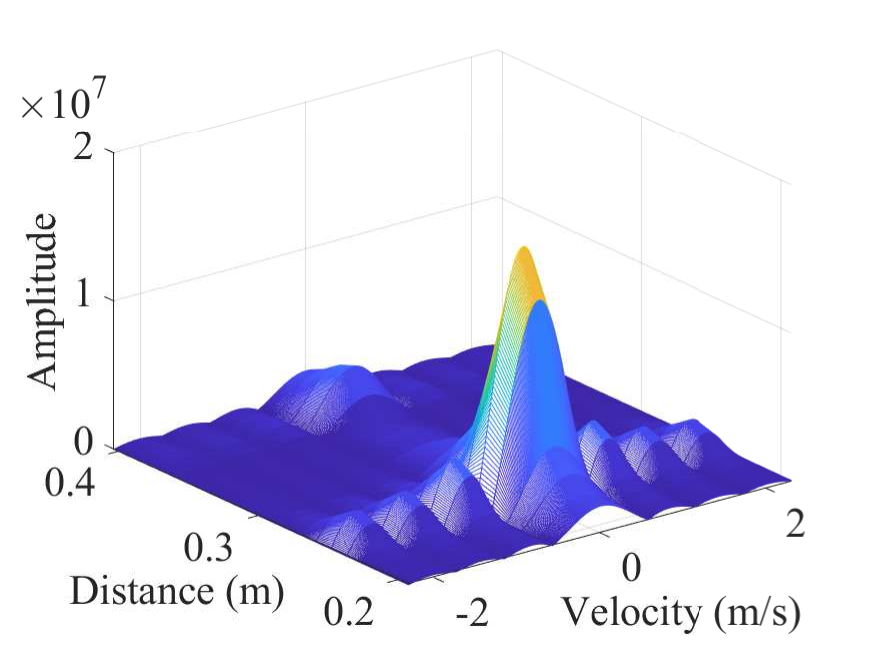}
    \hfill
    \includegraphics[width=0.48\linewidth]{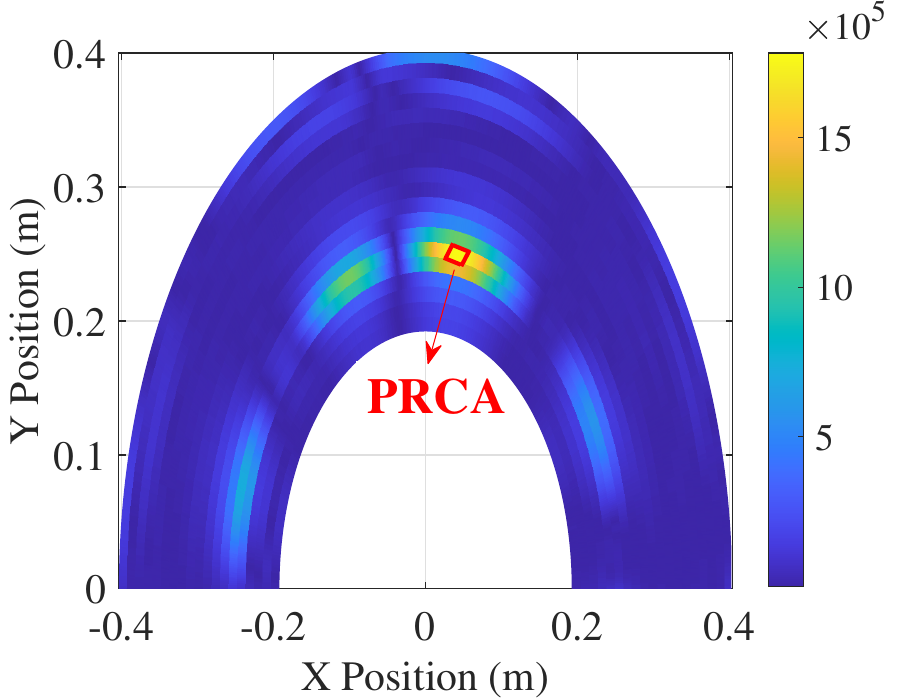}
    \caption{(Left) RD map; \\(Right) RA map.}
    \label{RD_RA}
\end{minipage}
\hfill 
\begin{minipage}{0.48\linewidth}
    \centering
    \includegraphics[width=0.48\linewidth]{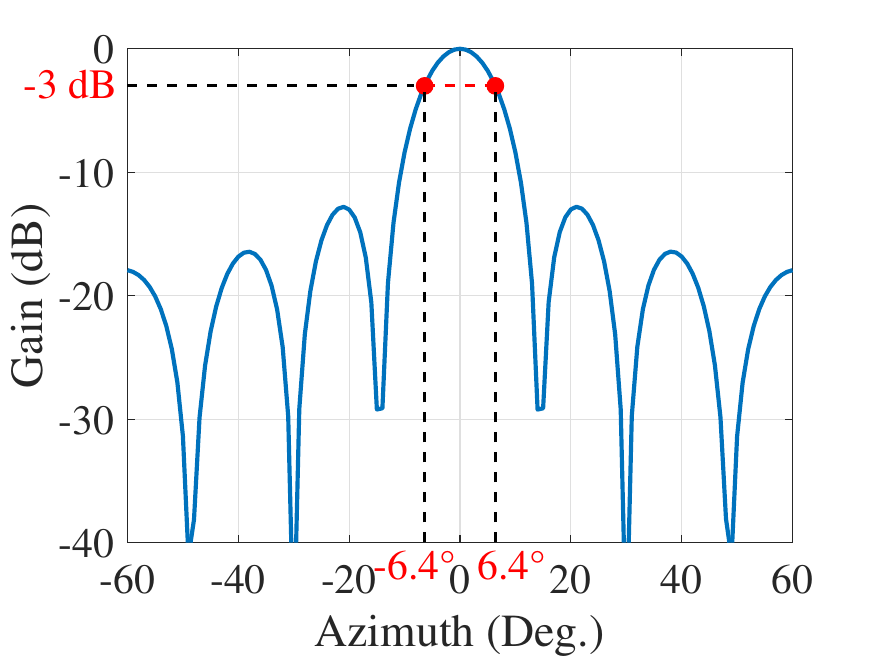}
    \hfill
    \includegraphics[width=0.48\linewidth]{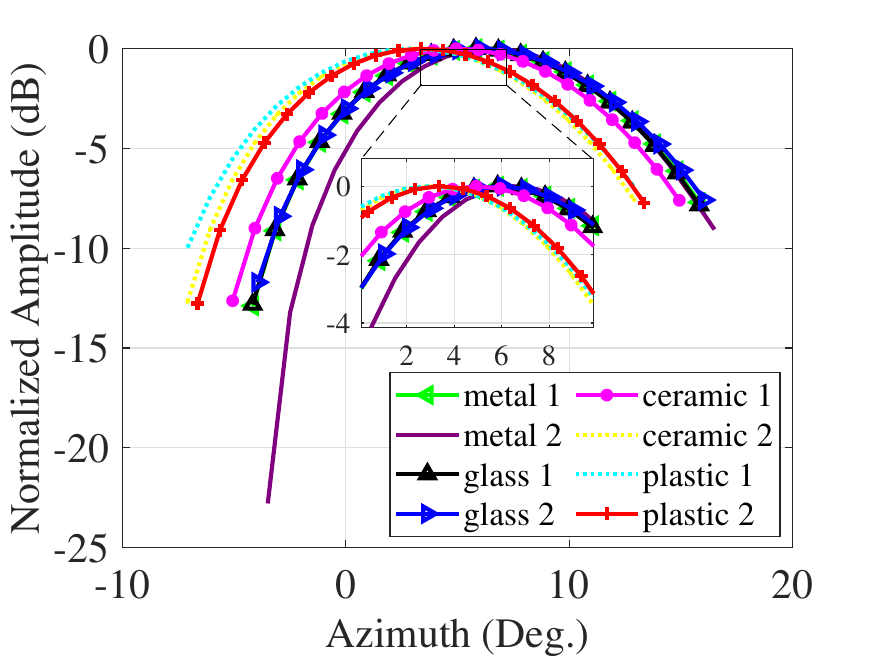}
    \caption{(Left) Beam pattern of the antenna array; (Right) Beam direction after beamforming.}
    \label{array_BF}
\end{minipage}
\end{figure*}
\begin{equation}
P_r = P_t G_t G_r \frac{\lambda^2 \sigma}{(4\pi)^3 R^4},
\label{eq1}
\end{equation}
where $P_r$ is the received power, $P_t$ is the transmitted power, $G_t$ and $G_r$ are the gains of the transmitting and receiving antennas, respectively, $\lambda$ is the signal wavelength, $\sigma$ is the target's RCS, and $R$ is the distance of the target relative to radar, derived from the RA map. Based on the $P_r$, the SNR of the target echo is:
\begin{equation}
SNR = \frac{P_r}{P_n} =  \frac{P_t G_t G_r \lambda^2 \sigma}{(4\pi)^3 k T_n B R^4} = K \frac{\sigma}{R^4},
\label{eq2}
\end{equation}
where $P_n$ is the total noise power at the receiver. For thermal noise, $P_n = k T_n B$, where $k$ is the Boltzmann constant, $T_n$ is the system noise temperature, and $B$ is the bandwidth. 
To determine the system constant $K$, we performed a calibration using a metal sphere ($d = 63$\, mm) as a standard target. Since the sphere's diameter is much larger than the signal wavelength ($\lambda = 5$ mm), it operates in the optical scattering region, thereby having an RCS equivalent to its physical cross-sectional area \cite{bernhardt2011radar}: $\sigma_c = \pi (d/2)^2 \approx 0.0031$ m$^2$.

After measuring the SNR of the sphere from its extracted signal, we computed $K$ by inversion of \Cref{eq2}. This calibrated constant was then used to calculate the RCS of an unknown target, which can be expressed as:

\begin{equation}
\sigma = SNR \frac{R^4}{K}.
\label{eq5}
\end{equation}

By leveraging the known $\sigma_c$, to eliminate the distance dependence of the received signal, the target RCS is:
\begin{equation}
    \sigma = \sigma_c \cdot \frac{SNR}{SNR_c} \cdot \left( \frac{R}{R_c} \right)^4,
    \label{rcs_cali}
\end{equation}
where $SNR_c$, $R_c$, and $\sigma_c$ represent the SNR, distance, and RCS of the calibration metal sphere, respectively.

Therefore, a one-time calibration is sufficient for VLMaterial to eliminate distance-dependent effects, enabling precise RCS measurement at arbitrary distances. Particularly, for $R = R_c$, $\sigma = \sigma_c \cdot {SNR}/{SNR_c}$.
For details on the SNR calculation, we refer to \Cref{SNR} within \Cref{Enhanced_SNR}.

\begin{figure*}[!t]
  \centering
  \includegraphics[width=\textwidth]{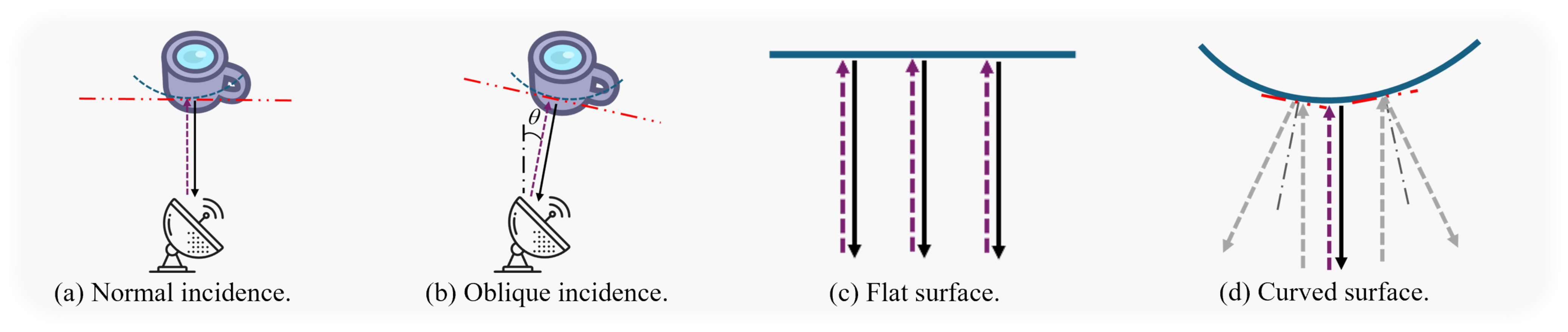}
  \caption{The impact of object positioning and surface curvature.}
  \label{object_positioning_surface_curvature}
\end{figure*}

\subsubsection{Enhanced SNR via Weighted Vector Synthesis}
\label{Enhanced_SNR}

So far, the RCS has been computed via SNR. For robust deployment of VLMaterial, we must address two primary geometric deviations that degrade the reflected signal in practice.

(1) \textit{Angular Misalignment.} The ideal setup places the object along the radar's normal for maximum beam gain (\Cref{object_positioning_surface_curvature}a). Practical positioning allows only approximate alignment (\Cref{object_positioning_surface_curvature}b), introducing unavoidable human-induced error.

(2) \textit{Surface Curvature Dispersion.} The PRCA in VLMaterial assumes near-specular reflection from flat surfaces (\Cref{object_positioning_surface_curvature}c). For curved surfaces (e.g., a cup), diffuse reflection scatters incident energy away from the receiver (\Cref{object_positioning_surface_curvature}d), causing energy dispersion, weaker echo, and reduced SNR.

To address these issues, we first correct angular misalignment via beamforming, ensuring the incident wave hits the PRCA. For surface curvature dispersion, we then develop a weighted vector synthesis strategy based on PRCA, which uses the full array and assigns coherence-derived weights to each antenna's spatial direction vector. Combining these weighted vectors generates a robust estimate of the reflected signal, encompassing both magnitude and direction, and reveals the target's physical orientation.

\begin{figure*}[!t]
  \centering
  \includegraphics[width=\textwidth]{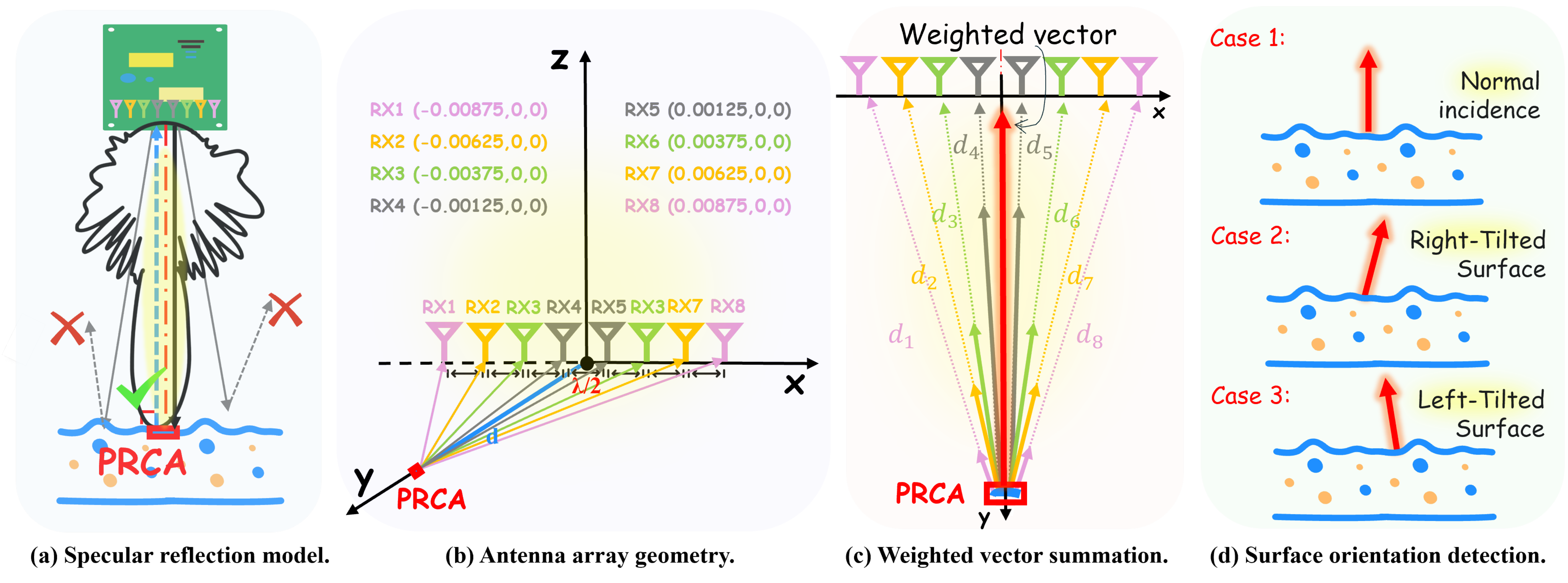}
  \caption{Weighted vector summation.}
  \label{weighted_vector}
\end{figure*}

This method operates on the specular reflection model in \Cref{weighted_vector}a. At 60 GHz, reflected signal energy peaks at antenna elements facing the object's surface orientation, corresponding to perpendicular (normal) incidence and reflection. Consider a linear antenna array consisting of $N$ elements (e.g., $N=8$) aligned along the x-axis, as depicted in \Cref{weighted_vector}b. Let the spatial coordinate of the $j$-th antenna be $\mathbf{p}_j \in \mathbb{R}^3$, and the coordinate of the PRCA voxel be $\mathbf{v} \in \mathbb{R}^3$. Prior to processing, we perform a calibration using a metallic sphere reference to eliminate hardware-induced phase inconsistencies. The calibration phasor $\mathcal{C}_j$ for the $j$-th antenna is derived by comparing the measured phase $\phi_{meas, j}$ with the theoretical calibration phase $\phi_{cali, j}$:
\begin{equation}
    \mathcal{C}_j = \exp\left(-j (\phi_{meas, j} - \phi_{cali, j})\right).
\end{equation}

We denote $I_j(\mathbf{v}) \in \mathbb{C}$ as the focused signal component for the $j$-th antenna. To ensure precise phase alignment, we apply the calibration phasor $\mathcal{C}_j$ to rectify hardware offsets and compensate for the geometric phase delay:
\begin{equation}
    I_j(\mathbf{v}) = x_j \cdot \mathcal{C}_j \cdot \exp\left(j \frac{4\pi \|\mathbf{p}_j - \mathbf{v}\|}{\lambda}\right),
\end{equation}
where $x_j$ is the raw received signal. The coherent backscatter from the PRCA to the radar array is then the coherent sum:
\begin{equation}
    S(\mathbf{v}) = \sum_{j=1}^{N} I_j(\mathbf{v}).
\end{equation}

To determine the surface orientation and improve the SNR, we first quantify the contribution of each antenna to the specular reflection. We define a scalar weight $w_j$ for the $j$-th antenna by projecting its individual signal vector onto the total coherent sum in the complex plane:
\begin{equation}
    w_j = \frac{\Re\{I_j(\mathbf{v})\} \cdot \Re\{S(\mathbf{v})\} + \Im\{I_j(\mathbf{v})\} \cdot \Im\{S(\mathbf{v})\}}{|S(\mathbf{v})|}.
\end{equation}
As visualized in \Cref{weighted_vector}c, this weighting mechanism acts as a spatial filter: antennas receiving phase-coherent specular reflections are assigned large positive weights to enhance the signal, while those receiving incoherent noise or multipath interference are suppressed.

Leveraging these weights, we perform the geometric synthesis shown in \Cref{weighted_vector}c. We define the geometric unit vector pointing from the target toward the $j$-th antenna as:
\begin{equation}
    \mathbf{u}_j = (\mathbf{p}_j - \mathbf{v}) / \|\mathbf{p}_j - \mathbf{v}\|.
\end{equation}

The estimated reflected signal vector is constructed by computing the weighted sum of these geometric vectors:
\begin{equation}
    \mathbf{S}_{\text{weighted}} = \sum_{j=1}^{N} w_j \cdot \mathbf{u}_j.
\end{equation}

This reconstructed vector serves a dual purpose: its magnitude indicates reflected signal strength, and its direction reveals the target's surface normal orientation (\Cref{weighted_vector}d). Symmetric weight distribution implies a surface normal pointing directly toward the radar (flat surface, case 1). Weights biased right or left tilt the vector accordingly, capturing the PRCA's geometric slope (case 2 or 3). To validate signal quality and suppress ghost targets, we propose the coherence factor $c(\mathbf{v})$, which quantifies the phase consistency across the array:
\begin{equation}
   c(\mathbf{v}) = \frac{|S(\mathbf{v})|^2}{N \cdot \sum_{j=1}^{N} |I_j(\mathbf{v})|^2}.
\end{equation}

Finally, the enhanced SNR is computed by weighting the raw power of the reconstructed vector with the $c(\mathbf{v})$:
\begin{equation}
    SNR = \frac{\|\mathbf{S}_{\text{weighted}}\|^2}{P_{noise}} \cdot c(\mathbf{v}),
    \label{SNR}
\end{equation}
where $P_{noise}$ represents the noise of the system. This weighted vector synthesis ensures that only spatially coherent reflections contribute to the final signal, resulting in a reliable SNR independent of angular and phase variations.

\subsubsection{Electromagnetic Parameter Estimation}
Based on the weighted reflected signal, we calculate SNR using background noise power from an empty scene, obtaining high-quality RCS.
However, the RCS itself is a composite parameter, conflating material properties with geometric shape. To isolate material characteristics, we introduce the backscattering coefficient $\sigma_0$ \cite{ulaby1986microwave}, defined as the average RCS per unit illuminated area $A_t$.
\begin{equation}
\sigma = \sigma_0 A_t.
\label{eq6}
\end{equation}

Inspired by this, we shift focus from total RCS to dominant reflection centers, defining a power reflection coefficient $\rho$ for the strong signals originating from a small but highly reflective region, which we term PRCA ($A_r$), as shown in \Cref{RD_RA}. Then $\rho$ becomes analogous to $\sigma_0$, and RCS can be expressed as:
\begin{equation}
\sigma = \rho A_r.
\label{eq7}
\end{equation}

PRCA calculation in Appendix~\ref{app:prca}. Based on the $\sigma$ and $A_r$, after normalizing $\rho$, we obtain the Fresnel reflection coefficient $r_p$. According to the relationship between $r_p$ and the dielectric constant $\varepsilon_r$ (see Appendix~\ref{app:fres_die}), we finally obtain:
\begin{equation}
r_p = \frac{\varepsilon_r \cos \theta - \sqrt{\varepsilon_r - \sin^2 \theta}}{\varepsilon_r \cos \theta + \sqrt{\varepsilon_r - \sin^2 \theta}}.
\label{r_p_varepsilon}
\end{equation}



Next, we define the variables as follows:
\begin{equation}
A = \varepsilon_r \cos \theta, \quad B = \sqrt{\varepsilon_r - \sin^2 \theta}.
\label{eq9}
\end{equation}








The final relationship between $\varepsilon_r$, $r_p$ and $\theta$ simplifies to:
\begin{equation}
\varepsilon_r = \frac{(r_p + 1)^2 \left[ 1 \pm \sqrt{1 - \left( \frac{\sin 2\theta (r_p - 1)}{r_p + 1} \right)^2} \right]}{2 \cos^2 \theta (r_p - 1)^2}.
\label{eq15}
\end{equation}

Specifically, for \(\theta = 0\), \Cref{eq15} becomes:
\begin{equation}
\varepsilon_r = \left( \frac{1 + r_p}{1 - r_p} \right)^2.
\label{eq16}
\end{equation}

\begin{figure*}[!t]
  \centering
  \includegraphics[width=0.95\textwidth]{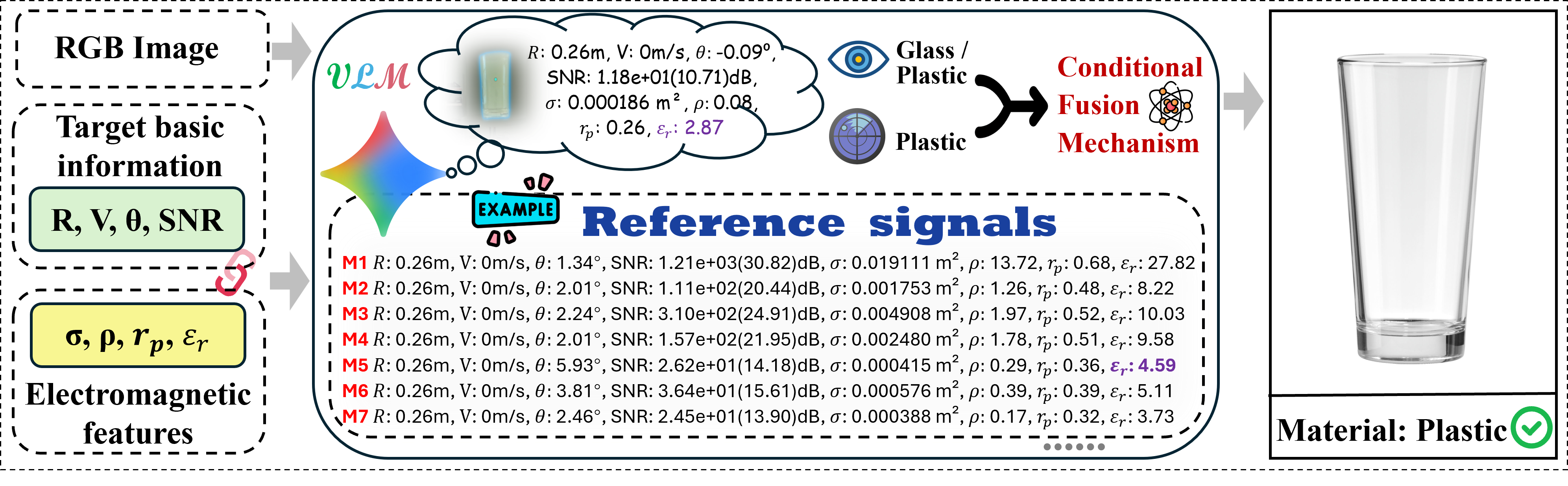}
  \caption{CAG-Enhanced VLMaterial.}
  \label{CAG}
\end{figure*}

As outlined in \Cref{electromagnetic_pipeline}, the workflow progresses from raw signal to enhanced SNR, through the critical analysis of RCS ($\sigma$), power reflection coefficient ($\rho$), and Fresnel reflection coefficient ($r_p$), to the final derivation of the relative dielectric constant ($\varepsilon_r$). This progression extracts physical characteristics from raw radar signals, enabling richer semantic understanding for material recognition.
Furthermore, our proposed VLMaterial is supported by \cite{raju2017dielectrics}, which confirms that the dielectric constant can effectively represent material, as different materials exhibit distinct values. See Appendix~\ref{app:grd_truth} for estimated vs. ground-truth dielectric constant comparison.

\subsection{CAG-Enhanced VLM}


As illustrated in \Cref{CAG}, eight radar-derived parameters are input into the CAG-enhanced VLM: basic target information (distance $R$, velocity $V$, angle $\theta$, SNR) and four electromagnetic features ($\sigma$, $\rho$, $r_p$, $\varepsilon_r$). Public data alone is insufficient, as it lacks 60GHz electromagnetic characterizations and may cause hallucination \cite{liu2024survey}. Thus, we constructed a specialized knowledge base by recording mmWave radar echoes from seven material boards (M1-M7) under ideal vertical reflection from flat surfaces, providing standardized reference signals for CAG to ground the VLM in domain-specific knowledge. Revisiting the cup heating task, the VLM initially narrows candidates to glass and plastic based on vision. Radar-derived features query the CAG module, and with a measured dielectric constant of 2.87, the CAG-enhanced VLMaterial evaluates candidates M5 (plastic) and M7 (wood), correctly identifying the material as plastic.

This integration bridges raw radar data and semantic understanding. Tracing outputs to knowledge base references ensures transparency and verifiability, often ignored by other methods, making answers correct and trustworthy. CAG also mitigates hallucination on out-of-domain queries, enabling VLMaterial to accurately distinguish materials (see Appendix~\ref{app:cag} for CAG reference signals across baselines). 

\section{Adaptive Fusion Mechanism}
\label{adaptive_fusion}
In this section, we present VLMaterial's final decision criterion for material identification. The camera and radar pipelines independently generate candidate sets $\mathcal{S}_{vis}$ (visual) and $\mathcal{S}_{rad}$ (electromagnetic). The final result is determined based on the relationship between these two sets.

\subsection{Intersection Case}
When $\mathcal{S}_{vis}$ and $\mathcal{S}_{rad}$ intersect ($\mathcal{S}_{vis} \cap \mathcal{S}_{rad} \neq 0$), the system selects the common element as the final material, confirming agreement between modalities. For example, while the camera cannot distinguish transparent glass from plastic, radar identifies glass via dielectric properties, so the intersection is glass. This resolves the visual ambiguity in \Cref{introduction}, enabling correct selection of a glass cup for microwaving milk.

\subsection{Conflict  Case} While camera-radar integration greatly enhances robustness, conflicts between the two modalities are inevitable due to their distinct physical limitations. For instance, VLMs degrade in low-light conditions, while radar becomes unreliable under low SNR or extreme incidence angles. A conflict arises when the candidate sets are disjoint ($\mathcal{S}_{vis} \cap \mathcal{S}_{rad} = 0$).

To resolve cross-modal conflicts, we propose a dynamic strategy termed uncertainty-aware adaptive conflict resolution. Instead of relying on static assumptions, this mechanism dynamically assigns a confidence score to each sensor based on real-time environmental and physical constraints.

\subsubsection{Visual Uncertainty Factors}
Visual system reliability is constrained by signal quality and model confidence. The visual uncertainty factors are formulated as:
\begin{equation}
    U_{vis} \propto \lambda_1 (1 - \mathcal{I}_{lum}) + \lambda_2 \mathcal{I}_{cplx} + \lambda_3 \mathcal{H}_{vlm}.
    \label{vis_uncertainty}
\end{equation}

We identify three factors compromising visual reliability, grounded in optical physics and information theory, with scaling coefficients $\lambda_1$, $\lambda_2$, $\lambda_3$ respectively. The first term accounts for signal degradation due to photon shot noise \cite{hasinoff2014photon}. As the normalized luminance $\mathcal{I}_{lum}$ approaches zero, the SNR drops proportionally to the photon flux, resulting in texture loss where information is physically missing rather than merely ambiguous. The second term addresses scene complexity and visual clutter \cite{rosenholtz2005feature} via $\mathcal{I}_{cplx}$, acknowledging that high background entropy increases the probability of segmentation errors and feature hallucination. Finally, the term $\mathcal{H}_{vlm}$ quantifies epistemic uncertainty \cite{kendall2017uncertainties}, capturing the internal cognitive hesitation of the VLM prediction distribution.

\subsubsection{Radar Uncertainty Factors} 
Radar reliability is governed by the physics of electromagnetic wave propagation and scattering. We describe the degradation factors as:
\begin{equation}
     U_{rad} \propto \gamma_1 \frac{1}{SNR} + \gamma_2 \left(\frac{d}{d_{max}}\right)^2 + \gamma_3 (1 - \cos \theta).
    \label{rad_uncertainty}
\end{equation}

We derive these three degradation factors from the fundamental radar equation \cite{barton2013radar} to identify unreliable signals, scaled by $\gamma_1$, $\gamma_2$, and $\gamma_3$ respectively. The first factor focuses on signal quality. Since measurement accuracy depends heavily on signal quality, we use the inverse form $1/SNR$ to ensure that system confidence drops drastically when the signal is drowned out by noise. The second factor addresses beam divergence \cite{richards2005fundamentals} using a squared term $(d/d_{max})^2$. Unlike lasers, radar beams widen as distance increases, capturing more background interference (clutter) which distorts the material-specific signature. The third factor, $(1 - \cos \theta)$, measures the energy loss caused by reflection angles. Since radar signals are strongest when hitting a surface straight on (normal incidence), this term captures the reliability drop when the radar views the object from an oblique angle.

\subsubsection{Reasoning via Domain Knowledge}
Crucially, our system does not explicitly calculate floating-point values for $\lambda$ and $\gamma$. Instead, the VLM implicitly interprets these coefficients through prompts that encode the mathematical relationships defined in \Cref{vis_uncertainty} and \Cref{rad_uncertainty} into the system prompt. This allows the model to dynamically infer the appropriate attention weights directly from the domain knowledge.

During inference, the VLM utilizes its self-attention mechanism \cite{vaswani2017attention} to process these rules. For instance, when the input metadata indicates low SNR, the model's attention heads focus on the corresponding rule (e.g., trust radar less if SNR is low). In this mechanism, the hyperparameters $\lambda$ and $\gamma$ are dynamically realized as the variable attention weights the model applies to specific physical constraints depending on the context. This capability matches recent findings on language model calibration \cite{kadavath2022language}, which demonstrate that VLMs can accurately estimate and articulate their own uncertainty when grounded in provided context, effectively performing qualitative reasoning without strict numerical computation.

\subsubsection{Final Confidence Estimation}
To resolve cross-modal conflicts, we employ a softmax-based gating mechanism to translate uncertainty scores into dynamic fusion weights. 

Analogous to the Boltzmann distribution \cite{russell1996boltzmann} in physics and maximum entropy \cite{gray2011entropy} in information theory, we employ a negative exponential mapping to transform uncertainty into probability. The weights are computed as:
\begin{equation}
    w_{vis} = \frac{e^{-U_{vis}}}{e^{-U_{vis}} + e^{-U_{rad}}}, \quad w_{rad} = 1 - w_{vis}.
    \label{fusion_weights}
\end{equation}

Given dynamic fusion weights $w_{vis}$ and $w_{rad}$ determined via the uncertainty gating mechanism, we derive the final confidence scores. Let $P_{vis}$ and $P_{rad}$ denote the material probability directly predicted by the VLM based on vision and radar parameters, respectively. The final confidence scores, $S_{vis}$ and $S_{rad}$, are then computed by modulating these input probabilities with their respective fusion weights:
\begin{equation}
    S_{vis} = w_{vis} \cdot P_{vis}, \quad S_{rad} = w_{rad} \cdot P_{rad}.
    \label{weighted_scores}
\end{equation}

This approach ensures robust adaptability: high visual uncertainty ($U_{vis}$) suppresses $S_{vis}$, thereby shifting decision dominance to the radar branch, whereas a noisy radar signal suppresses $S_{rad}$, leading to a greater reliance on visual information. Crucially, the final decision is not merely a rigid numerical comparison of these values. Instead, the VLM relies on its internal reasoning capabilities to comprehensively synthesize these weighted scores, acting as an intelligent arbitrator to generate a logically grounded material recognition (details in \Cref{results_analysis}: Impact of Fusion Method).

\section{Results and Analysis}
\label{results_analysis}

\begin{figure*}[!t]
  \centering
  \subfloat[Cups.]{
    \includegraphics[width=0.17\textwidth]{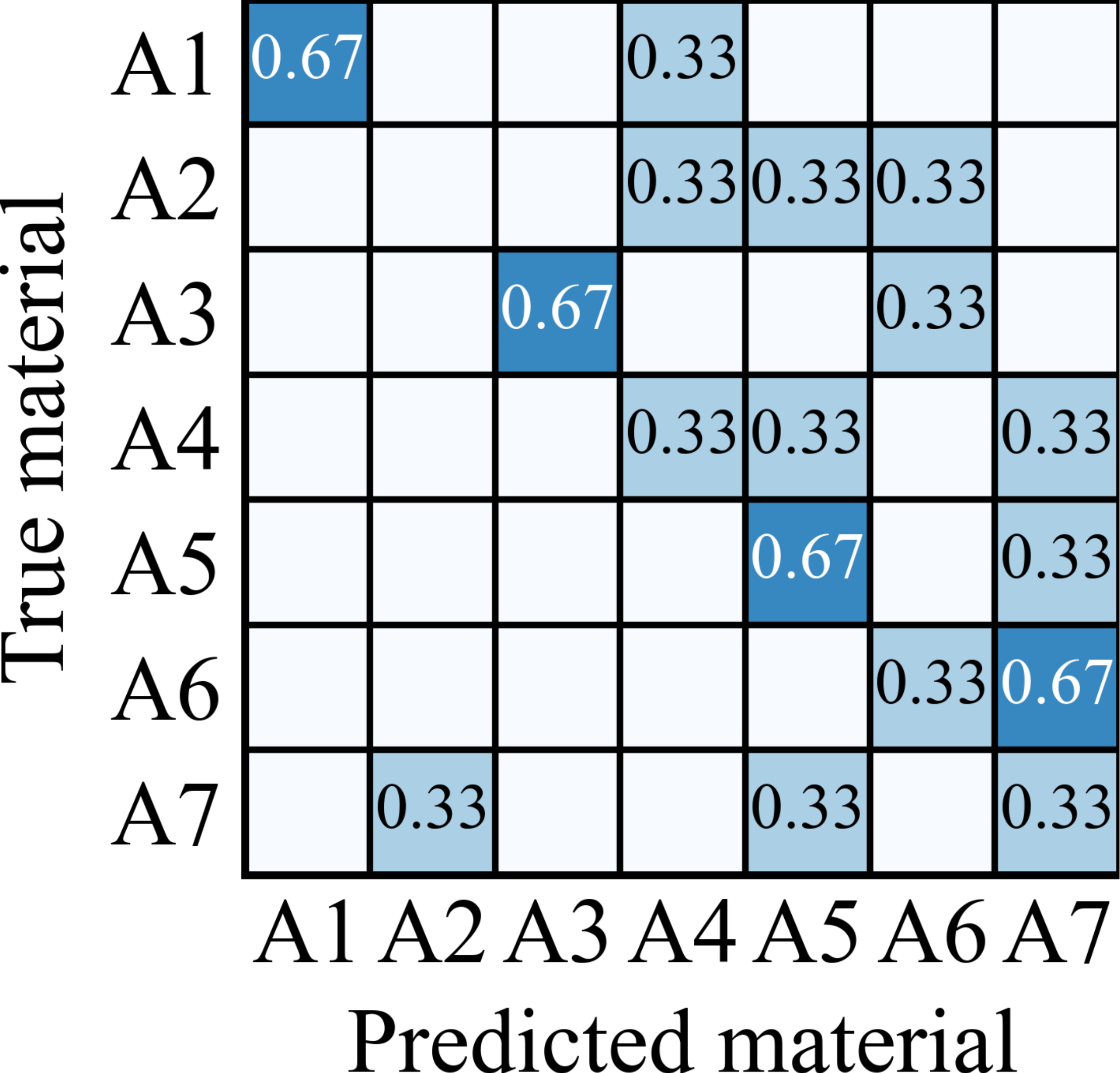}
    \label{LLMaterial_41_v1-1}
  }\hfill
  \subfloat[Bottles.]{
    \includegraphics[width=0.17\textwidth]{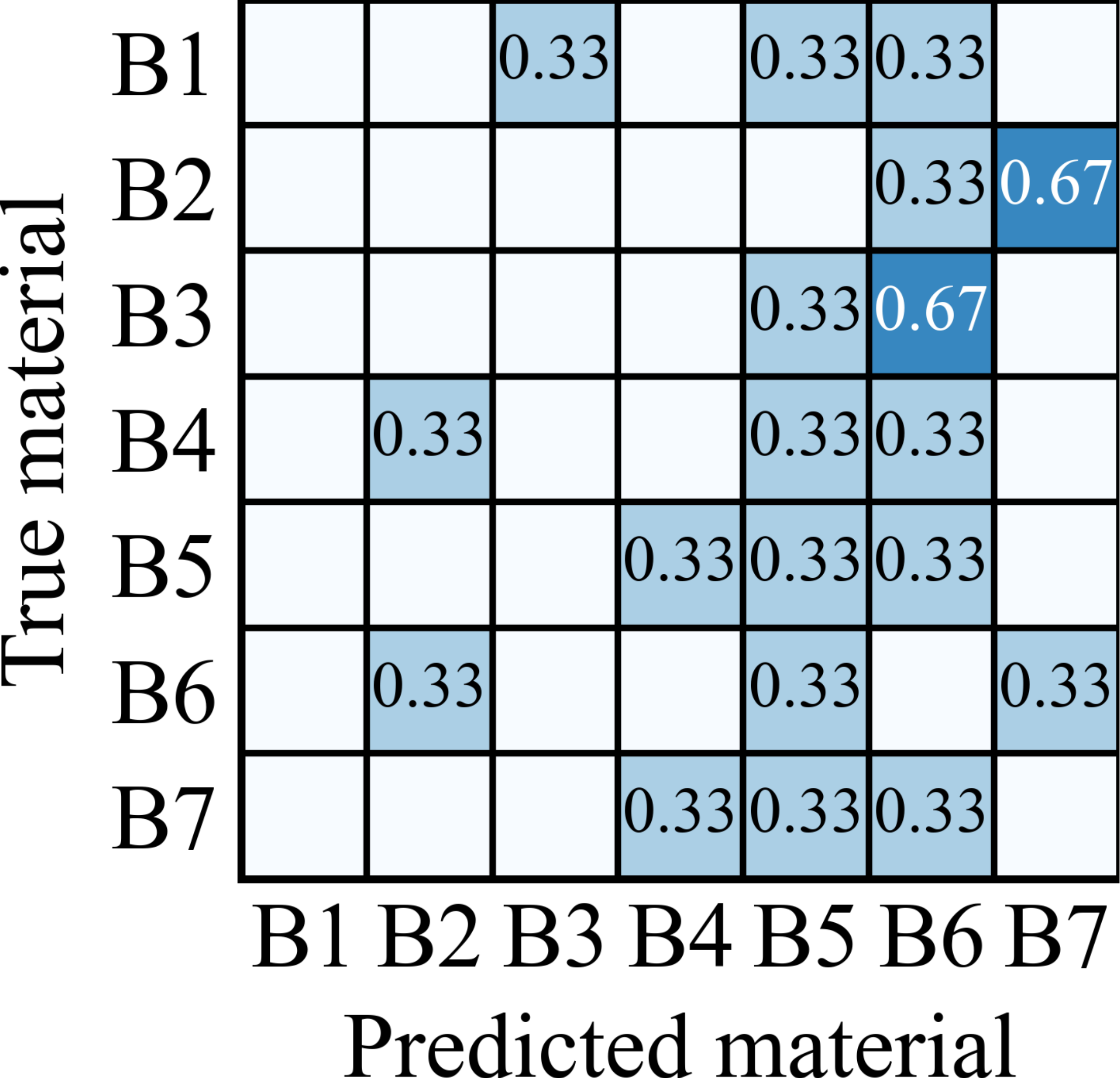}
    \label{LLMaterial_41_v1-2}
  }\hfill
  \subfloat[Kettles.]{
    \includegraphics[width=0.17\textwidth]{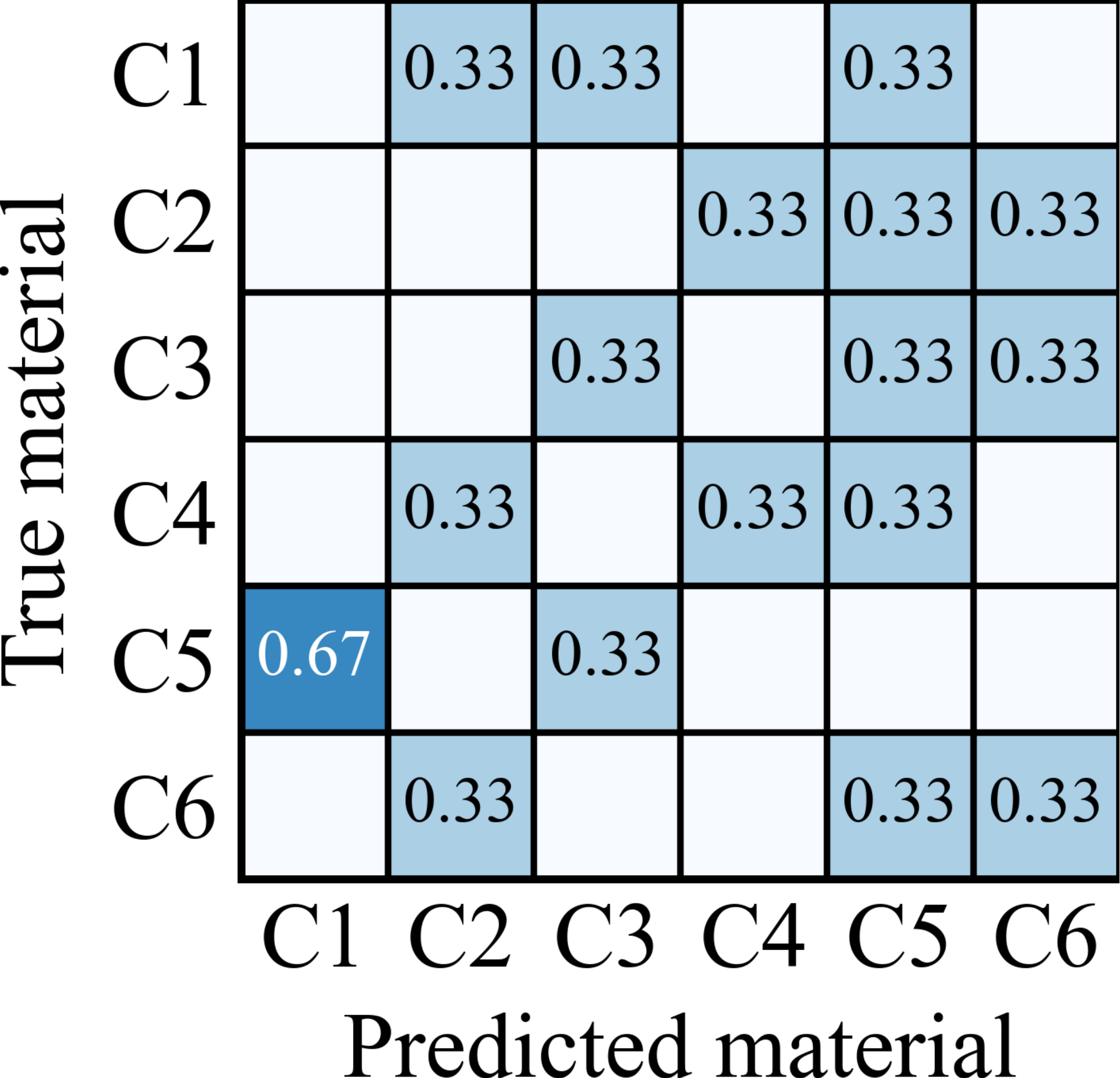}
    \label{LLMaterial_41_v1-3}
  }\hfill
  \subfloat[Boxes.]{
    \includegraphics[width=0.17\textwidth]{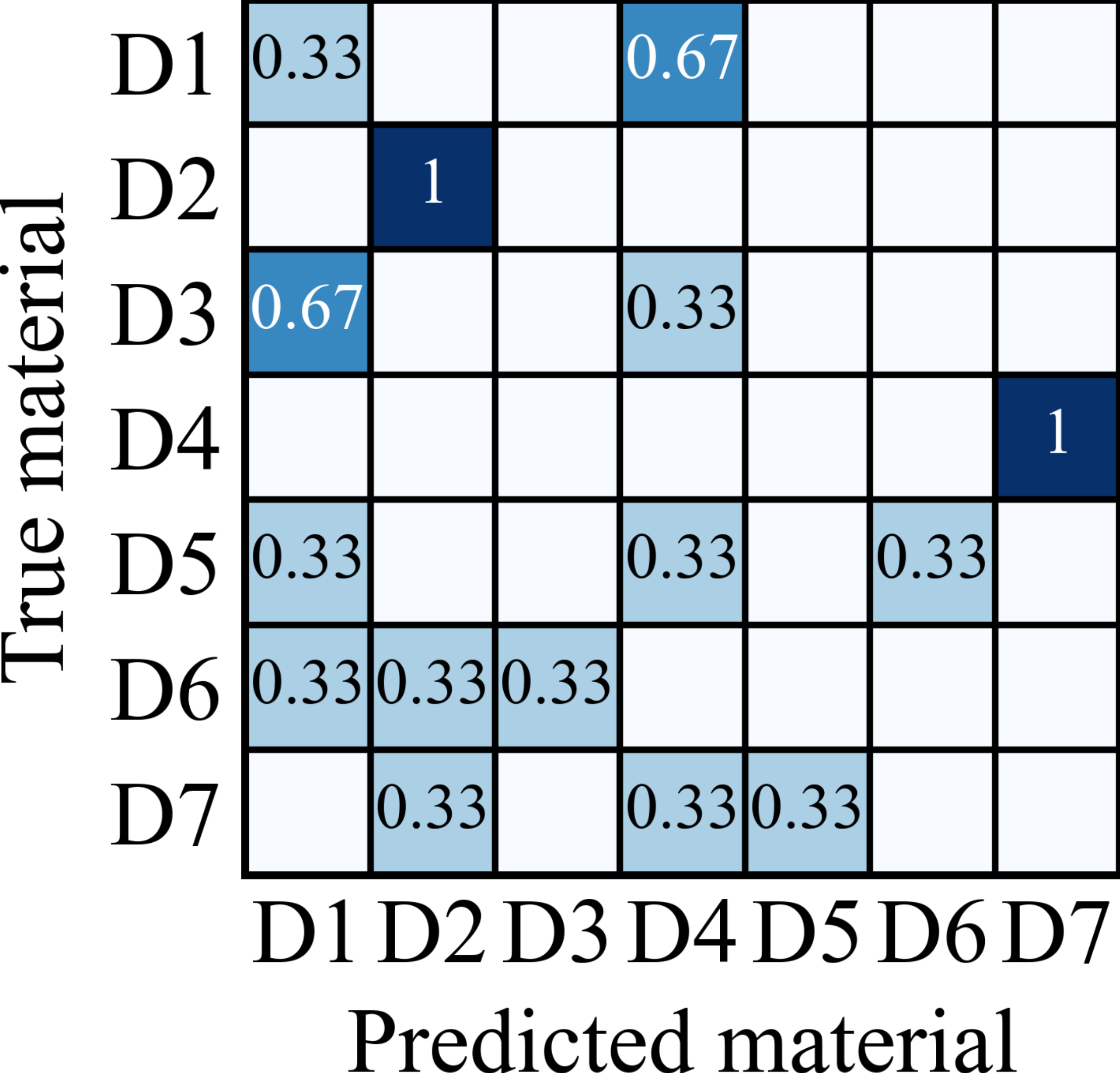}
    \label{LLMaterial_41_v1-4}
  }\hfill
  \subfloat[Plates.]{
    \includegraphics[width=0.17\textwidth]{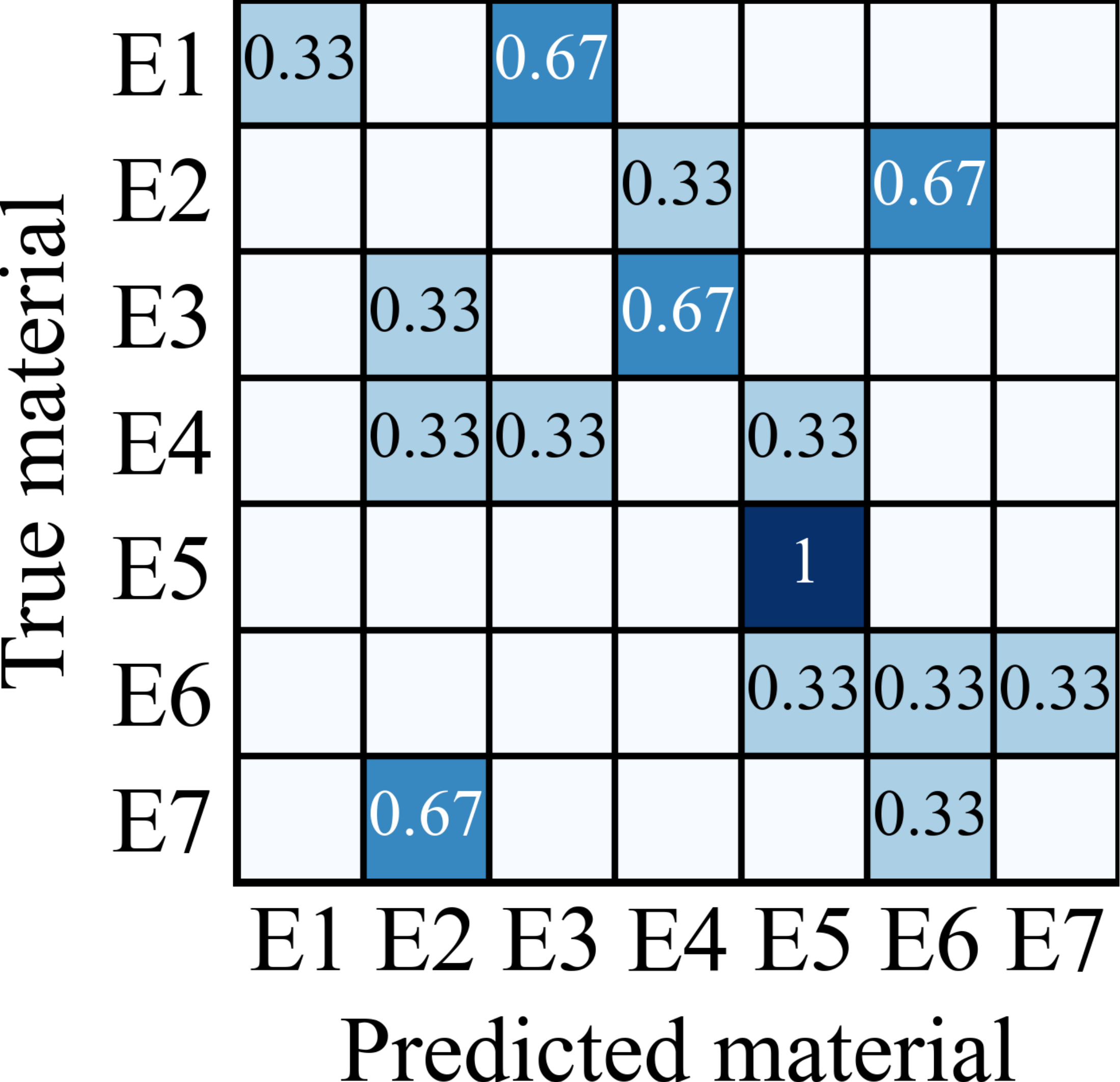}
    \label{LLMaterial_41_v1-5}
  }\\
  \caption{Overall object material recognition (LLMaterial).}
  \label{LLMaterial_41}
\end{figure*}

\begin{figure*}[!t]
  \centering
  \subfloat[Cups.]{
    \includegraphics[width=0.17\textwidth]{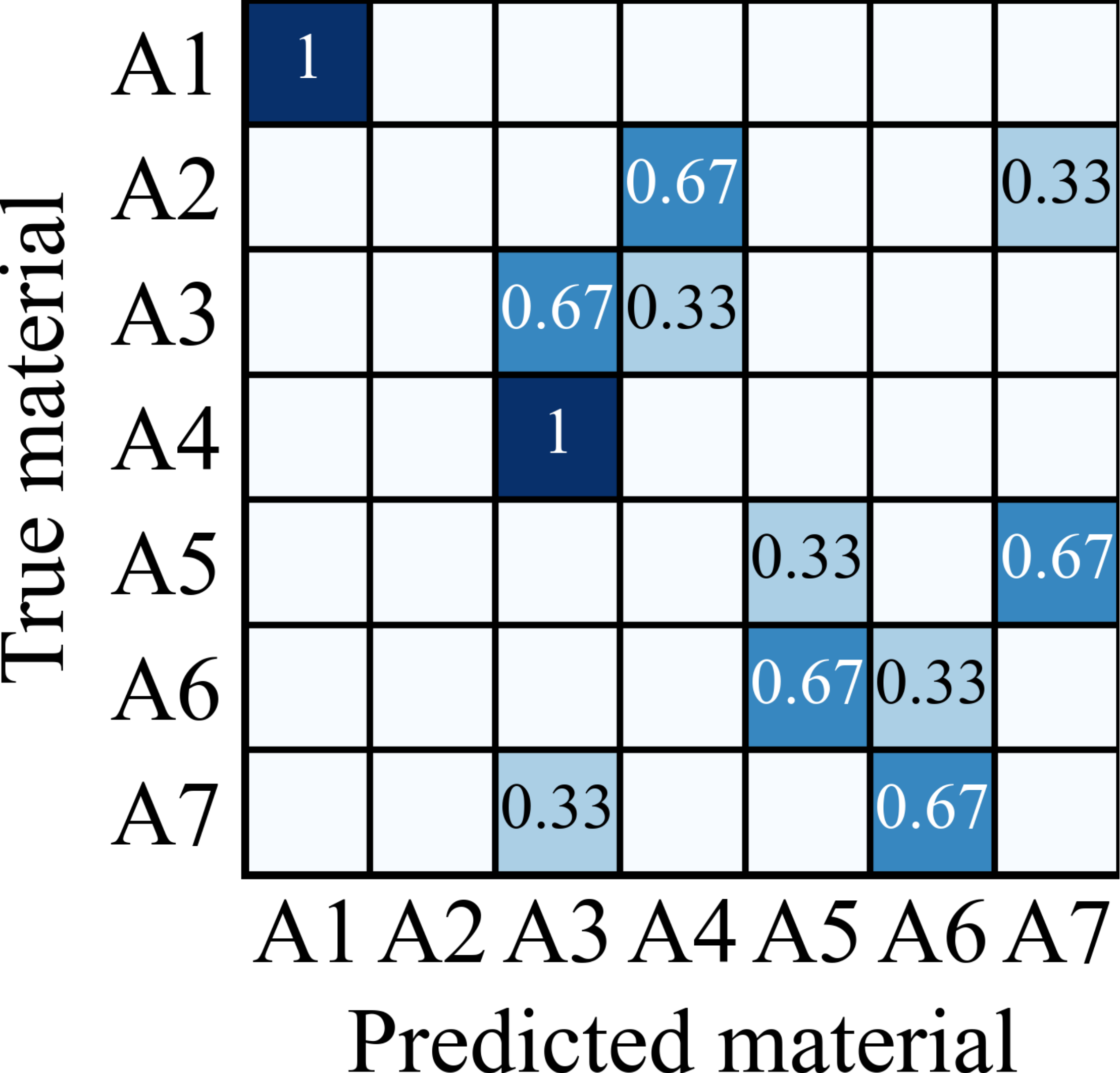}
    \label{Radar_only_41_v1-1}
  }\hfill
  \subfloat[Bottles.]{
    \includegraphics[width=0.17\textwidth]{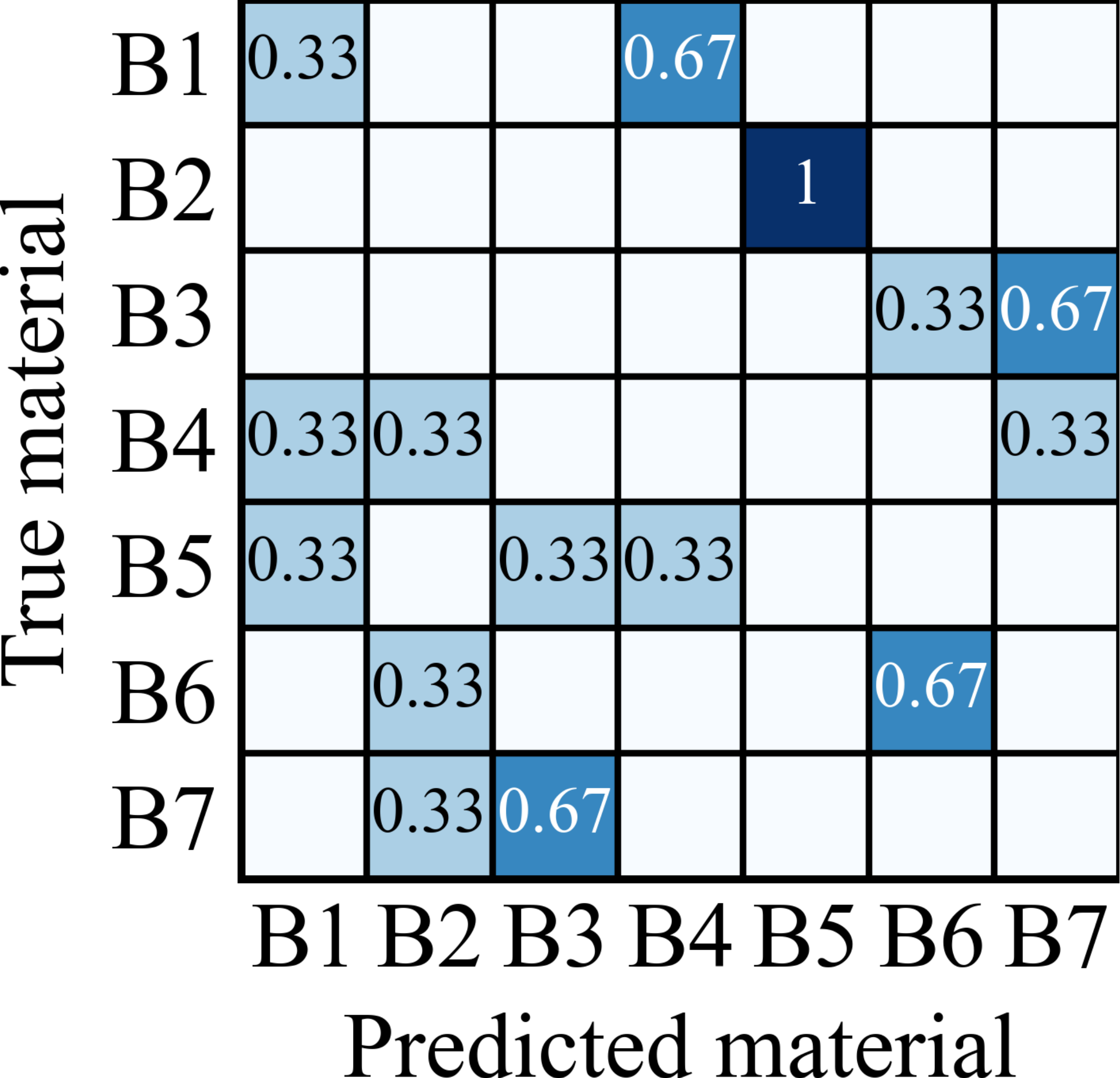}
    \label{Radar_only_41_v1-2}
  }\hfill
  \subfloat[Kettles.]{
    \includegraphics[width=0.17\textwidth]{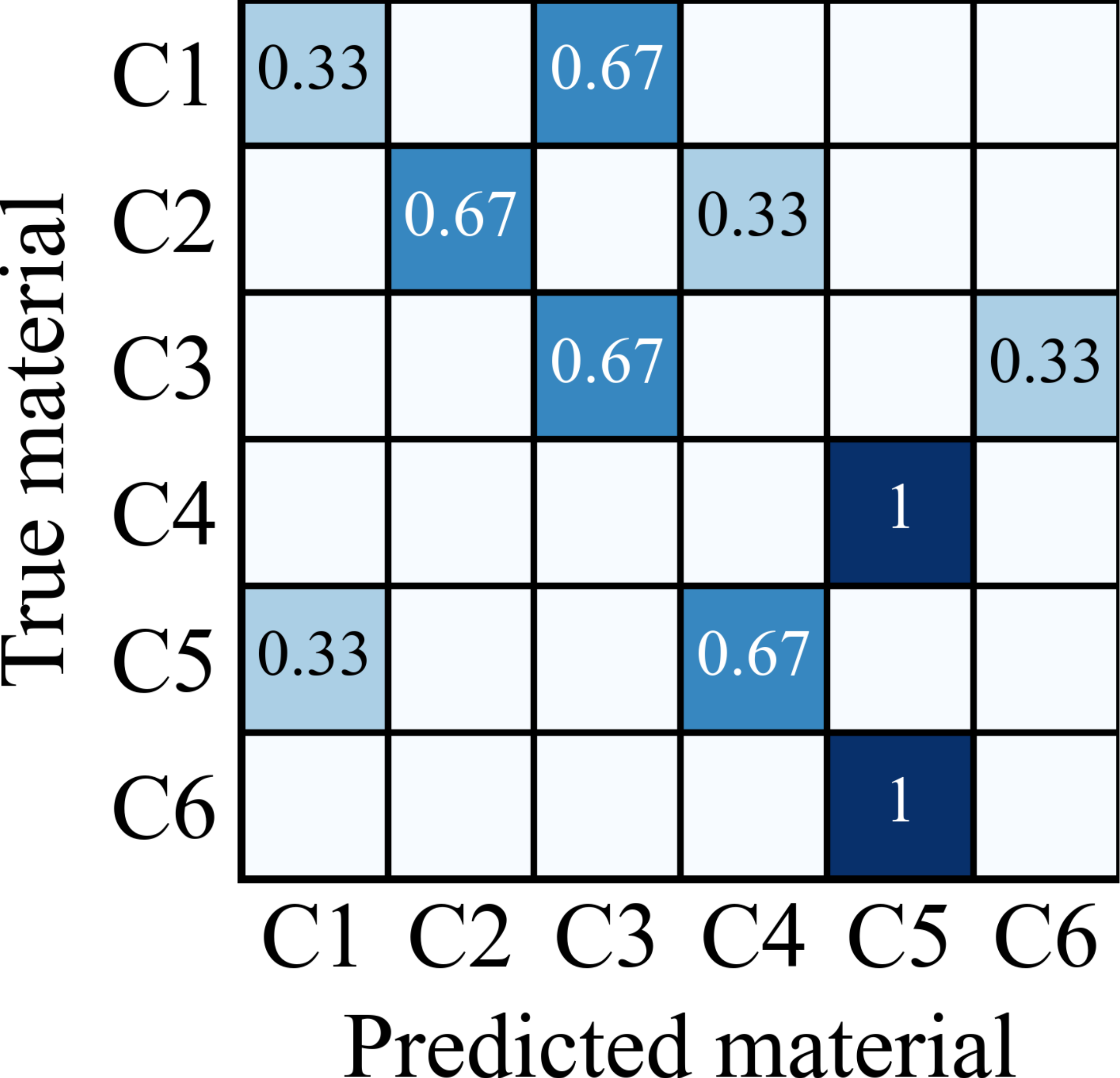}
    \label{Radar_only_41_v1-3}
  }\hfill
  \subfloat[Boxes.]{
    \includegraphics[width=0.17\textwidth]{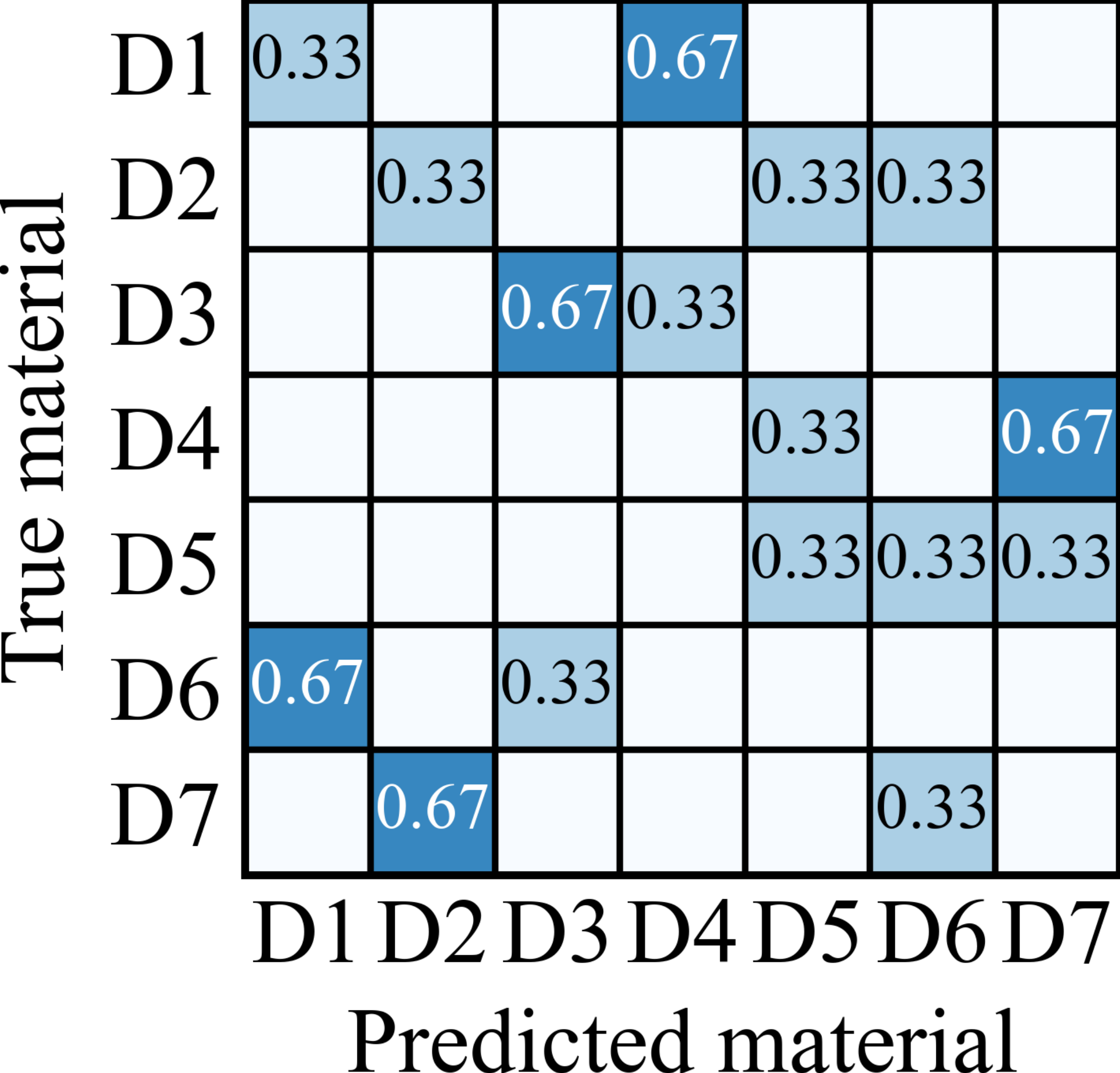}
    \label{Radar_only_41_v1-4}
  }\hfill
  \subfloat[Plates.]{
    \includegraphics[width=0.17\textwidth]{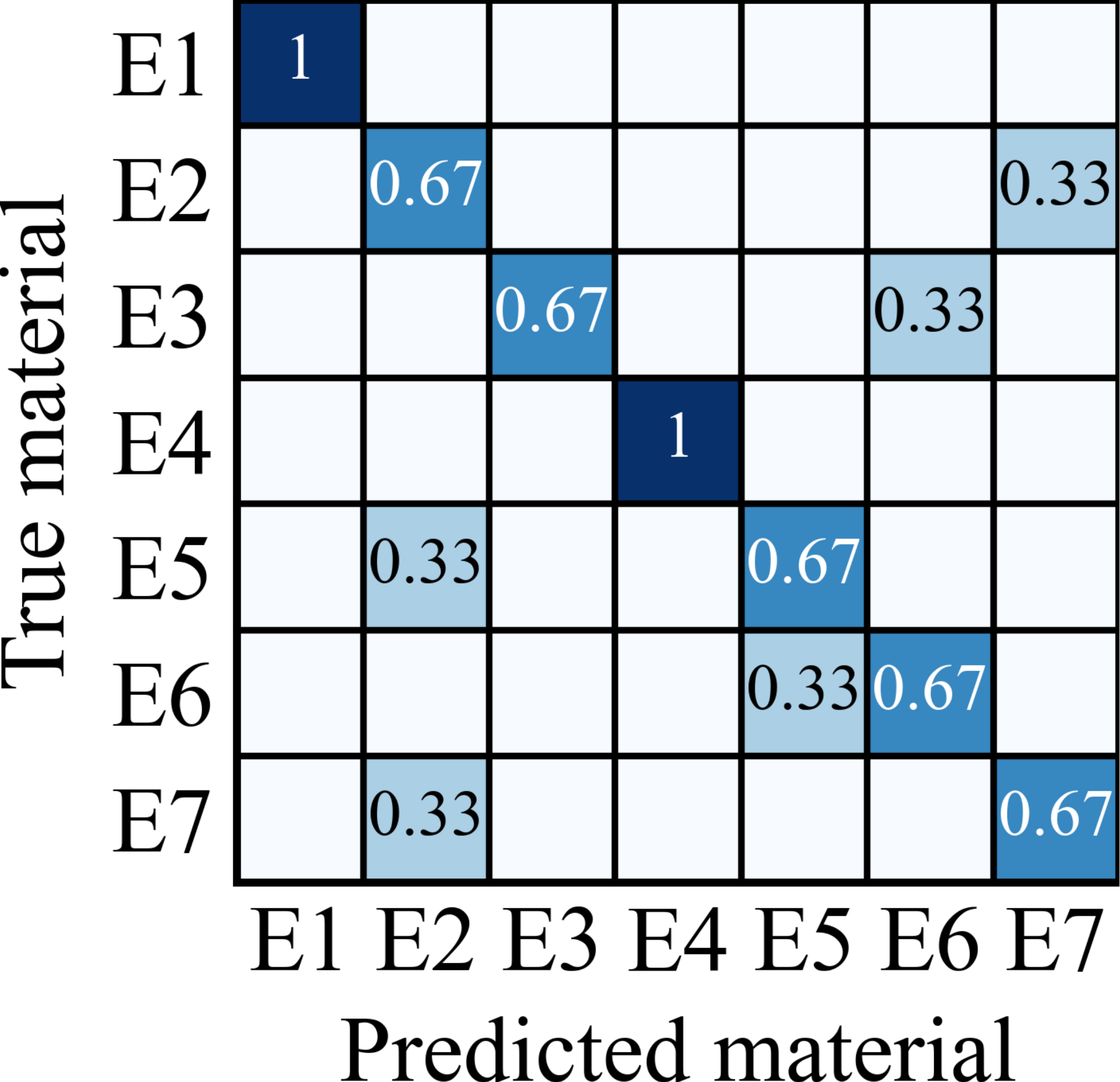}
    \label{Radar_only_41_v1-5}
  }\\
  \caption{Overall object material recognition (Radar-only (VLMaterial)).}
  \label{Radar_only_41}
\end{figure*}

\begin{table}[!t]
  \centering
  \caption{Radar parameter settings.}
  \label{Parameter_table}
  \resizebox{0.4\textwidth}{!}{%
  \begin{tabular}{c c l}
    \toprule
    Parameter & Value & Explanation \\
    \midrule
    $f_0$ & $60\,\mathrm{GHz}$ & Carrier Frequency \\
    $B$ & $3.96\,\mathrm{GHz}$ & Bandwidth \\
    $S$ & $66\,\mathrm{MHz}/\mu\mathrm{s}$ & Modulation Slope \\
    $f_s$ & $10\,\mathrm{MHz}$ & Sampling Rate \\
    \bottomrule
  \end{tabular}
  }
  \vspace{\baselineskip}
\end{table}
In our work, we first collect raw target signals using an mmWave radar sensor (IWR6843ISK) manufactured by Texas Instruments (Dallas, TX, USA), along with RGB images captured by an Intel RealSense (Santa Clara, CA, USA) D435i RGB-D camera. Both sensing modules are co-mounted on a custom-designed 3D-printed stand. The parameters of mmWave radar are listed in \Cref{Parameter_table}. Our signal processing pipeline was implemented in MATLAB R2023a. All experiments were conducted on a Windows desktop equipped with a 2.60 GHz Intel Core i7-9750H CPU and 16 GB of RAM. We conducted the experiments using the SOTA Gemini-3-Pro.

\subsection{Material Recognition Performance} 

To evaluate VLMaterial on real-world objects, we used 41 everyday items (\Cref{all_targets}) across three realistic environments (\Cref{three_scenarios}). Objects were oriented with relatively smooth surfaces toward the radar (e.g., A1's flat bottom facing the radar) at approximately $25$ cm distance. The laboratory setup is shown in \Cref{three_scenarios}d. For each object, we obtained synchronized RGB and radar signals across the three scenarios. We evaluate on cups (Group A), bottles (Group B), kettles (Group C), boxes (Group D), and plates (Group E), then compare three baselines: VLM-only (\Cref{Feasibility}), LLMaterial, and Radar-only (VLMaterial).
Notably, Radar-only (VLMaterial) uses only the eight radar-derived parameters.


{\textbf{LLMaterial.}} Our evaluation across three scenarios reveals suboptimal LLMaterial performance, with frequent misclassifications (\Cref{LLMaterial_41}). Metal was confused with frosted ceramic (e.g., A1, D1) or smooth glass (e.g., B1, C1, E1); frosted glass with wood (e.g., A2, E2); ceramic with glass (e.g., B4, C4, E4); plastic with paper (e.g., A5) or wood (e.g., B5); wood with frosted glass (B6); and paper with frosted glass (e.g., A7, D7, E7). These failures stem from two limitations. First, conventional processing struggles with overlapping dielectric constants, especially for smaller objects. Second, LLMaterial relies on RAG from theoretical online data (unlike CAG), lacking empirical grounding detached from real-world complexities. Thus, LLMaterial achieves only 19.69\% accuracy, insufficient for robust physics-grounded material recognition.

\begin{figure*}[!t]
  \centering
  \subfloat[Cups.]{
    \includegraphics[width=0.17\textwidth]{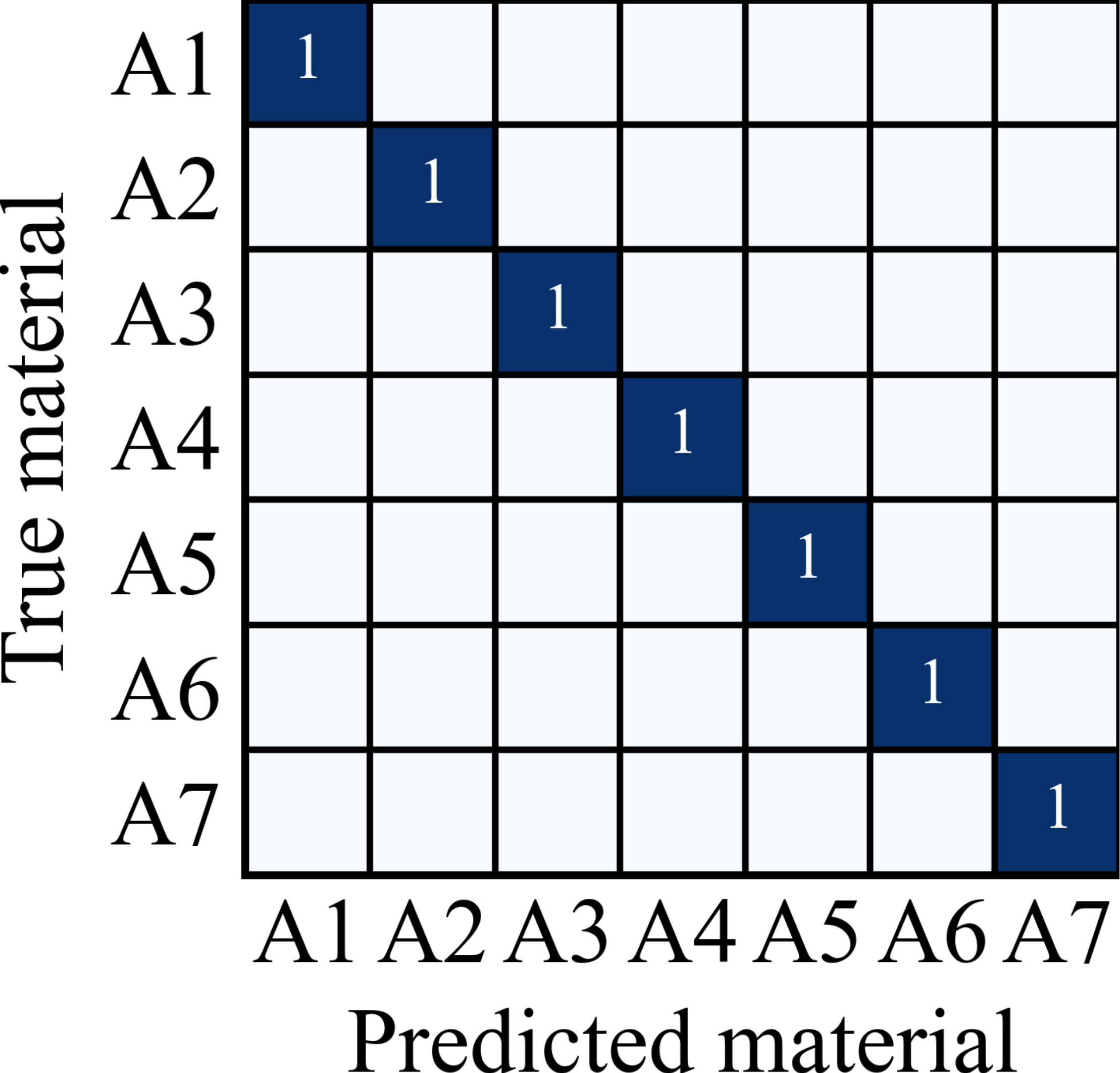}
    \label{VLMaterial_41_v1-1}
  }\hfill
  \subfloat[Bottles.]{
    \includegraphics[width=0.17\textwidth]{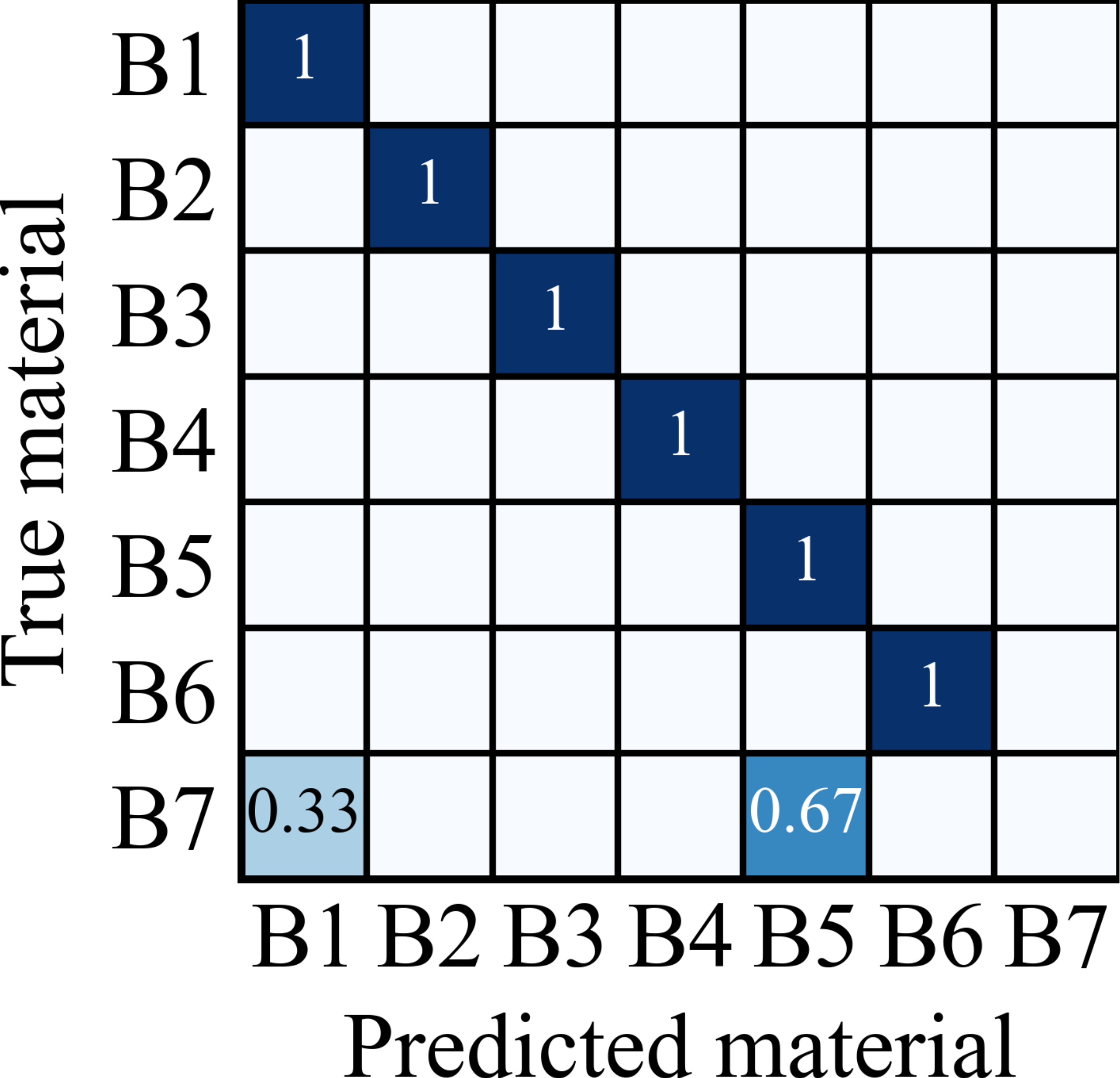}
    \label{VLMaterial_41_v1-2}
  }\hfill
  \subfloat[Kettles.]{
    \includegraphics[width=0.17\textwidth]{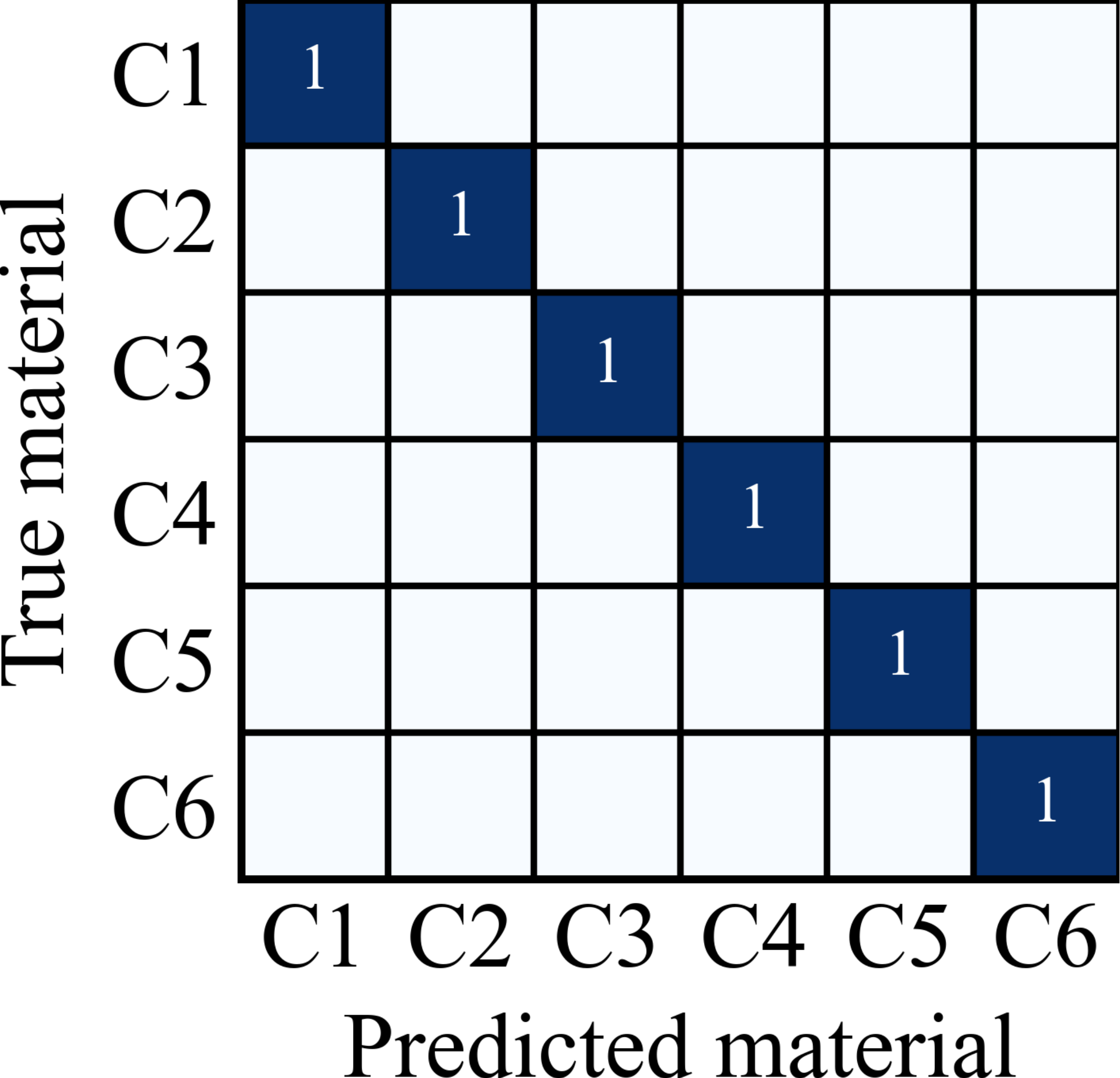}
    \label{VLMaterial_41_v1-3}
  }\hfill
  \subfloat[Boxes.]{
    \includegraphics[width=0.17\textwidth]{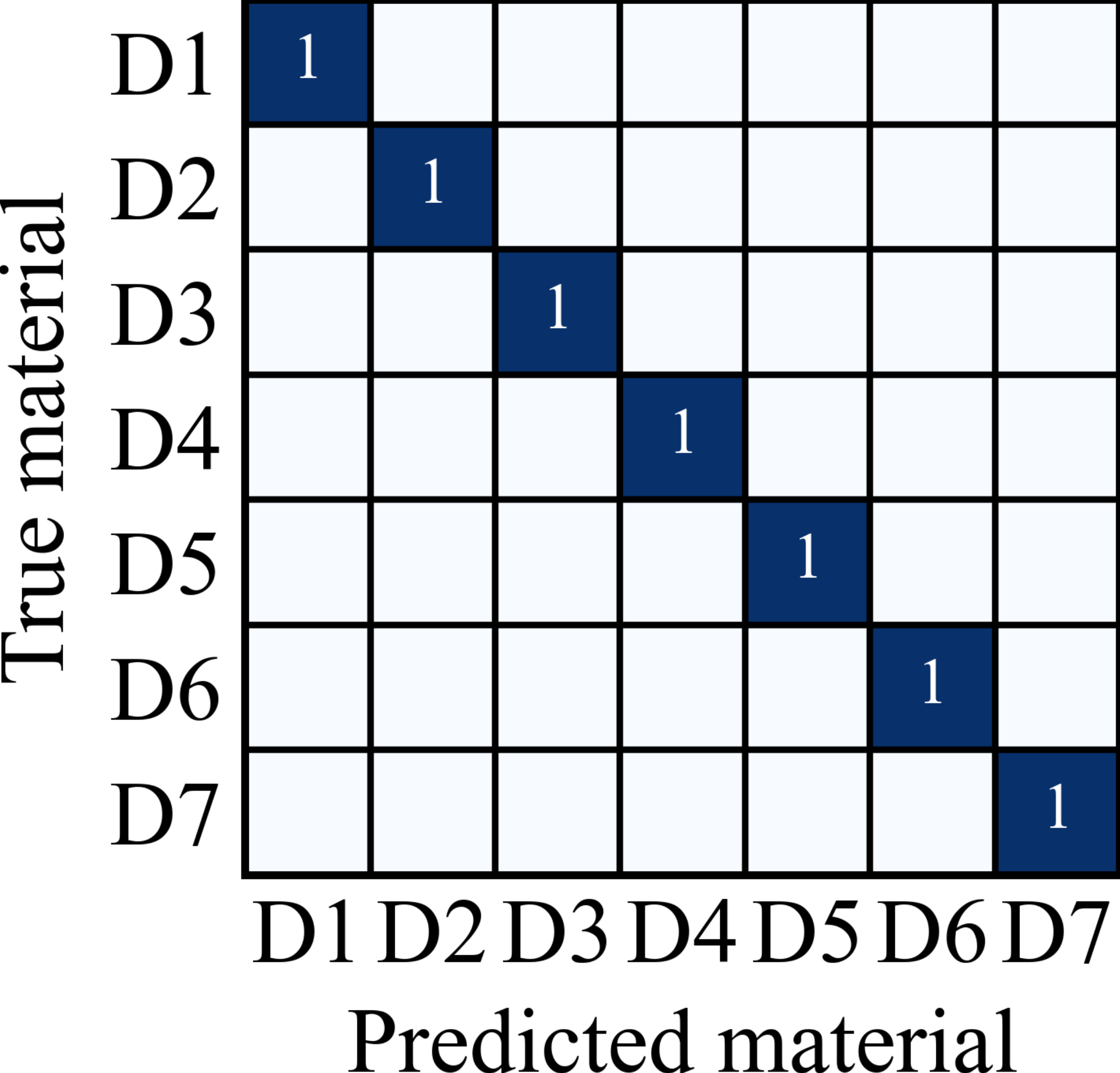}
    \label{VLMaterial_41_v1-4}
  }\hfill
  \subfloat[Plates.]{
    \includegraphics[width=0.17\textwidth]{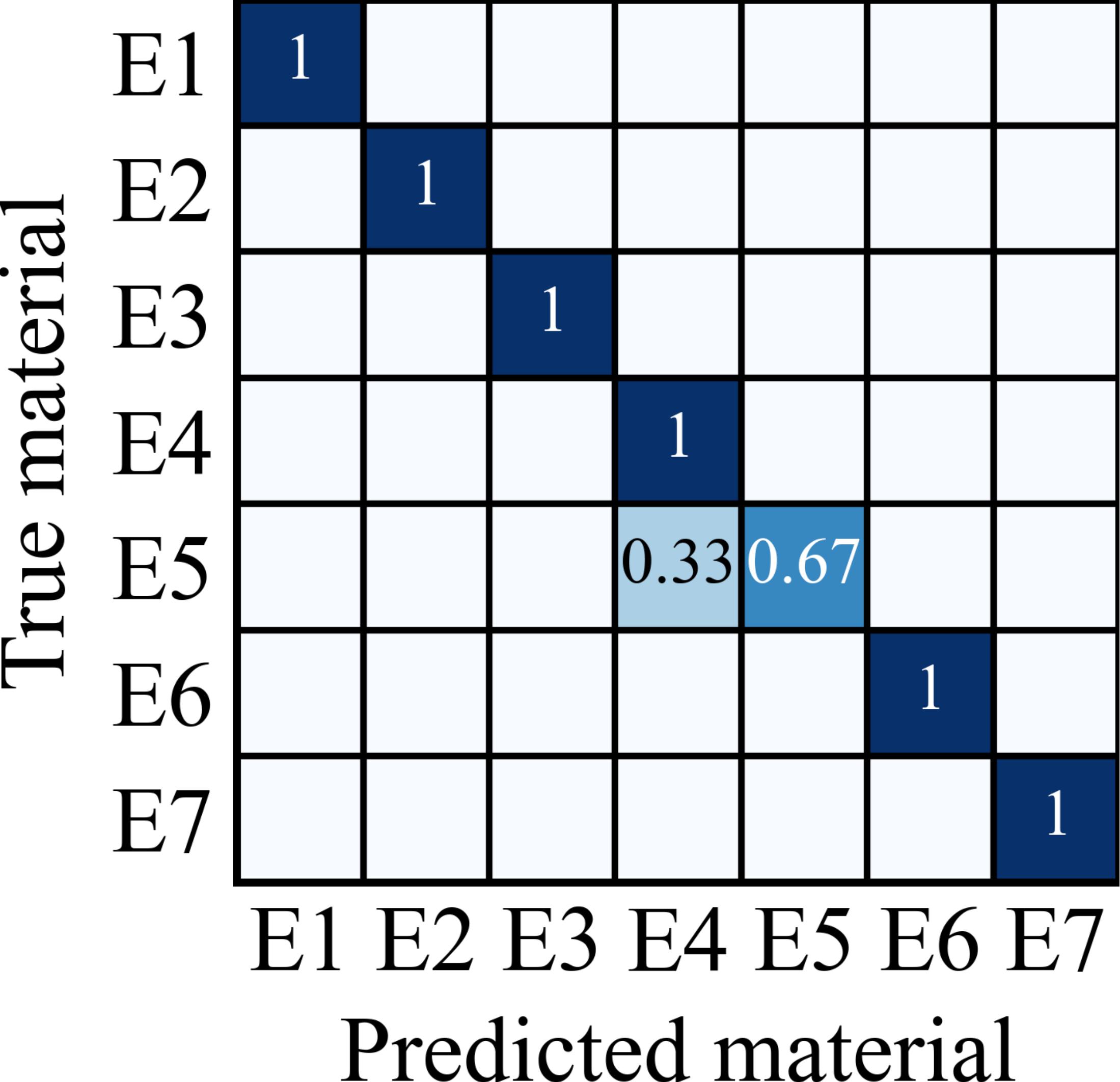}
    \label{VLMaterial_41_v1-5}
  }\\
  \caption{Overall object material recognition (VLMaterial).}
  \label{VLMaterial_41}
\end{figure*}

{\textbf{Radar-only (VLMaterial).} The recognition results are detailed in \Cref{Radar_only_41}. This system captures fundamental electromagnetic features, especially for smaller objects and materials with distinct dielectric properties, significantly improving over LLMaterial. However, classification errors persist due to dielectric constant ambiguity. Metals were misidentified as specially processed ceramic (e.g., B1, D1) or smooth glass (e.g., C1). Glass confusion occurred: frosted glass as ceramic (e.g., A2, C2), mirror glass as ceramic (e.g., A3, D3). Ceramic was misclassified as glass (e.g., A4) or plastic (e.g., D4) due to surface roughness causing signal scattering rather than specular reflection. Plastic was mislabeled as paper (e.g., A5, D5) or glass (e.g., E5), wood as plastic (e.g., A6, C6), and paper as glass (e.g., B7, D7, E7). Signal loss lowers the calculated dielectric constant, leading to paper misclassification, while exceptionally smooth wood (D6) and polished paper packaging are misidentified as glass. These issues, prevalent in \Cref{Radar_only_41}, show radar sensing is susceptible to overlapping dielectric constants, surface roughness, and positioning errors. Despite these challenges, Radar-only (VLMaterial) achieves 41.73\% accuracy, substantially outperforming LLMaterial.

{\textbf{VLMaterial.}} 
Given that vision fails under visual similarity or poor lighting, while radar suffers from overlapping dielectric constants, we explore and validate that integrating both modalities is essential to overcome their inherent limitations. 
Detailed results are in \Cref{VLMaterial_41}. Our fusion strategy operates on two principles. First, when visual and radar candidate sets intersect, we select the common category. For example, for the heart-shaped plastic box (D5), the VLM suggests mirror glass or plastic, while radar-based analysis identifies it as plastic, wood, or paper. The intersection gives plastic. Second, when predictions do not overlap, we cross-verify using physical properties. Consider the frosted glass cup (A2): the VLM proposes frosted glass or plastic; the radar classifies it as ceramic. Although the radar classification is incorrect, its physical parameter is critical. Since the measured dielectric constant exceeds that of plastic, VLMaterial removes the plastic candidate. Then, integrating the high-confidence visual input, the system correctly identifies the object as frosted glass.
Ultimately, VLMaterial achieves an accuracy of 96.08\%. 

\begin{figure*}[!t]
  \centering
  \includegraphics[width=0.8\linewidth]{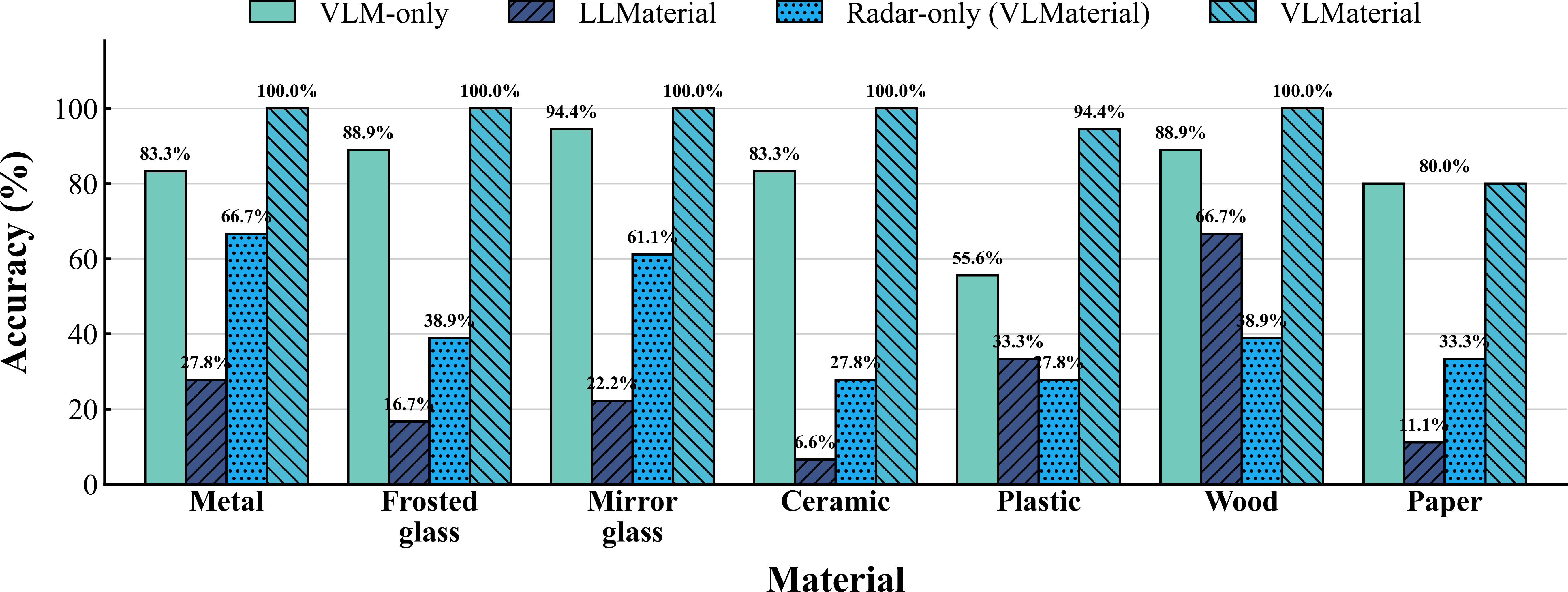}
    \caption{Material recognition accuracy.}
  \label{VLM_Radar_VLMaterial_41_bar_acc}
\end{figure*}

{\textbf{Baseline Comparisons.}} Recognition accuracies for seven material types are summarized in \Cref{VLM_Radar_VLMaterial_41_bar_acc}. VLMaterial achieves 100\% accuracy for five material categories, while achieving 94.44\% for plastic and 80.00\% for paper. The paper drop comes from bottle B7, whose smooth surface generates strong reflections, causing misclassification as metal (33.33\%) or plastic (66.67\%). A random error also occurred on plastic plate E5 (scenario 2), misidentified as ceramic.
VLM-only baseline ranges from 55.56\% to 94.44\%, showing vision alone is unreliable for plastic and paper. Radar-only (VLMaterial) performs worst, dropping to 27.78\% for ceramic and plastic, with a maximum of 66.67\% for metal.
VLMaterial achieves the highest accuracy for every material type, demonstrating mutual complementarity between modalities. Crucially, beyond validating on idealized large flat surfaces from prior work, we show VLMaterial's robustness and high accuracy on small, irregular everyday objects with diverse geometries.

\begin{table*}[!t]
\centering
\caption{Performance comparison of existing methods.}
\label{tab_comparison}
\renewcommand{\arraystretch}{1.2}
\resizebox{\textwidth}{!}{%
\begin{tabular}{c|c|c|c|c|c|c}
\hline
\textbf{Method} & LLMaterial \cite{zhu2025can} & VLM-only & Radar-only (VLMaterial) & \textbf{VLMaterial} & mSense \cite{wu2020msense} & CRFUSION \cite{xiao2025crfusion}\\
\midrule 
Training-free & \multicolumn{4}{c|}{\cmark}  & \multicolumn{2}{c}{\xmark} \\
\midrule
Recognition accuracy & 19.69\% & 78.74\% & 41.73\% & \textbf{96.08\%} & 93.00\% & 96.00\%   \\
\midrule
\end{tabular}
}
\end{table*}

\begin{figure*}[!t]
  \centering
  \begin{minipage}[t]{0.48\linewidth}
    \centering
    \includegraphics[width=\textwidth]{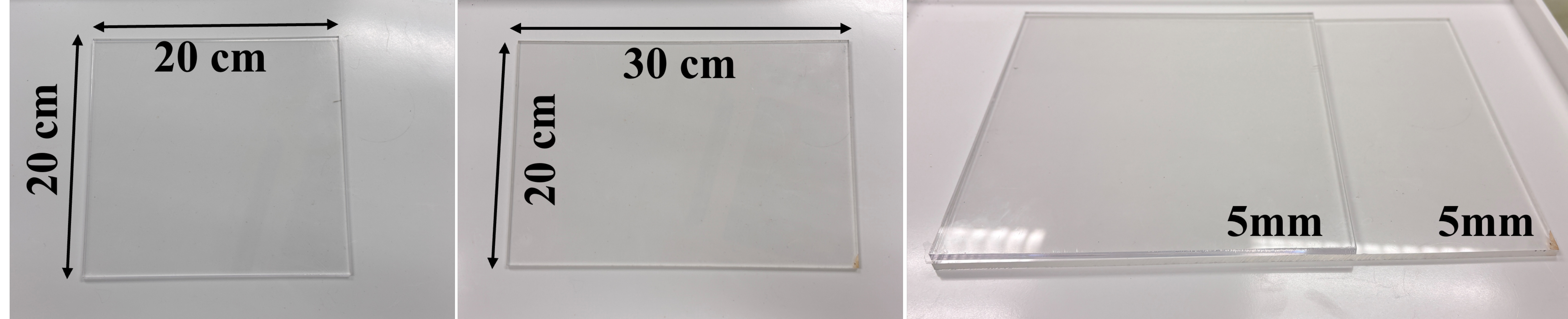}
  \caption{Plastic boards (different sizes, same thickness).}
  \label{dif_sickness_plastic}
  \end{minipage}
  \hfill
  \begin{minipage}[t]{0.48\linewidth}
    \centering
    \includegraphics[width=\textwidth]{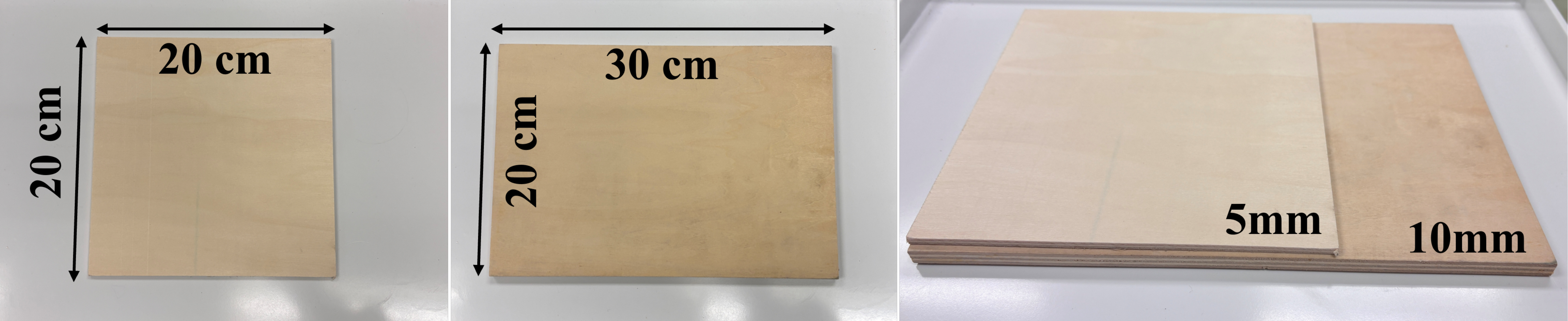}
  \caption{Wood boards (different sizes, different thicknesses).}
  \label{dif_sickness_wood}
  \end{minipage}
\end{figure*}

So far, to the best of our knowledge, only one existing work, CRFUSION \cite{xiao2025crfusion}, has demonstrated the capability to simultaneously detect both category and material with top accuracy. Regarding physics-grounded material recognition using radar signals, LLMaterial \cite{zhu2025can} is, as far as we know, the only existing approach that analyzes mmWave radar reflection signals. Nevertheless, we conducted a comparative evaluation of recognition performance, as shown in \Cref{tab_comparison}. The results demonstrate that our proposed VLMaterial, leveraging camera–radar fusion, achieves a physics-grounded recognition accuracy of 96.08\%. This performance is comparable to that of mSense \cite{wu2020msense} (93\%) and CRFUSION \cite{xiao2025crfusion} (96\%), both of which require dedicated closed-set training. Furthermore, our method significantly outperforms single-modality baselines in physics-grounded material recognition, surpassing VLM-only and Radar-only (VLMaterial) approaches, which achieved accuracies of 78.74\% and 41.73\%, respectively.

\subsection{Micro-benchmarks}
\label{microbenchmarks}
To comprehensively evaluate VLMaterial's contribution, we conduct a series of micro-benchmarks, utilizing the laboratory setting as a representative environment for precise analysis.

{\textbf{Impact of Size and Thickness.}} 
To evaluate VLMaterial's robustness against variations in object size and thickness, we conducted experiments on plastic and wood boards. Two plastic boards with identical thickness (5 mm) but different sizes (20 cm $\times$ 20 cm and 30 cm $\times$ 20 cm) were used to assess size effects (\Cref{dif_sickness_plastic}). Two wooden boards differing in both size and thickness (20 cm $\times$ 20 cm $\times$ 5 mm and 30 cm $\times$ 20 cm $\times$ 10 mm) were used to examine the combined impact of size and thickness, as shown in \Cref{dif_sickness_wood}. 
The results correctly identify both plastic and wood, showing VLMaterial remains unaffected by size or thickness due to our proposed PRCA.

\begin{figure*}[!t]
  \centering
  \begin{minipage}[b]{0.28\linewidth}
    \centering
    \includegraphics[width=\linewidth]{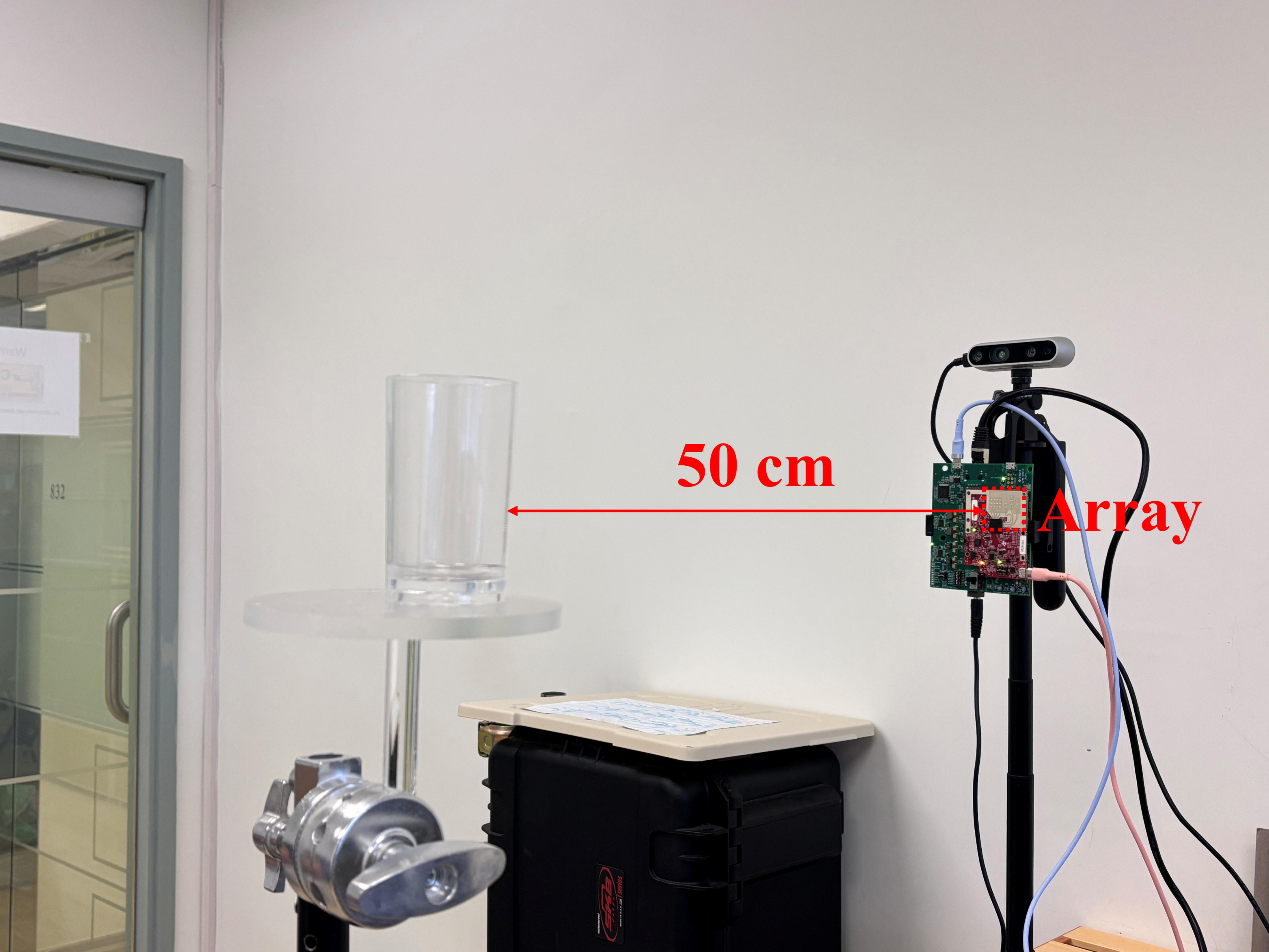}
    \caption{Scene ($50~\mathrm{cm}$, bright light).}
  \label{scenario_dif_distance}
  \end{minipage}
  \hfill
  \begin{minipage}[b]{0.35\linewidth}
    \centering
    \includegraphics[width=\linewidth]{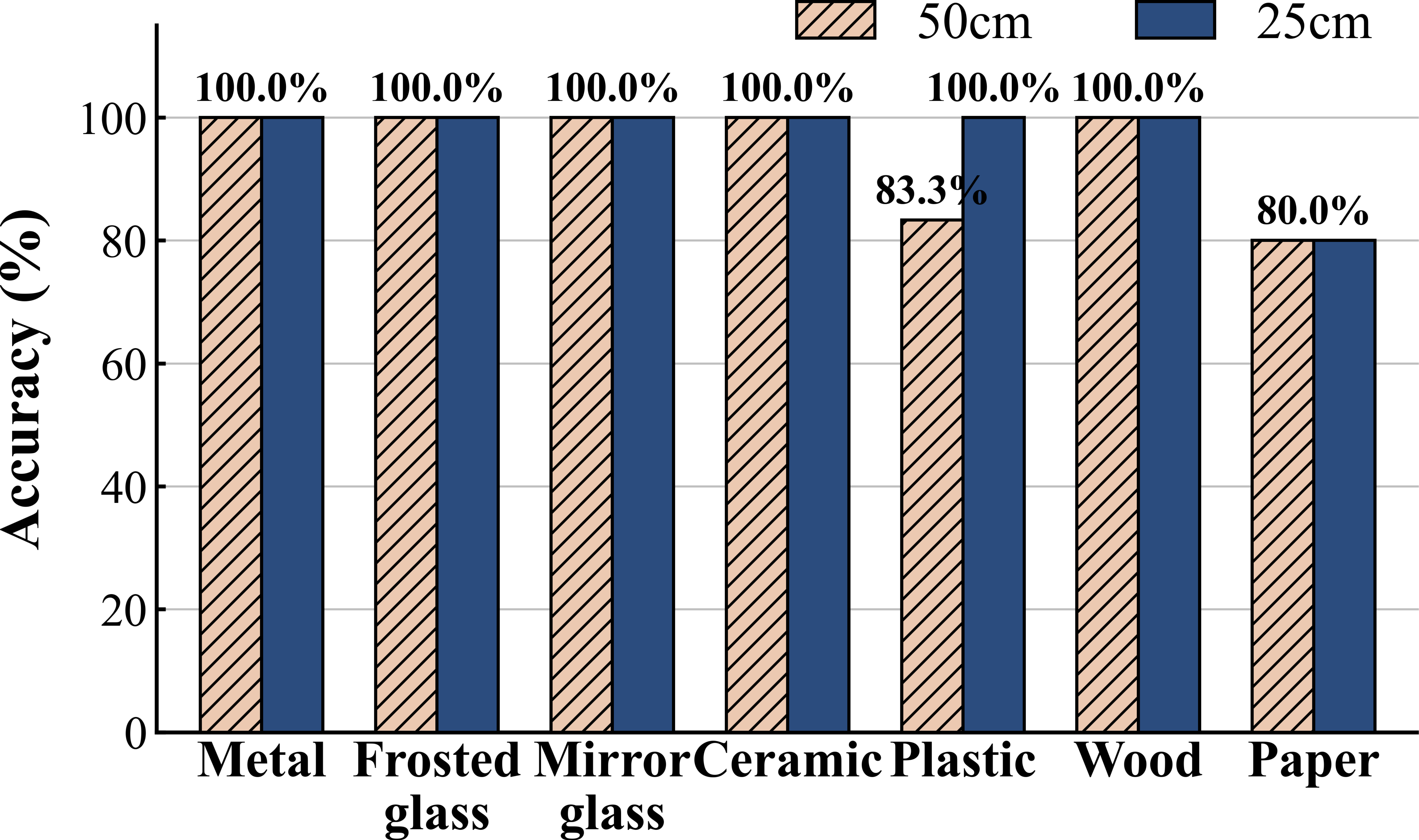}
    \caption{VLMaterial (different distances).}
  \label{dif_distance}
  \end{minipage}
  \hfill
  \begin{minipage}[b]{0.35\linewidth}
    \centering
    \includegraphics[width=\linewidth]{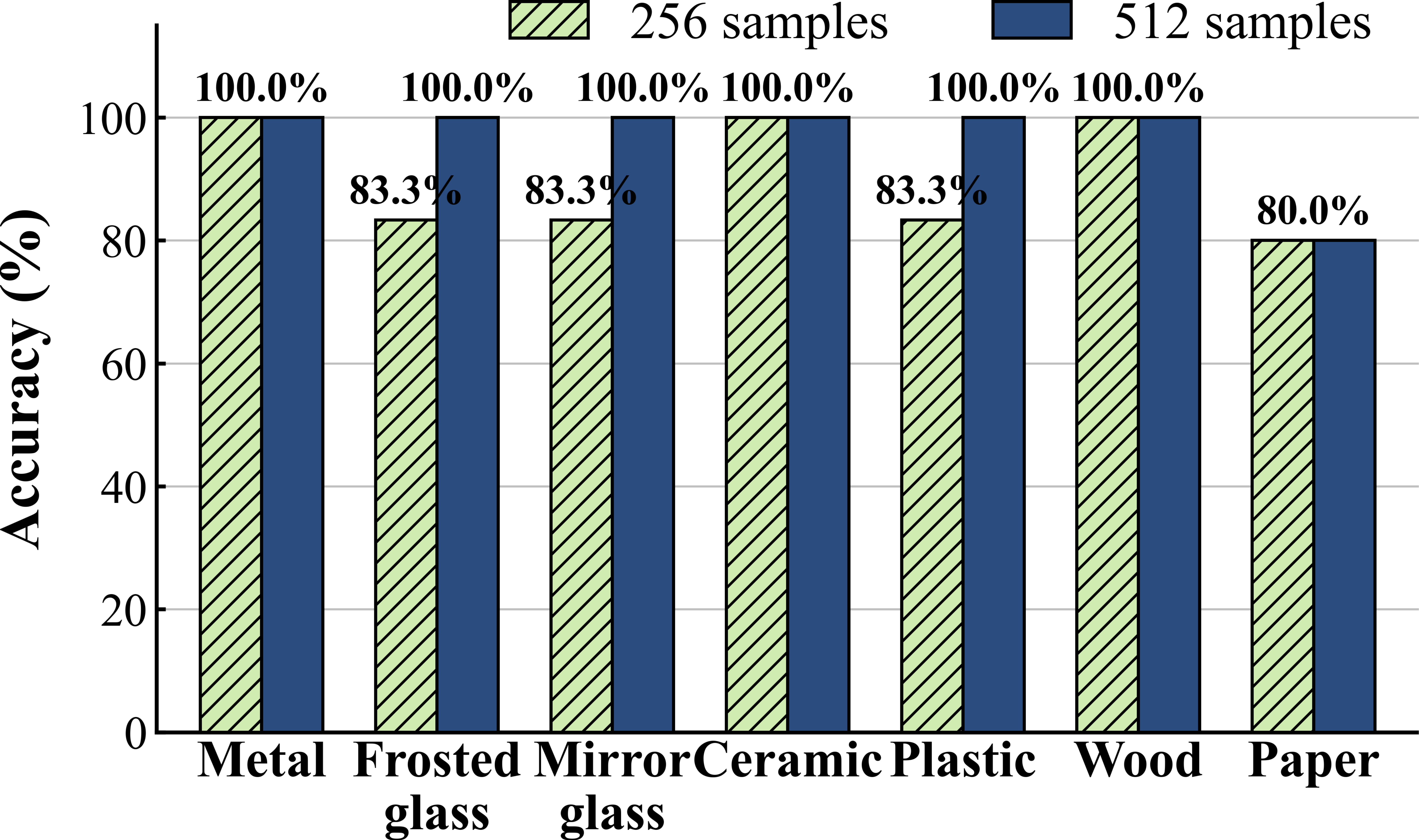}
    \caption{VLMaterial (different samples).}
  \label{dif_adc}
  \end{minipage}
\end{figure*}

{\textbf{Impact of Distance.}}
We also evaluate VLMaterial at different target distances ($25$ cm and $50$ cm, incidence angle $\approx 0^\circ$), with the experimental scene in \Cref{scenario_dif_distance}. As shown in \Cref{dif_distance}, accuracy does not fluctuate significantly with distance. The only exception is the heart-shaped plastic box (D5). For smaller targets, maintaining perpendicular beam incidence becomes harder at longer distances, while visual image quality also degrades. D5's small size reduces its occupied region within the camera's field of view (FoV) and its weak radar echoes produce a low confidence score (10\%), leading to a fused prediction of glass. Although signal attenuation causes a minor performance reduction at longer distances, the overall variation is minimal, validating VLMaterial's reliability for dynamic real-world applications.
 
{\textbf{Impact of the Number of Samples per Frame.}} 
This section analyzes the effect of sampling frequency on system performance. We compared recognition accuracy at 512 ADC sampling points per frame (maximum hardware setting) against 256 points. Reducing samples from 512 to 256 degrades theoretical distance resolution, compromising signal extraction quality and echo characteristics, particularly for smaller targets where SNR is already critical. As shown in \Cref{dif_adc}, the reduced resolution caused misclassifications: frosted glass cup (A2) as plastic, plastic cup (A5) as glass, and mirror glass cup (B3) as plastic. These errors are not random but directly attributable to the loss of fine-grained signal features. Therefore, a higher ADC sampling rate is essential for accurate material identification, especially for small objects. Moreover, 512 sampling points represents the best achievable performance for our system, making it the optimal design choice to maximize VLMaterial's robustness.

{\textbf{Impact of Ambient Illuminance.}}  
Compared to bright lighting, low-light environments (\Cref{scenario_dif_light}) introduced different recognition errors.
\textbf{(1) VLM-only.} Recognition accuracies are in \Cref{dif_light_VLM_only}. Metal bottle (B1) and metal board (F1) were misclassified as plastic; metal plate (E1) as mirror glass. Reduced visibility caused ceramic plate (E4) and board (F4) to be misclassified as plastic, plastic cup (A5) and box (D5) as glass. In extreme low light, plastic plate (E5) became paper, wooden box (D6) and wood board became paper and plastic. Conversely, dim lighting reduced specular reflections, benefiting some objects: metal board (F1), frosted glass board (F2), mirror glass board (F3), and paper bottle (B7) were correctly identified. Despite these improvements, the overall accuracy dropped from 75.61\% (bright) to 70.73\% (dim). 
\textbf{(2) VLMaterial.} Results are in \Cref{dif_light}. Dim lighting degrades camera performance, but radar fusion provides adaptability. Dim lighting mitigates specular reflections, correcting paper bottle (B7) from metal to paper, improving paper accuracy. However, wooden box (D6) is misidentified as metal, reducing wood accuracy to 83.33\%. Overall, radar corrects most visual errors, demonstrating VLMaterial's robustness to lighting variations. While sufficient light is generally preferred, excessively bright environments introduce specular reflections; moderate lighting may offer a better balance.

\begin{figure*}[!t]
  \centering
  \begin{minipage}[b]{0.32\linewidth}
    \centering
    \includegraphics[width=0.82\linewidth]{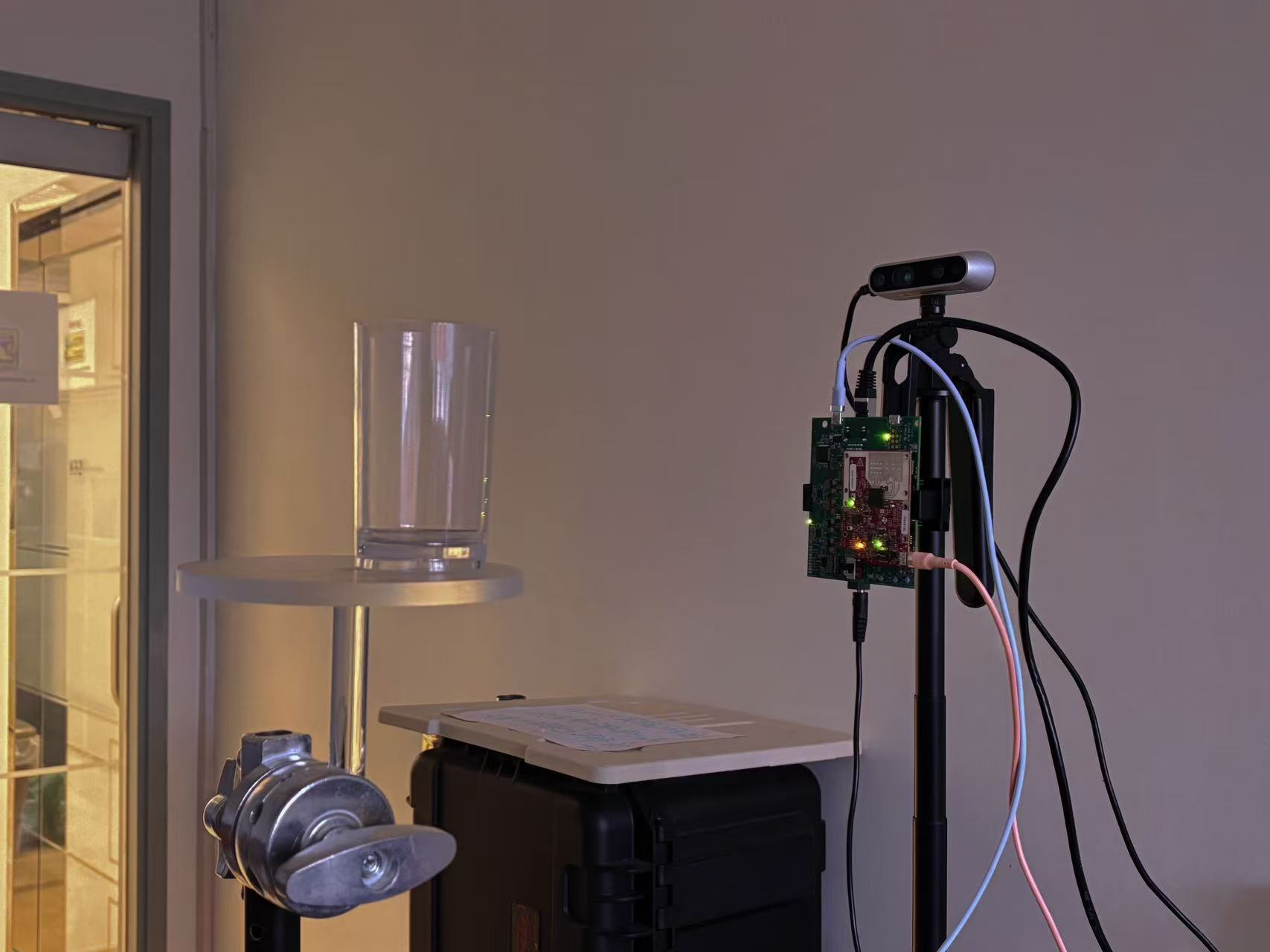}
    \caption{Scene ($25~\mathrm{cm}$, dim light).}
  \label{scenario_dif_light}
  \end{minipage}
  \hfill
  \begin{minipage}[b]{0.32\linewidth}
    \centering
    \includegraphics[width=\linewidth]{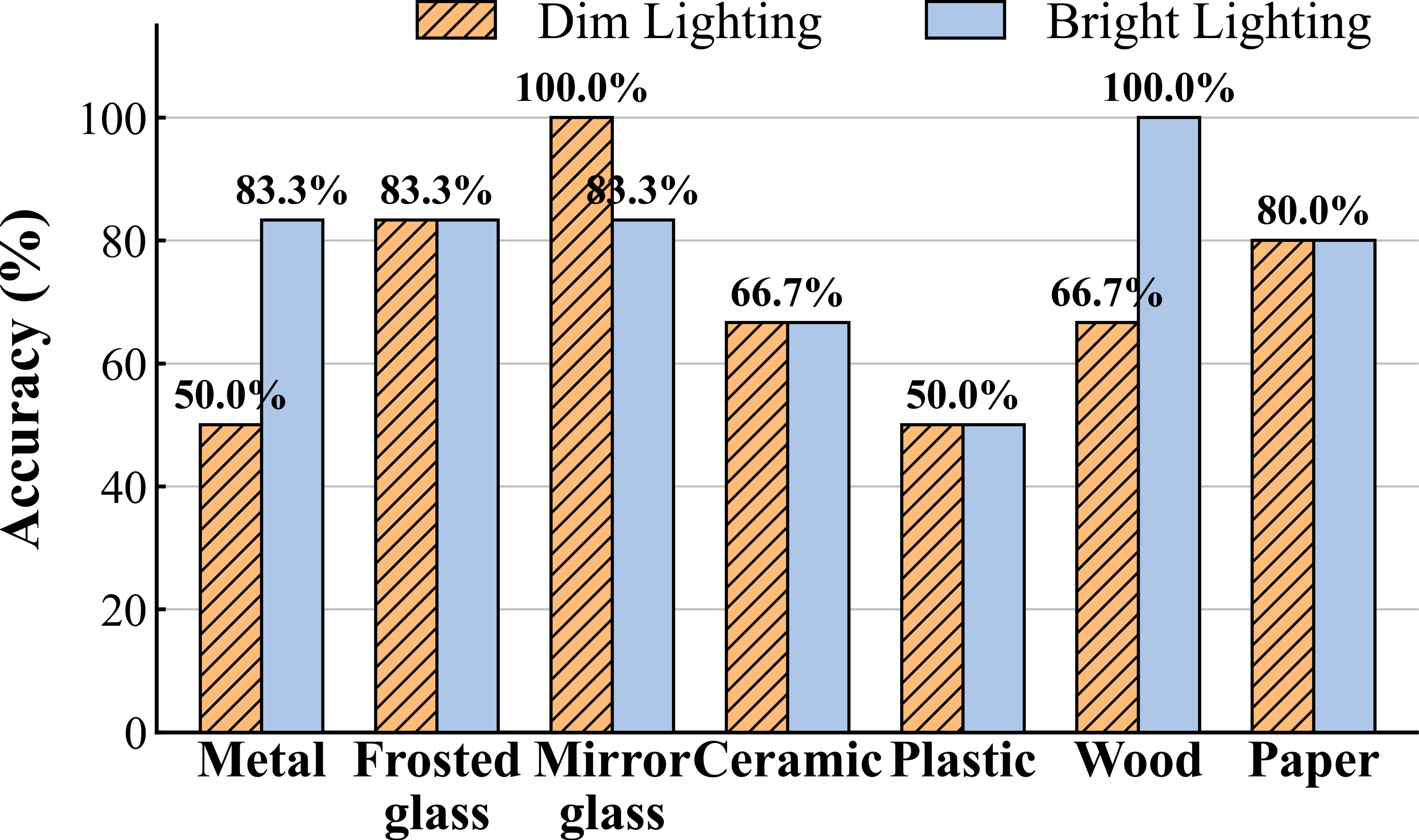}
    \caption{VLM-only (different lighting).}
  \label{dif_light_VLM_only}
  \end{minipage}
  \hfill
  \begin{minipage}[b]{0.32\linewidth}
    \centering
    \includegraphics[width=\linewidth]{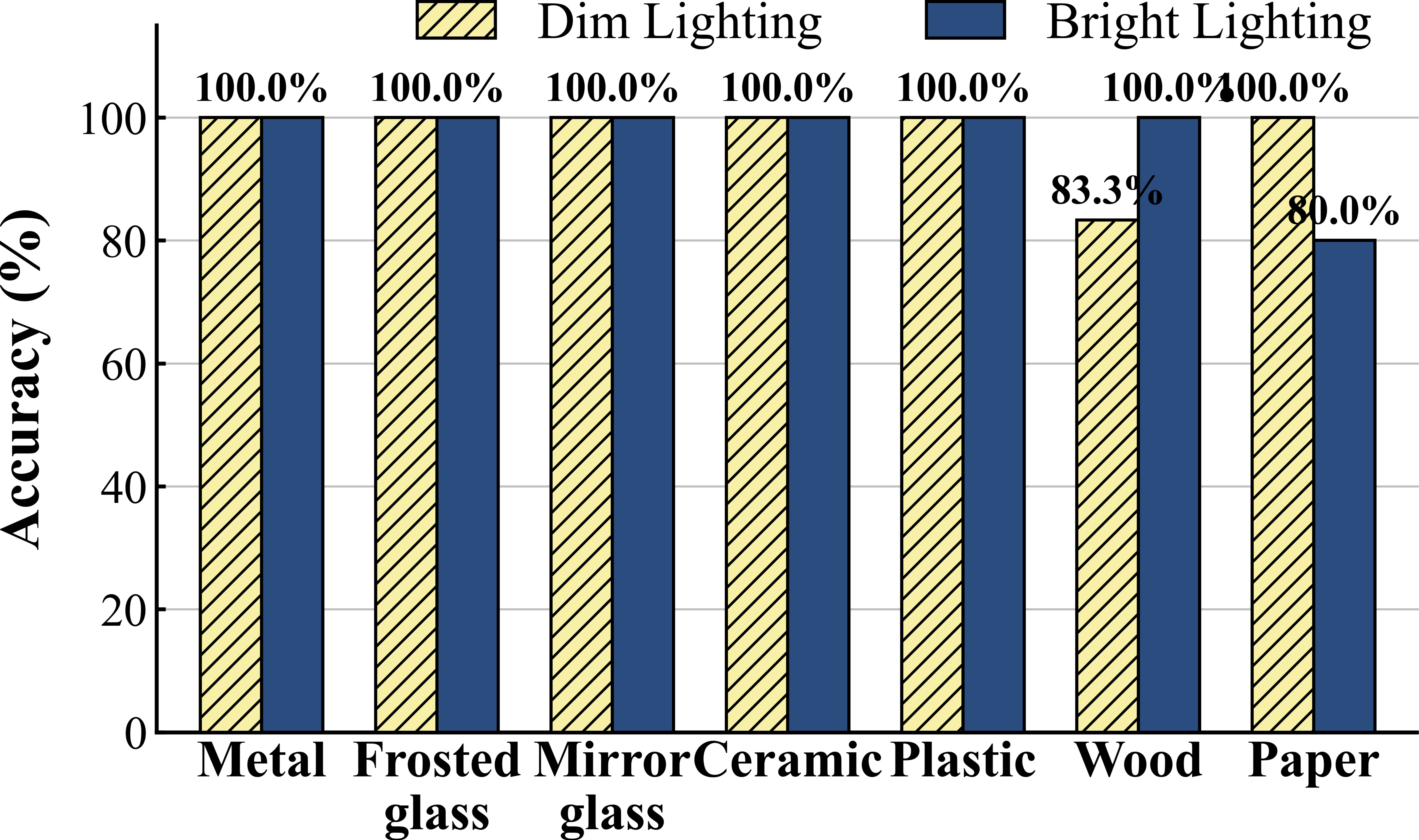}
    \caption{VLMaterial (different lighting).}
  \label{dif_light}
  \end{minipage}
\end{figure*}

\begin{figure*}[!t]
  \centering
  \begin{minipage}[b]{0.32\linewidth}
    \centering
    \includegraphics[width=\linewidth]{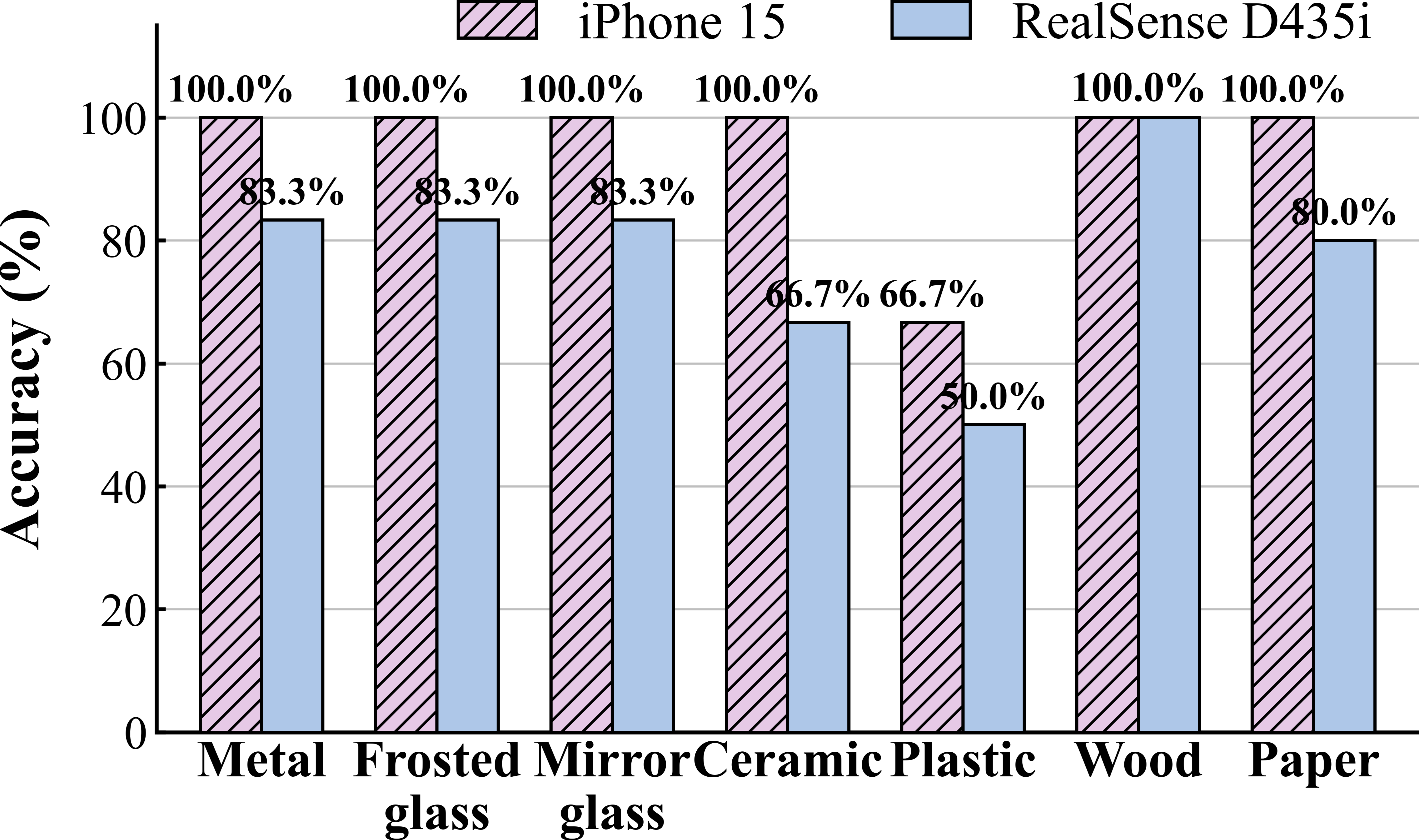}
    \caption{VLM-only (different camera).}
  \label{dif_camera_VLM_only}
  \end{minipage}
  \hfill
  \begin{minipage}[b]{0.32\linewidth}
    \centering
    \includegraphics[width=\linewidth]{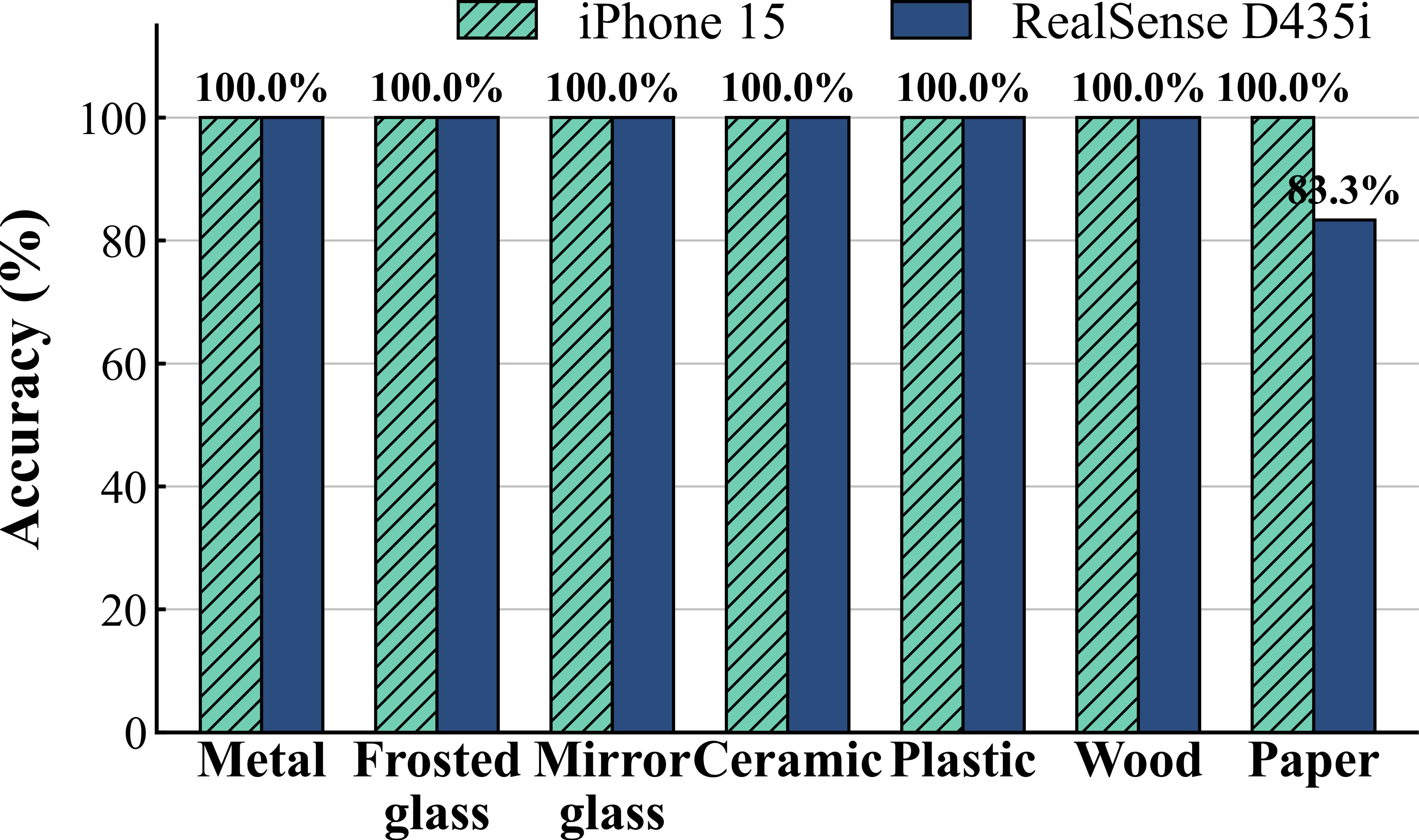}
    \caption{VLMaterial (different cameras).}
  \label{dif_camera}
  \end{minipage}
  \hfill
  \begin{minipage}[b]{0.32\linewidth}
    \centering
    \includegraphics[width=\linewidth]{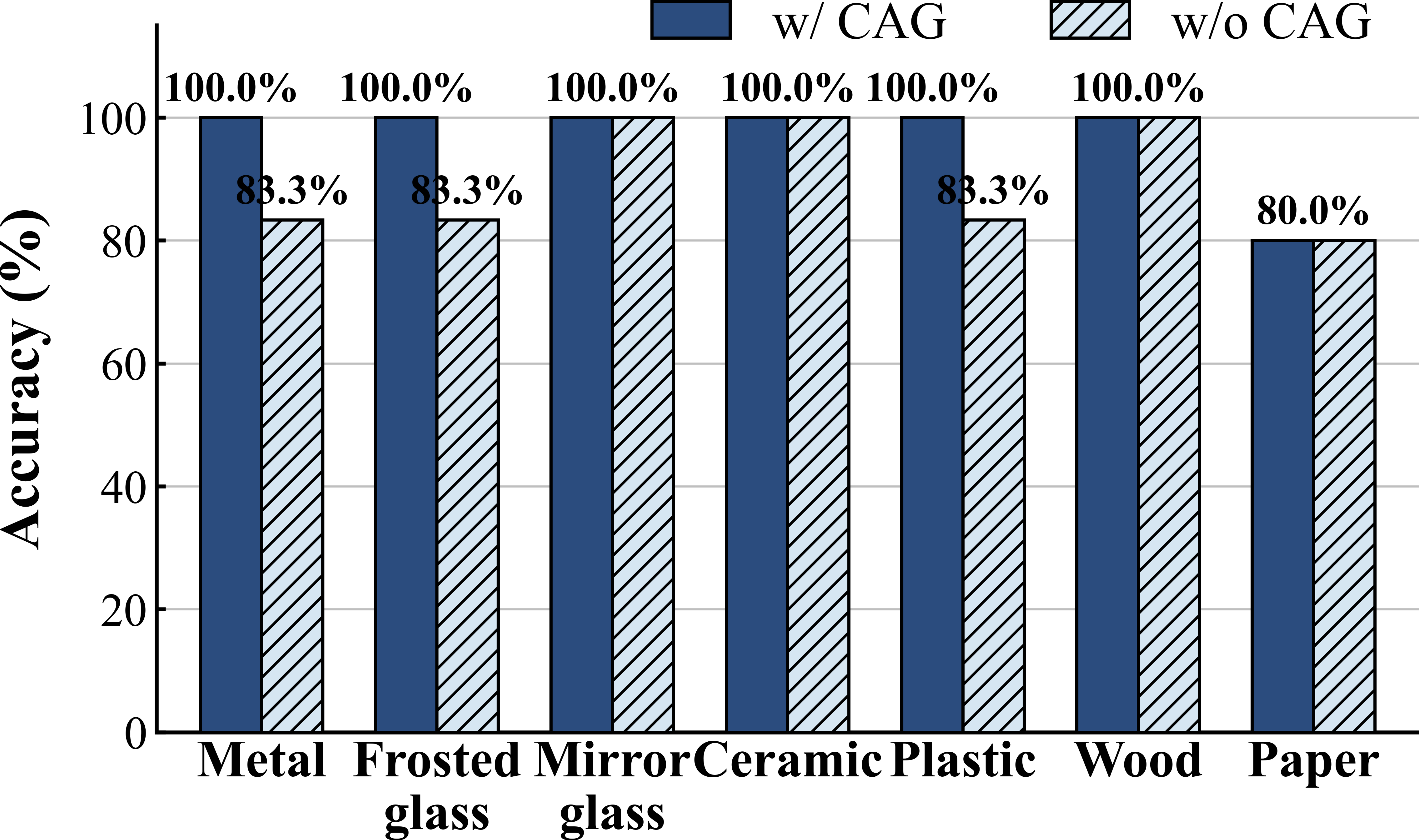}
    \caption{VLMaterial (w/ vs. w/o CAG).}
  \label{CAG_ablation}
  \end{minipage}
\end{figure*}

\begin{figure*}[!t]
  \centering
  \includegraphics[width=0.9\textwidth]{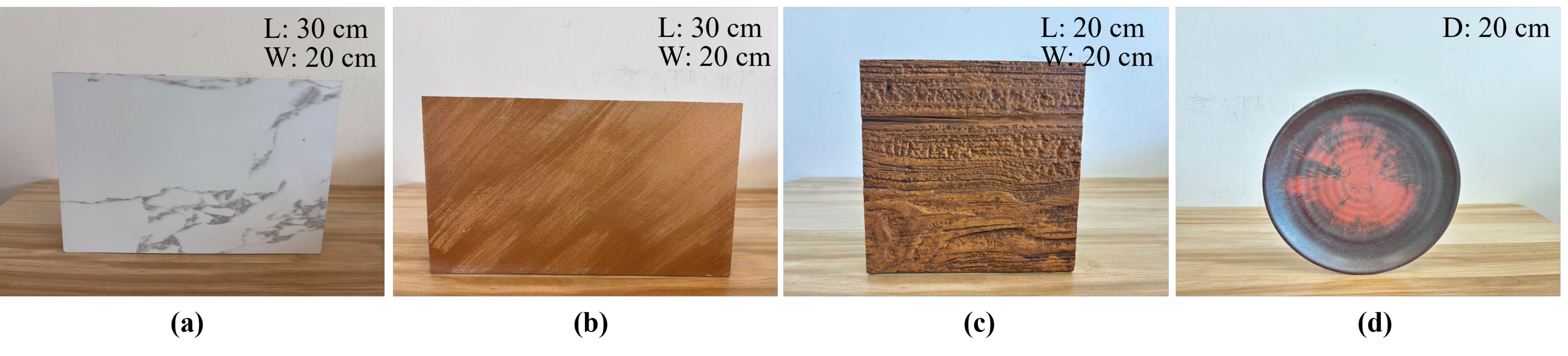}
  \caption{Counterfeit objects. (a)~Aluminum-plastic panel (ceramic-like finish). (b)~Ceramic board (wood-like appearance). (c)~Plastic board (wood-like texture). (d)~Plastic plate (ceramic-like finish).}
  \label{mimic_objs}
\end{figure*}

{\textbf{Impact of Camera Resolution.}} 
To unify the resolution of RGB images, the Intel RealSense D435i was configured to its maximum supported resolution of \(1280 \times 720\). Since camera resolution is critical for VLM-based recognition, we further analyzed its impact using a high-resolution smartphone. Specifically, an iPhone 15 with a 48-megapixel main camera was employed to capture ultra-high-resolution images at 24 MP and 48 MP.
\textbf{(1) VLM-only.} Results are in \Cref{dif_camera_VLM_only}. Even with a 25-fold resolution increase, VLM-only still fails on visually similar objects, e.g., plastic box (D5) as glass and plastic plate (E5) as ceramic. This confirms higher resolution alone cannot solve visual ambiguity, underscoring the value of VLMaterial's vision-radar fusion. VLM-only naturally improved with better data quality (iPhone: $\sim$3.5 MB vs. RealSense: $\sim$1 MB per image).
\textbf{(2) VLMaterial.} Overall performance shows slight improvement (\Cref{dif_camera}), mainly for paper type (100\% accuracy). Higher resolution corrected paper bottle (B7) from metal to paper. However, given the iPhone 15 costs roughly four times more than RealSense, expensive hardware provides limited benefits relative to cost and data size.

{\textbf{Impact of CAG.}}
We compared our full system (VLMaterial w/ CAG) against an ablated version without this module (VLMaterial w/o CAG). A key issue is that the VLM relies on general prior knowledge about dielectric constants, which often mismatches the actual values derived from our 60 GHz radar measurements. For example, while the VLM knows that metal's dielectric constant is theoretically infinite, our derived value for metal cup (B1) is 30.51. Without CAG, the model misclassifies B1 as ceramic. However, in our system, 30.51 is significantly higher than other materials and should clearly indicate metal. 
Similar errors occurred: frosted glass bottle (C2) as glass, and plastic box (D5) as mirror glass. To address this, we used different material boards as reference targets (last row of \Cref{all_targets}) for CAG. Only B7 remains an exception, confirming that CAG is needed to bridge theoretical priors and measured data. Ablation results are in \Cref{CAG_ablation}.
This domain-specific context, unavailable from standard online sources, constitutes the core contribution of VLMaterial.

{\textbf{Impact of Fusion Method.}}
In addition to our proposed adaptive fusion mechanism, we also explored the potential of exploiting the inherent reasoning capabilities of VLMs through a structured approach rather than blind inference. To this end, we implemented a prompt-guided fusion strategy, which encapsulates RGB images, radar parameters (SNR, dielectric constant), and physical rules into a unified VLM prompt to guide the decision-making process. The results reveal significant instability of VLMaterial (see Appendix~\ref{app:fusion} for detailed analysis of prompt-guided fusion results).



{\textbf{Impact of Counterfeit Objects.}} To further validate VLMaterial against visually deceptive targets, we collected imitation objects: an aluminum-plastic panel (ceramic-like finish), a ceramic board (wood-like), a plastic board (wood-like texture), and a plastic plate (ceramic-like finish) (\Cref{mimic_objs}). 
Their ground truth materials are metal, ceramic, and two types of plastic: plastic 1 (wood-like texture) and plastic 2 (ceramic-like finish). 
VLMaterial achieves perfect counterfeit detection by resolving conflicts between visual and radar modalities, overcoming deceptions that defeat either modality alone (see Appendix~\ref{app:mimic} for detailed analysis of counterfeit objects results).




\section{Limitations and Future Work}
\label{limitations_discussion}
While our proposed VLMaterial system demonstrates promising results, it is important to acknowledge its current limitations and discuss potential areas for future improvement.


\textbf{Multi-Object Scenarios.} Our VLMaterial currently uses SAM for single-object segmentation. For multi-object scenes, SAM identifies all objects, but radar must measure them sequentially, limiting throughput. Future work will adopt advanced radar imaging for material identification.

\textbf{Fusion-Based Material Recognition.}
Our hierarchical decision-level fusion strategy effectively combines VLM semantics with radar representation, achieving robust performance. However, under extreme visual degradation (e.g., total darkness), VLM cannot provide reliable cues, limiting system recovery. Future work will integrate lighting-invariant modalities (tactile, acoustic) to enhance reliability.

\textbf{Inference Latency and Model Efficiency.} Our current framework uses SOTA VLMs to maximize reasoning accuracy, but their large size introduces inference latency, limiting real-time applicability. As this work focuses on establishing a multi-modal material sensing benchmark rather than real-time optimization, future work will use knowledge distillation to transfer supervision from a large model to a smaller one, balancing efficiency and robustness.

Although our current framework is static, it validates the core sensing methodology as a foundation for future deployment on a robotic arm, enabling active perception, multi-object evaluation, and smarter interactions in home robotics.

\section{Conclusion}
\label{conclusion}
In conclusion, this paper introduces VLMaterial, the first framework to successfully leverage camera-radar fusion based on VLMs for physics-grounded material identification. We address this task through a visual-radar fusion framework whereby radar assists vision in disambiguating specular reflections, visual similarities, and visually deceptive objects, while conversely, vision supplements radar in distinguishing materials with similar dielectric constants to effectively overcome the overlap in intrinsic parameters. These complementary representations are integrated within a reference-based inference enhanced VLM for advanced multi-modal recognition. Our experiments confirm that VLMaterial achieves a final recognition accuracy of 96.08\% in a training-free setting and maintains stable performance under the influence of varying distances, numbers of samples per frame, ambient illuminance levels, camera resolutions and so on. We further validated the system's reliability on four additional types of imitation materials where vision-only VLMs failed completely, demonstrating for the first time that VLMs can effectively interpret object materials by fusing intrinsic physical properties from raw radar signals with visual information.


\bibliographystyle{IEEEtran}
\bibliography{IEEEabrv,main}

\clearpage
\appendices

\section{Calculation of PRCA.} 
\label{app:prca}
To quantify the coverage of the beam, we calculate the -3dB beam area as $A_r$. The process involves identifying the main lobe region and integrating the area of its grid cells.

First, the main lobe region $\mathcal{R}$ is identified by thresholding. Let $P_{peak}$ be the peak amplitude of the RA map. The half-power threshold is defined as $P_{th} = P_{peak} / \sqrt{2}$. We extract the connected component of grid points containing the peak where the amplitude exceeds $P_{th}$, filtering out isolated sidelobes.

Second, the total area is approximated by summing the areas of the discrete quadrilateral grid cells within $\mathcal{R}$. For a grid cell located at index $(i,j)$, the Cartesian coordinates of its four vertices are defined as $v_1=(x_{i,j}, y_{i,j})$, $v_2=(x_{i+1,j}, y_{i+1,j})$, $v_3=(x_{i+1,j+1}, y_{i+1,j+1})$, and $v_4=(x_{i,j+1}, y_{i,j+1})$. The area of this specific cell, $S_{i,j}$, is computed using the Shoelace formula (Polygon Area Formula) \cite{lee2017shoelace}:

\begin{equation}
    S_{i,j} = \frac{1}{2} \left| \sum_{k=1}^{4} (x_k \cdot y_{k+1} - x_{k+1} \cdot y_k) \right|,
\end{equation}
where $(x_5, y_5)$ is defined as $(x_1, y_1)$ to close the loop.

Third, the total -3dB coverage area (i.e., PRCA) is obtained by aggregating the individual cell areas:
\begin{equation}
    A_r = \sum_{(i,j) \in \mathcal{R}} S_{i,j}.
\end{equation}

Finally, comparing the power reflection coefficient of the target with the reference signal from a smooth metal plate, which acts as a perfect reflector, we derive the normalized reflection coefficient (i.e., Fresnel reflection coefficient $r_p$). 

\section{Relationship Between Fresnel Reflection Coefficient and Dielectric Constant.} 
\label{app:fres_die}
Theoretically, Fresnel equations describe the reflectance at an interface, as derived from Maxwell's equations \cite{huray2009maxwell}. The reflection coefficient is defined as a function of the incidence angle and the polarization state of the incident beam. Our radar system employs vertical polarization, meaning the transmitted electric field vector is oriented perpendicular to the ground. 
In this antenna arrangement, the electric field is parallel to the plane of incidence.
Therefore, the incident wave corresponds to the p-polarization case (also known as transverse magnetic or TM polarization). For p-polarized waves, the reflection amplitude coefficient, defined as the ratio of the reflected to the incident electric field \cite{salik2009optical}, can be expressed as:
\begin{equation}
r_p = \frac{n^2 \cos \theta - \sqrt{n^2 - \sin^2 \theta}}{n^2 \cos \theta + \sqrt{n^2 - \sin^2 \theta}},
\label{r_p_n}
\end{equation}
where $r_p$ denotes the reflection amplitude coefficient, commonly known as the Fresnel reflection coefficient, $\theta$ is the angle of incidence, $n = n_t / n_i$, $n$ is refractive index, $n_i$ and $n_t$ are the refractive indices of the incident and transmitted media, respectively.

This relationship originates from Maxwell's equations \cite{huray2009maxwell}, which describe how electric and magnetic fields are generated and interact. From these equations, we can derive the propagation speed $v$, which is determined by the medium's dielectric constant $\varepsilon$ and permeability $\mu$:
\begin{equation}
v = \frac{1}{\sqrt{\varepsilon\mu}}.
\label{v}
\end{equation}

By definition, the refractive index is the ratio of the speed of light in a vacuum to its speed in the medium:
\begin{equation}
n = \frac{c}{v}.
\label{n_ti}
\end{equation}

Substituting \Cref{v} into \Cref{n_ti}, and noting that the speed of light $c$ is determined by the dielectric constant $\varepsilon_0$ and permeability $\mu_0$ of free space, we obtain:




\begin{equation}
n = \frac{1}{\sqrt{\varepsilon_0\mu_0}} \cdot \sqrt{\varepsilon\mu} = \sqrt{\left(\frac{\varepsilon}{\varepsilon_0}\right) \cdot \left(\frac{\mu}{\mu_0}\right)}.
\end{equation}

Here, we define two important relative quantities.
The first parameter is relative dielectric constant, defined as $\varepsilon_r = \varepsilon / \varepsilon_0$. This dimensionless quantity represents the factor by which the electric field is reduced inside the material compared to a vacuum.
The second one is relative permeability, defined as $\mu_r = \mu / \mu_0$.

Thus, we derive the general equation connecting the refractive index, dielectric constant, and permeability:
\begin{equation}
    n = \sqrt{\varepsilon_r \mu_r}.
\end{equation}

For common materials such as non-magnetic metals (e.g., copper, aluminum, gold, and silver), plastics, and wood, we assume a relative permeability of $\mu_r \approx 1$. The general equation can be significantly simplified:

\begin{equation}
    \varepsilon_r = n ^2.
\end{equation}

Therefore, for the vast majority of general materials encountered in daily scenarios, the square of the refractive index is equal to the relative dielectric constant. 

\section{Estimated dielectric constants vs. ground truth.} 
\label{app:grd_truth}
\Cref{gt} validates the dielectric constant estimated by our method against theoretical ground truth derived from the ITU-R P.2040 standard \cite{series2015effects}. 

Due to the unavailability of tabulated values for the 60 GHz band, we derived theoretical baselines using the standard's frequency-dependent model, $\varepsilon_r(f) = a \cdot f^b$, where $f$ is frequency and $a, b$ are material-specific coefficients. The validation results are presented in \Cref{gt}. In this plot, the standard reference denotes the dielectric constant of the reference signal we use to construct the CAG, while mean and std. dev. represent the average measurement (excluding outliers) and the standard deviation, respectively. The red interval indicates the theoretical range of dielectric constants calculated based on the ITU-R standard.  
Crucially, the results demonstrate high physical consistency. The estimated $\varepsilon_r$ values follow a correct descending order: from reflective metal ($\approx 27.8$) and glass ($\approx 8.2\text{--}10.0$) to low-permittivity materials like wood ($\approx 5.1$) and paper ($\approx 3.7$). This clear distinction validates that VLMaterial captures intrinsic electromagnetic properties rather than superficial signal strength.

\begin{figure}[!t]
  \centering
  \includegraphics[width=0.45\textwidth]{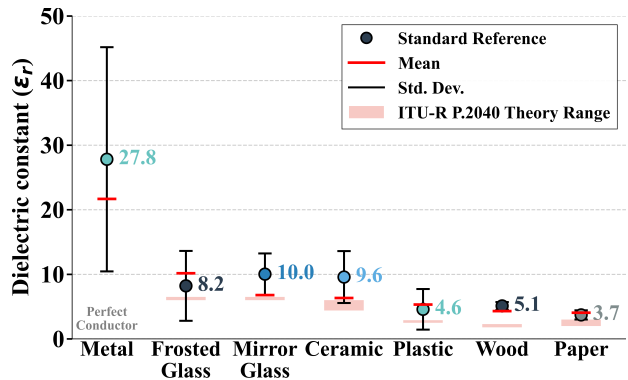}
  \caption{Comparison of estimated dielectric constants with  standard ITU-R reference ranges.}
  \label{gt}
\end{figure}
Despite overall consistency with theoretical baselines, the results show a slight positive bias for porous materials such as wood and paper. These discrepancies are primarily attributed to environmental moisture; whereas ITU models assume dry conditions, organic materials are hygroscopic, and absorbed water significantly increases the dielectric constant. Additionally, specific additives in composite materials (e.g., plastics) may differ from the generic categories in the standard. Therefore, these deviations reflect the system's sensitivity to actual material states. Crucially, this highlights why the CAG is indispensable: relying solely on theoretical dielectric constants is insufficient for complex real-world scenarios. By constructing our CAG with extra electromagnetic properties, we ensure VLMaterial adapts to these variations, guaranteeing reliable performance in practical applications.

\section{Reference Signals for CAG.} 
\label{app:cag}
\begin{figure}[!t]
  \centering
  \subfloat[VLM-only.]{
    \includegraphics[width=0.22\textwidth]{photo/VLM_only_41_v1-6.pdf}
    \label{VLM_only_41_v1-6}
  }\hfill
  \subfloat[LLMaterial.]{
    \includegraphics[width=0.22\textwidth]{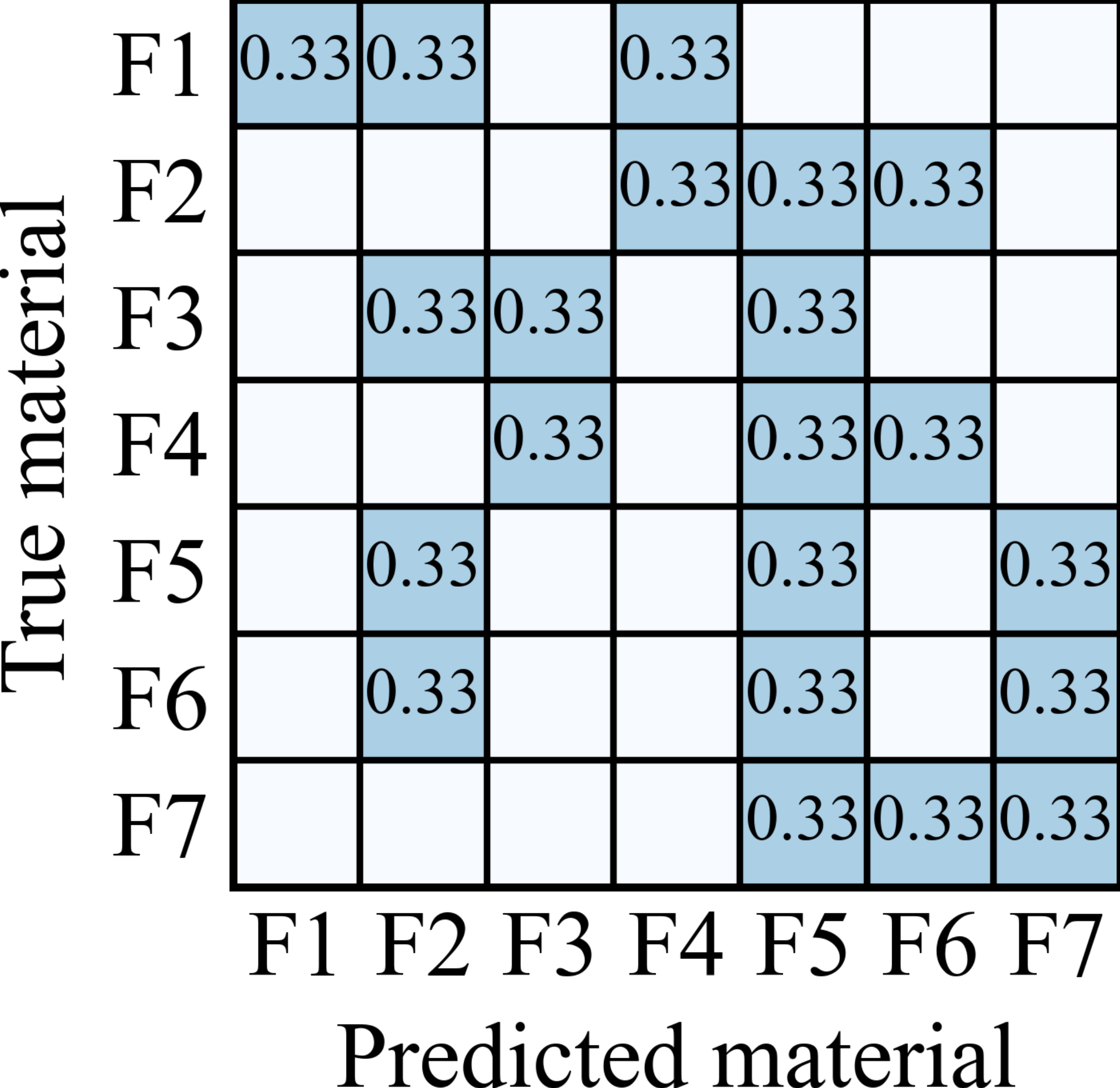}
    \label{LLMaterial_41_v1-6}
  }\\
  \subfloat[Radar-only (VLMaterial).]{
  \centering
    \includegraphics[width=0.22\textwidth]{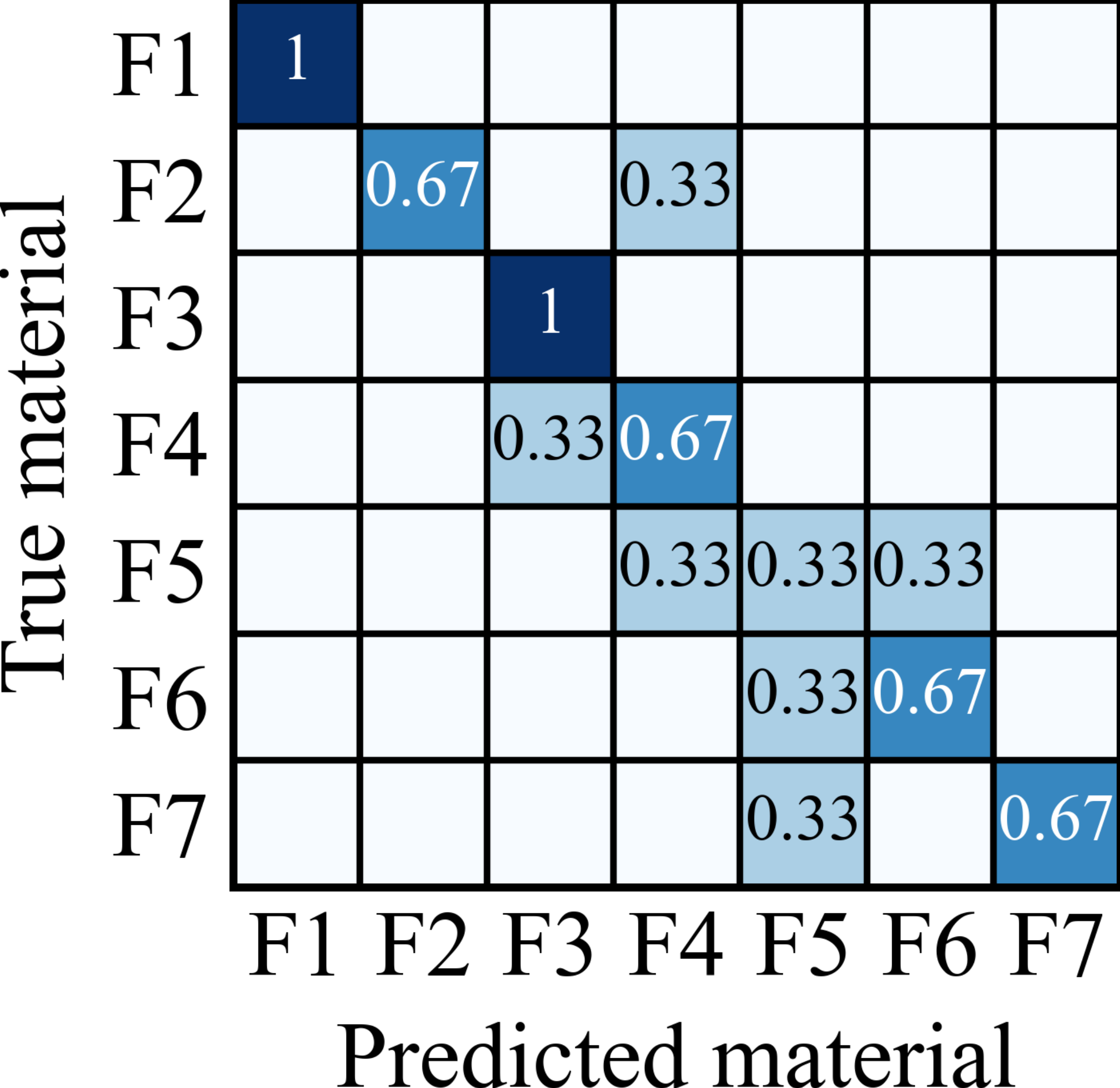}
    \label{Radar_only_41_v1-6}
  }\hfill
  \subfloat[VLMaterial.]{
  \centering
    \includegraphics[width=0.22\textwidth]{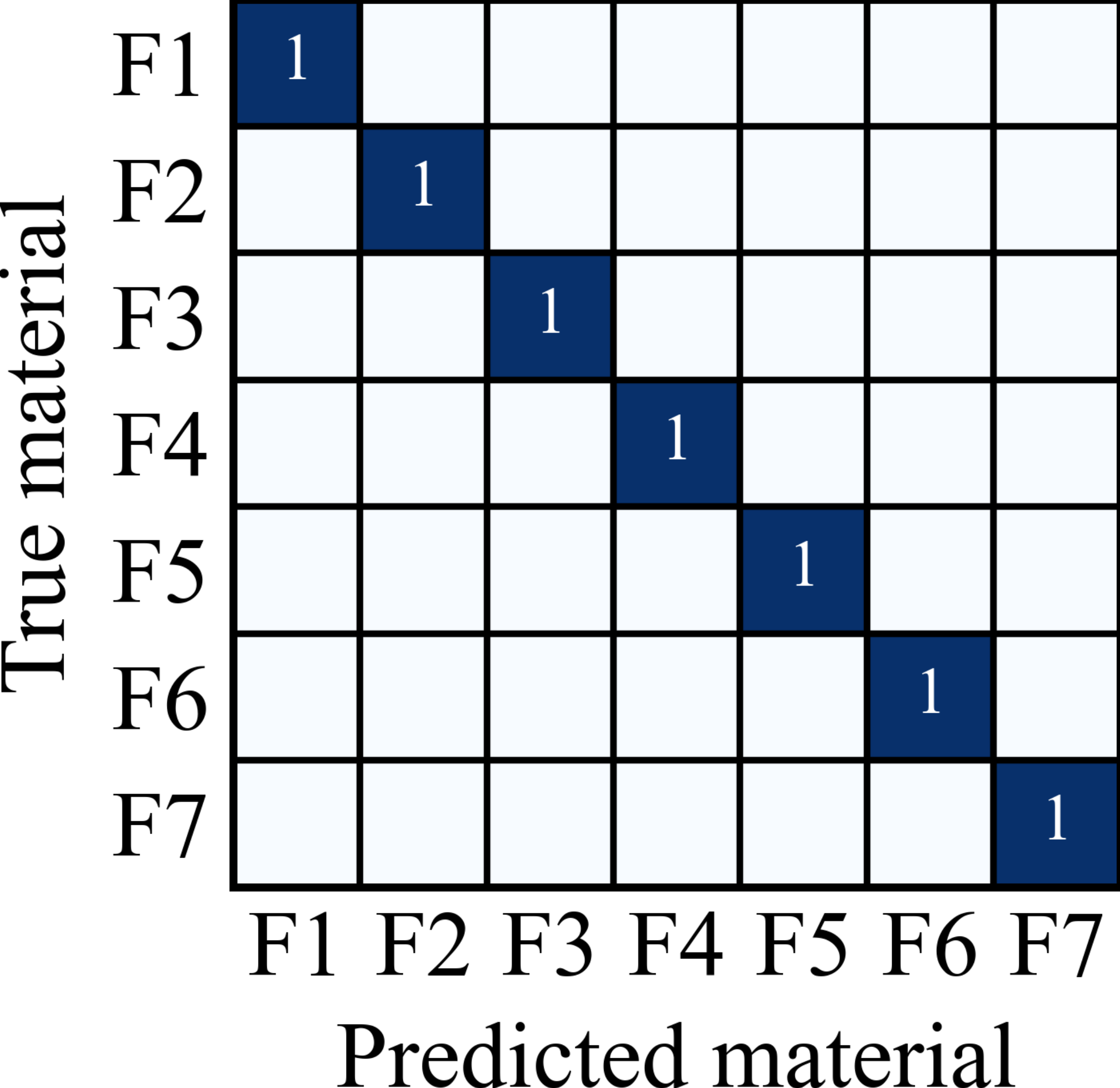}
    \label{VLMaterial_41_v1-6}
  }\\
  \caption{Comparison of material recognition results for the board category (reference signals).}
  \label{reference_signals}
\end{figure}

We validate the feasibility of this approach on ideal flat boards (Group F): metal, frosted glass, mirror glass, ceramic, plastic, wood, and paper (see the bottom row of \Cref{all_targets}). As shown in \Cref{VLM_only_41_v1-6}, the vision-only approach frequently failed due to specular reflections, misclassifying metal (e.g., F1) and ceramic (e.g., F4) as glass or plastic, and wood (e.g., F6) as ceramic. Furthermore, visual similarities posed significant challenges; for instance, glass (e.g., F2, F3) was misidentified as plastic, while plastic (e.g., F5) was mistaken for glass.
Regarding LLMaterial, its performance suffers in standard environments as it relies on radar-absorbing backgrounds. As clearly shown in \Cref{LLMaterial_41_v1-6}, recognition results are poor across most categories. Since these widespread failures are immediately apparent from the figure, we omit a detailed case-by-case analysis.
Conversely, the Radar-only approach failed when materials shared similar dielectric constants (\Cref{Radar_only_41_v1-6}). For instance, it consistently misclassified frosted glass as plastic (F2), ceramic as mirror glass (F4), plastic as ceramic or wood (F5), wood as plastic (F6), and paper as plastic (F7). Crucially, however, we observe that these radar-confused materials often exhibit distinct visual appearances, suggesting that a camera can effectively aid in disambiguation.
Finally, the fusion results shown in \Cref{VLMaterial_41_v1-6} demonstrate that all materials were correctly identified. This leads to a key preliminary conclusion: radar effectively compensates for visual failures caused by reflections or similarity, while the camera corrects radar errors arising from similar dielectric constants. This mutual complementarity validates VLMaterial and supports our CAG implementation.

\section{Impact of Fusion Method.} 
\label{app:fusion}

As illustrated in \Cref{simpleVLMaterial_41}, the results reveal significant instability. While the method achieves accurate identification when both sensors match (e.g., the cups group), its performance deteriorates markedly in conflicting scenarios.

\begin{figure*}[!t]
  \centering
  \subfloat[Cups.]{
    \includegraphics[width=0.14\textwidth]{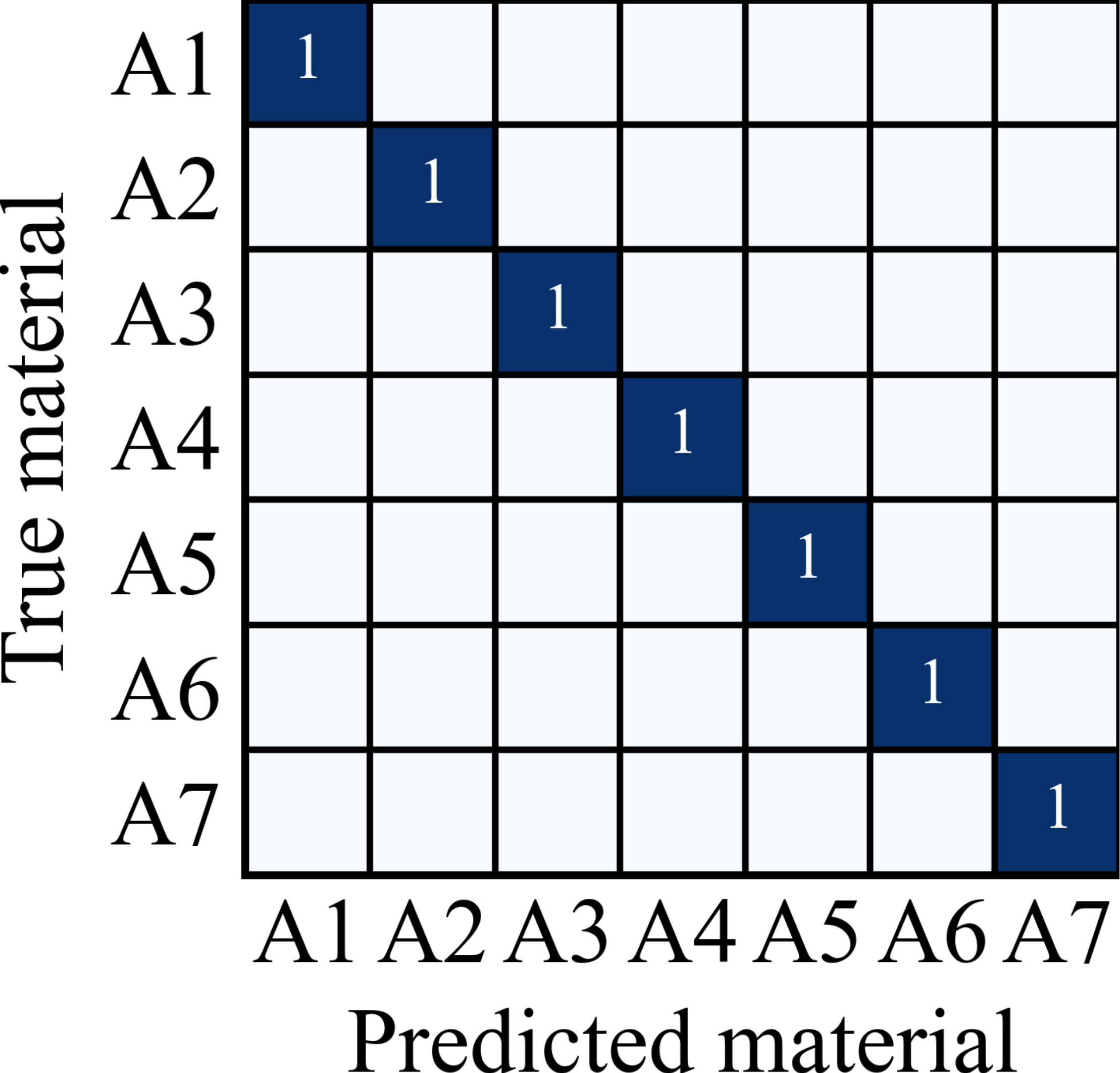}
    \label{simpleVLMaterial_41_v1-1}
  }\hfill
  \subfloat[Bottles.]{
    \includegraphics[width=0.14\textwidth]{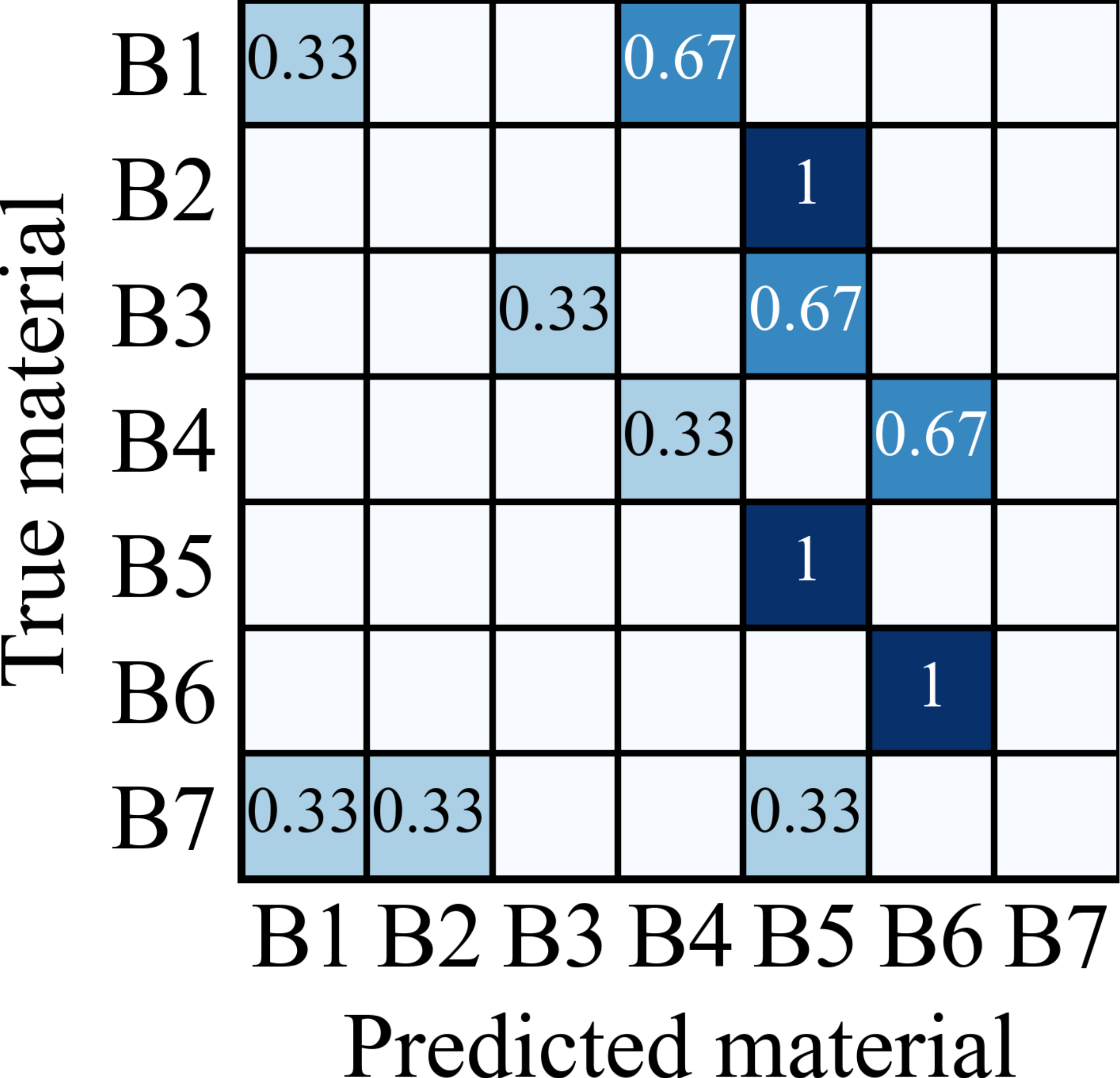}
    \label{simpleVLMaterial_41_v1-2}
  }\hfill
  \subfloat[Kettles.]{
    \includegraphics[width=0.14\textwidth]{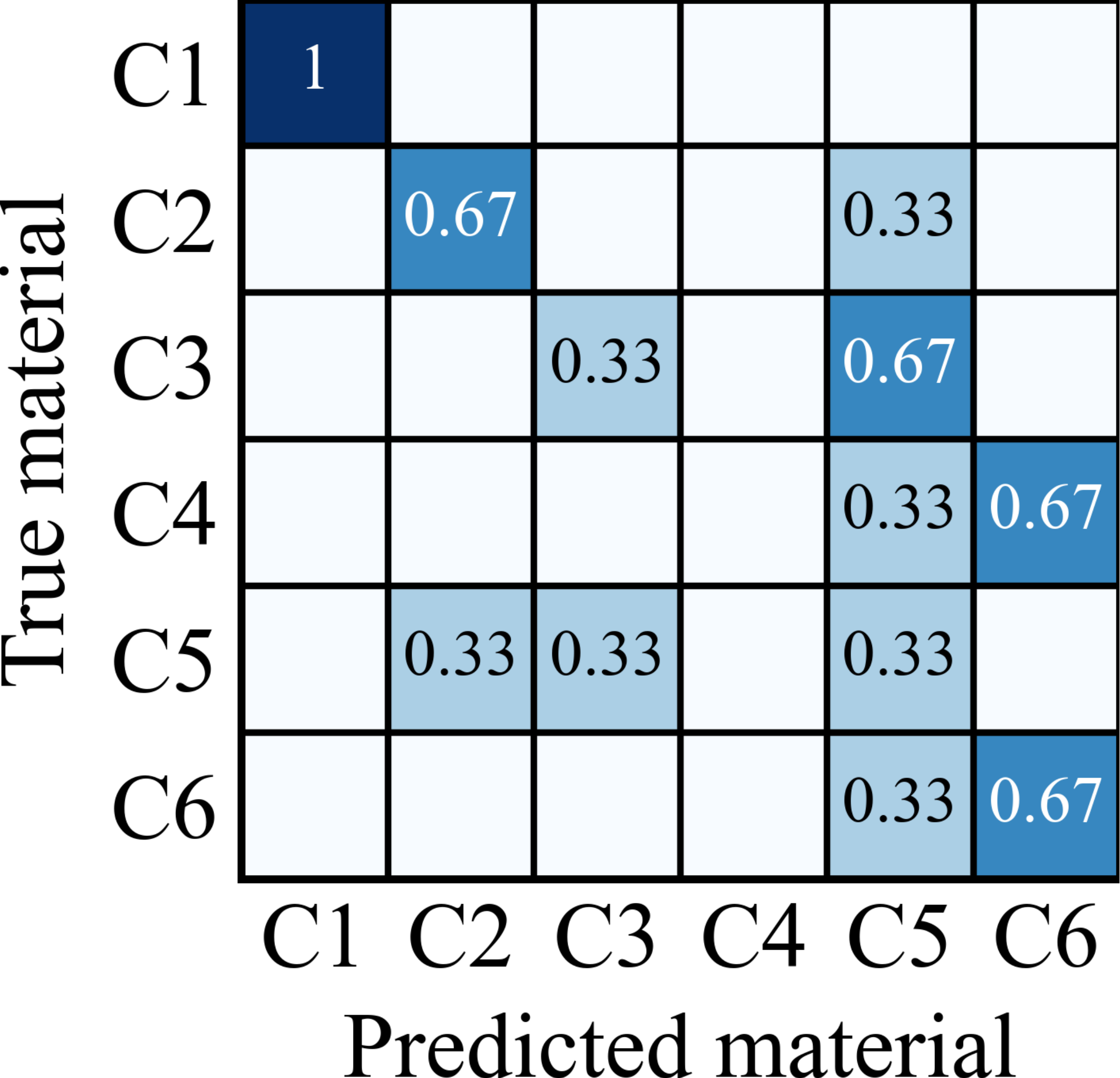}
    \label{simpleVLMaterial_41_v1-3}
  }\hfill
  \subfloat[Boxes.]{
    \includegraphics[width=0.14\textwidth]{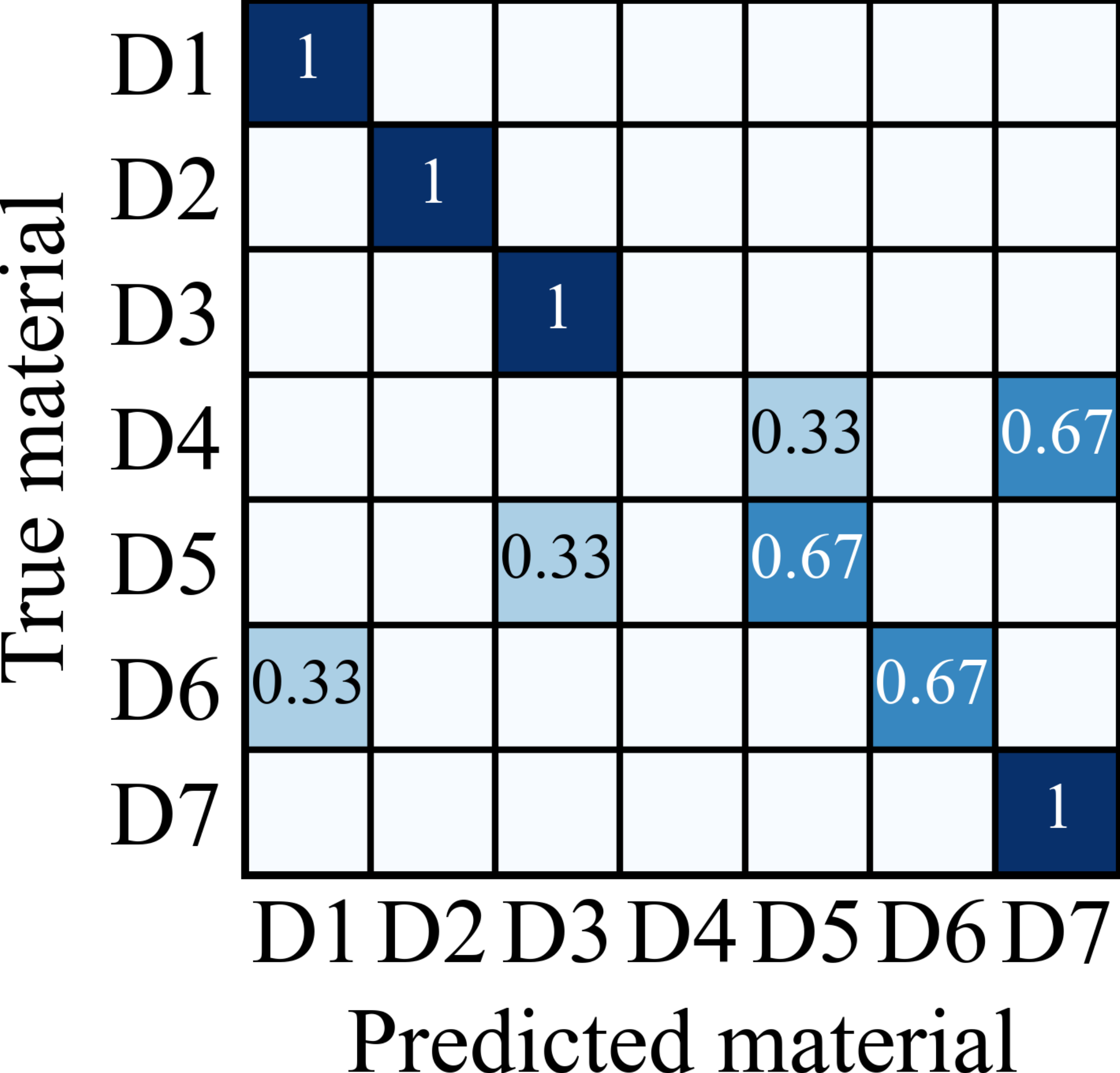}
    \label{simpleVLMaterial_41_v1-4}
  }\hfill
  \subfloat[Plates.]{
    \includegraphics[width=0.14\textwidth]{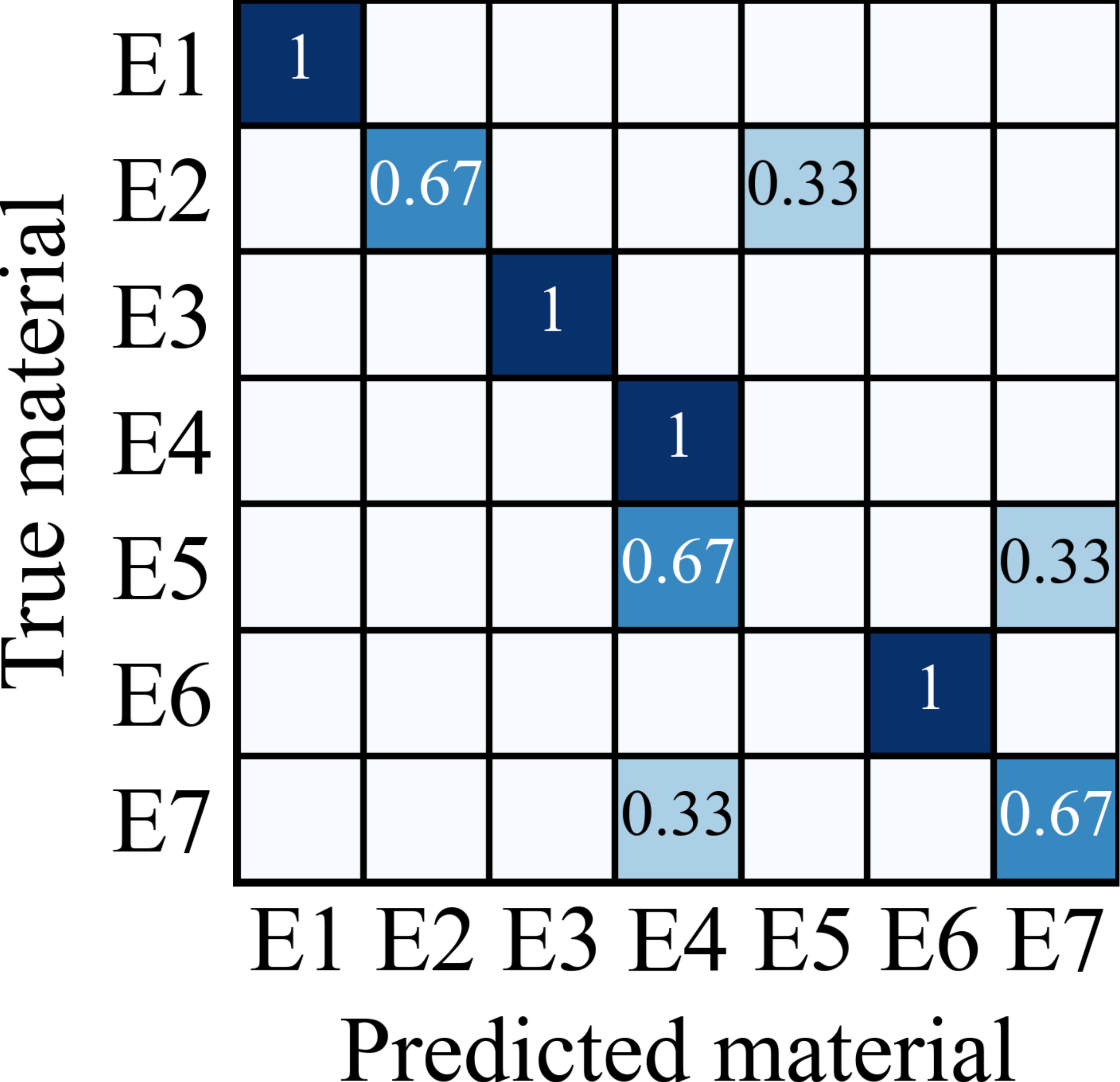}
    \label{simpleVLMaterial_41_v1-5}
  }\hfill
  \subfloat[Boards.]{
  \includegraphics[width=0.14\textwidth]{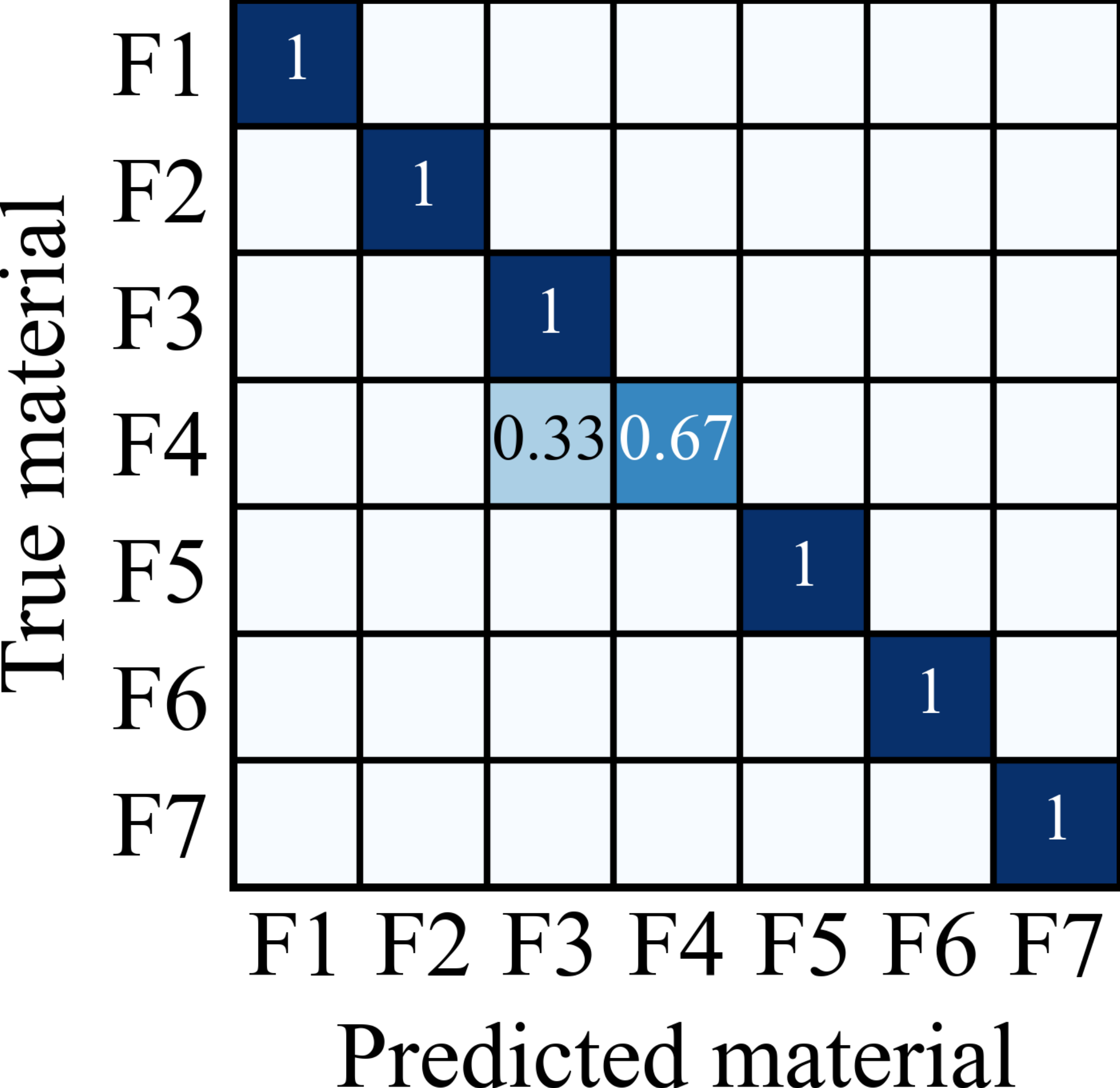}
    \label{simpleVLMaterial_41_v1-6}
  }\\
  \caption{Overall object material recognition via prompt-guided camera-radar fusion.}
  \label{simpleVLMaterial_41}
\end{figure*}

The errors manifest in two distinct patterns of misjudgment. On the one hand, when the VLM over-relies on visual features, it misclassifies ceramic as glass (F4), plastic as glass (D5) or ceramic (E5), and paper as metal or frosted glass (B7). On the other hand, when the VLM incorrectly prioritizes radar evidence without sufficient context, it misidentifies metal as ceramic (B1), frosted glass and mirror glass as plastic (B2, C2, E2, B3, C3), and ceramic as wood, plastic, or paper (B4, C4, D4). Furthermore, plastic is confused with glass (C5) or paper (E5), wood with plastic (C6) or metal (D6), and paper with plastic (B7) or ceramic (E7). These extensive misclassifications demonstrate that relying solely on the VLM's intrinsic reasoning is insufficient; a robust system must explicitly evaluate the signal quality of both camera and radar to dynamically allocate fusion weights.

In VLMaterial, the optical characterization pipeline not only outputs precise object categories but also provides all possible material candidates based on the RGB image. Simultaneously, the electromagnetic characterization pipeline provides a distinct material identification based on a precise calculation of the object's dielectric constant. Our decision-level fusion logic operates under two primary scenarios (see \Cref{VLMaterial_41} for details):
\textbf{(1) Consensus through Intersection.} If there is an intersection between the material candidates proposed by the optical pipeline and the material identified by the electromagnetic pipeline, the final result is determined by this shared consensus. For example, in the case of the plastic cup (A5), the camera might suggest {glass, plastic} as possibilities, while the radar confidently identifies {plastic}. The intersection of the predicted sets generates plastic, which is correctly accepted as the final result. This represents the most common scenario encountered in our experiments, where both modalities corroborate each other.
\textbf{(2) Conflict Resolution with Weighted Trust.} 
In situations where there is no intersection between the outputs of the two pipelines, we resolve the conflict using a weighted decision logic based on uncertainty estimation. We present two representative examples to illustrate how radar assists vision and, conversely, how vision assists radar.
First, regarding radar-assisted vision: for the plastic box (D5), visual ambiguity led to incorrect candidates. However, our VLMaterial framework, after evaluating multiple factors, derived confidence scores ($S_{vis}$ and $S_{rad}$) of 25\% and 75\%, respectively. Consequently, the system prioritized the radar prediction, correctly outputting plastic. This case demonstrates VLMaterial's capability to leverage radar for correcting visual misclassifications caused by appearance similarities.
Second, regarding vision-assisted radar: the wooden box (D6) features a smooth surface that reflects strong radar signals, causing the radar to misclassify it as metal. However, the object possesses distinct visual textures, leading the visual model to correctly identify it as wood. In this instance, our uncertainty-aware adaptive conflict resolution module calculated $S_{vis}$ at 88\% and $S_{rad}$ at 12\%. Thus, the visual confidence dominated, successfully correcting the radar's error and generating the correct result of wood.
Notably, the only failure case for VLMaterial occurred with the paper bottle (B7). While the camera identified it as plastic (also incorrect but closer in texture), the radar erroneously classified it as metal due to its specially processed, smooth surface which generated a high SNR. Unfortunately, the system placed excessive trust in the robust radar signal, leading to a final misclassification. 

\section{Impact of Counterfeit Objects.} 
\label{app:mimic}

\begin{figure}[!t]
  \centering
  \subfloat[VLM-only.]{
    \includegraphics[width=0.22\textwidth]{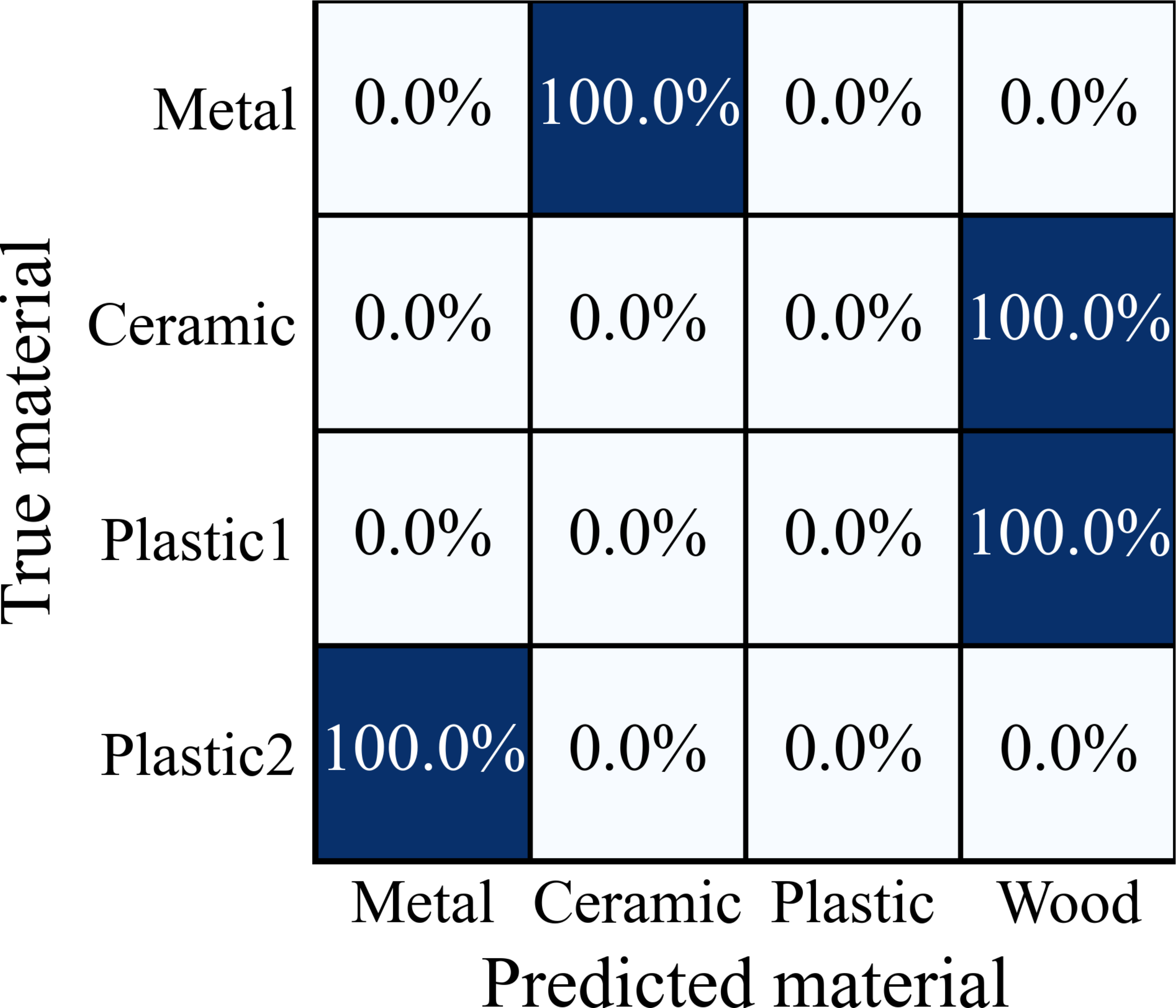}
    \label{mimic4_VLM_only}
  }\hfill
  \subfloat[LLMaterial.]{
  \centering
    \includegraphics[width=0.22\textwidth]{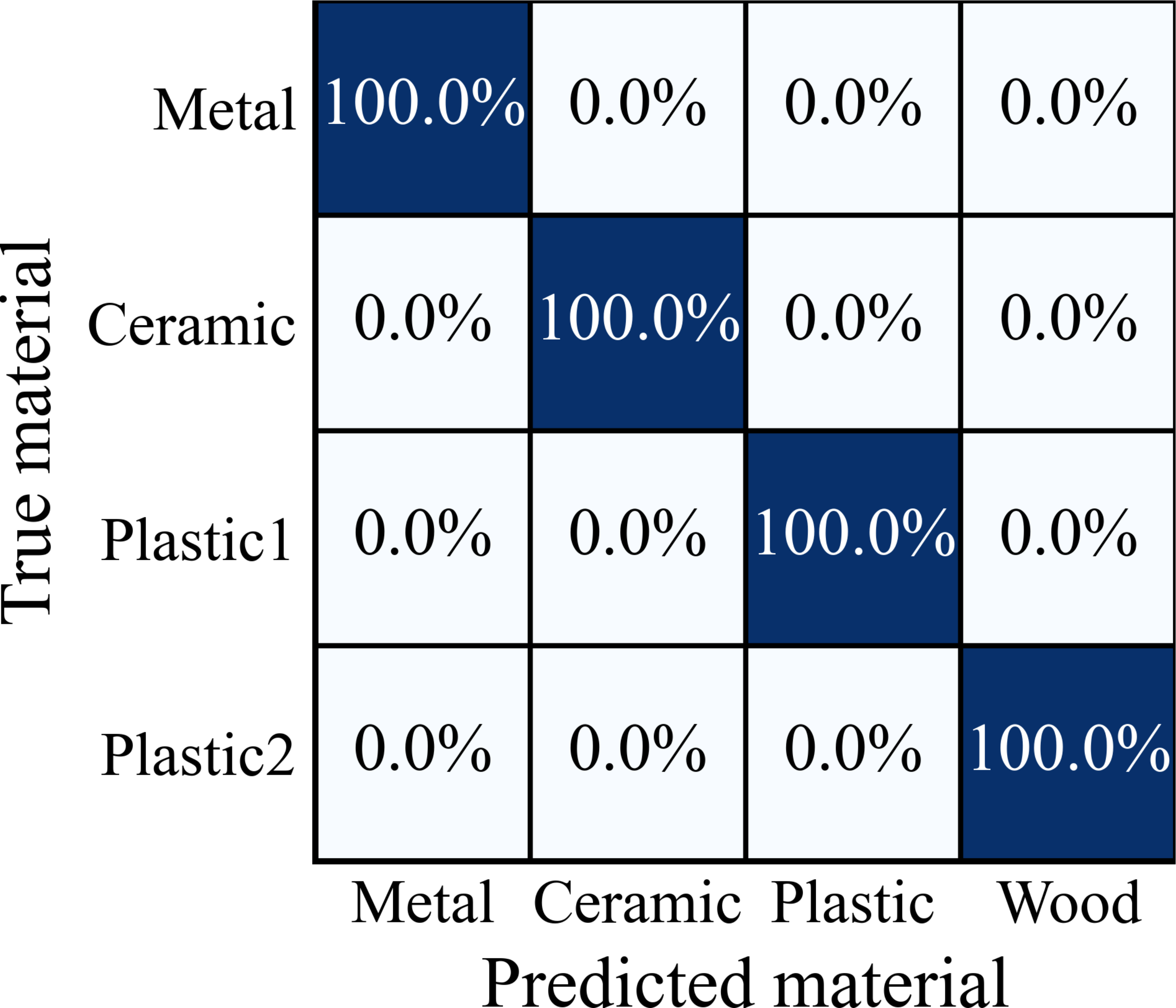}
    \label{mimic4_LLMaterial}
  }\\
  \subfloat[Radar-only (VLMaterial).]{
  \centering
    \includegraphics[width=0.22\textwidth]{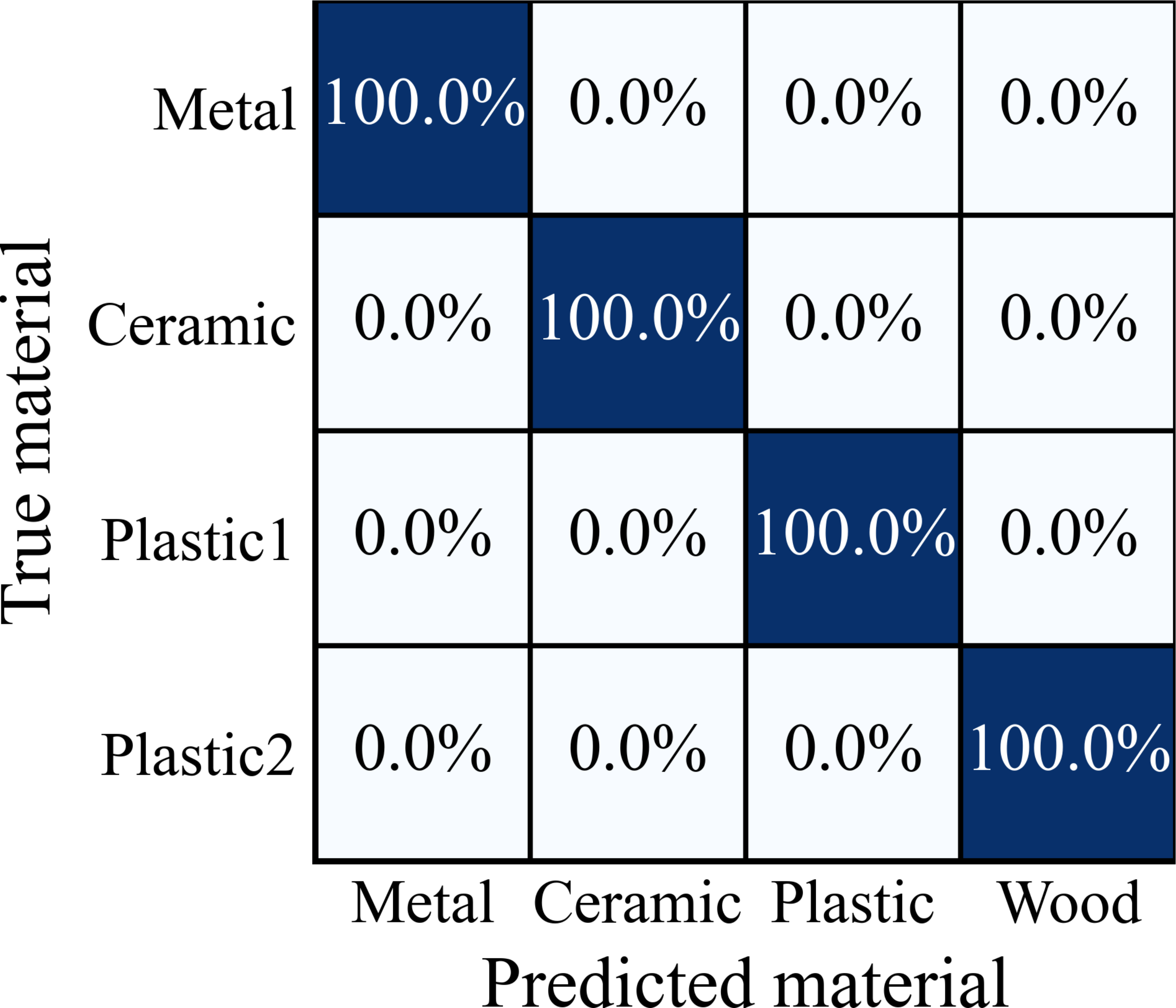}
    \label{mimic4_radar_only}
  }\hfill
  \subfloat[VLMaterial.]{
  \centering
    \includegraphics[width=0.22\textwidth]{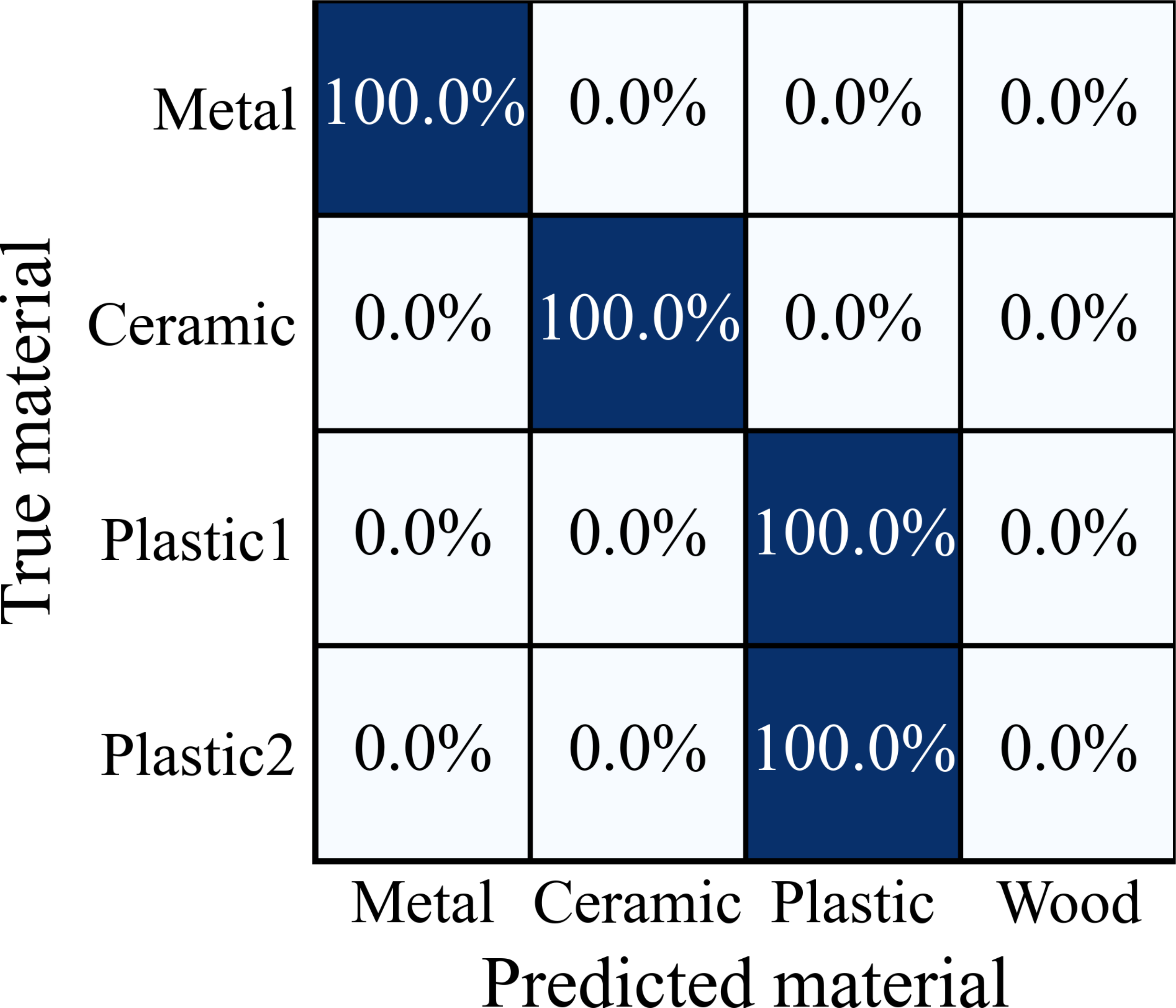}
    \label{mimic4_VLMaterial}
  }
  \caption{Comparison of material recognition results for four counterfeit objects.}
  \label{Comparison_4_mimic_acc}
\end{figure}

\textbf{(1) VLM-only.} Results in \Cref{Comparison_4_mimic_acc}a show the VLM was completely misled by deceptive textures, failing on all counterfeit objects.
\textbf{(2) LLMaterial \& Radar-only (VLMaterial).} Both methods performed well on these objects with distinct dielectric constants, producing identical results. However, plastic 2's smooth surface raised its dielectric constant slightly above typical plastic, causing misclassification as wood (see \Cref{Comparison_4_mimic_acc}b and \Cref{Comparison_4_mimic_acc}c). This indicates that radar alone is insufficient for reliable material identification, as surface variations can influence dielectric measurements.
{\textbf{(3)~VLMaterial.}} With uncertainty-aware adaptive conflict resolution, VLMaterial correctly identified all counterfeit objects (\Cref{Comparison_4_mimic_acc}d), including those that deceived both the visual model and the human eye. This demonstrates VLMaterial's exceptional material identification and error correction capabilities. Crucially, it exhibits robustness against visual deception, penetrating superficial appearances to reveal intrinsic material composition, a capability critical for real-world applications with visual ambiguity, reflective surfaces, or counterfeit materials.

\end{document}